\documentclass[preprint]{aastex}
\usepackage[section] {placeins}
\usepackage{color}

\include{abbrev}
\include{abbr-engl}

\slugcomment{DRAFT: \today}

\shorttitle{ ($^{12}$C + $^{12}$C) and s$-$process.}
\shortauthors{}

\begin{document}

\title{The $^{12}$C + $^{12}$C reaction and the impact on nucleosynthesis in massive stars.}
\author{M. Pignatari\altaffilmark{1,10},
R. Hirschi\altaffilmark{2,9,10},
M. Wiescher\altaffilmark{3,4},
R. Gallino\altaffilmark{5},
M. Bennett\altaffilmark{2,10},
M. Beard\altaffilmark{3,4,10},
C. Fryer\altaffilmark{6,10},
F. Herwig\altaffilmark{4,7,10}
G. Rockefeller\altaffilmark{6,10}
F.X. Timmes\altaffilmark{4,8,10}
 \\
}
\altaffiltext{1}{Department of Physics, University of Basel, Klingelbergstrasse 82, CH-4056 Basel, Switzerland}
\altaffiltext{2}{Keele University, Keele, Staffordshire ST5 5BG, United Kingdom.}
\altaffiltext{3}{Department of Physics, University of Notre Dame, Notre Dame, Indiana 46556, USA.}
\altaffiltext{4}{The Joint Institute for Nuclear Astrophysics, USA.}
\altaffiltext{5}{Universita' di Torino, Torino, Via Pietro Giuria 1, 10126 Italy.}
\altaffiltext{6}{Computational Physics and Methods (CCS-2), LANL, Los Alamos, NM, 87545, USA.}
\altaffiltext{7}{Department of Physics \& Astronomy, University of Victoria, Victoria, BC,
V8P5C2 Canada.}
\altaffiltext{8}{Arizona State University (ASU), School of Earth and Space Exploration (SESE), PO Box 871404, Tempe, AZ, 85287-1404, USA.}
\altaffiltext{9}{Institute for the Physics and Mathematics of the Universe, University of Tokyo, 5-1-5 Kashiwanoha, Kashiwa 277-8583, Japan}
\altaffiltext{10}{NuGrid collaboration}
\email{marco.pignatari@unibas.ch}

\begin{abstract}
Despite much effort in the past decades, the C-burning reaction rate is uncertain
by several orders of magnitude, and the relative strength
between the different channels $^{12}$C($^{12}$C,$\alpha$)$^{20}$Ne,
$^{12}$C($^{12}$C,p)$^{23}$Na and $^{12}$C($^{12}$C,n)$^{23}$Mg is poorly determined.
Additionally, in C-burning conditions a high $^{12}$C+$^{12}$C rate may lead to
lower central C-burning temperatures and to
$^{13}$C($\alpha$,n)$^{16}$O emerging as a more dominant
neutron source than $^{22}$Ne($\alpha$,n)$^{25}$Mg,
increasing significantly the $s$-process production.
This is due to the rapid decrease
of the $^{13}$N($\gamma$,p)$^{12}$C with decreasing temperature, causing
the $^{13}$C production via $^{13}$N($\beta$$^+$)$^{13}$C.

Presented here is the impact
of the $^{12}$C+$^{12}$C reaction uncertainties on the $s$-process and on
explosive $p$-process nucleosynthesis in massive stars, including also fast
rotating massive stars at low metallicity.
Using various $^{12}$C+$^{12}$C rates, in particular an upper and lower rate
limit of $\sim$ 50000 higher and $\sim$ 20 lower than the standard rate
at 5$\times$10$^8$ K, five 25~M$_{\odot}$ stellar models are calculated.
The enhanced $s$-process signature due to $^{13}$C($\alpha$,n)$^{16}$O
activation is considered, taking into account the impact of the
uncertainty of all three C-burning reaction branches. Consequently, we show
that the $p$-process abundances have an average production factor increased
up to about a factor of 8 compared to the standard case, efficiently producing
the elusive Mo and Ru proton-rich isotopes.
We also show that an $s$-process being
driven by $^{13}$C($\alpha$,n)$^{16}$O is a secondary process, even though the
abundance of $^{13}$C does not depend on the initial metal content.
Finally, implications for the Sr-peak elements inventory in the Solar
System and at low metallicity are discussed.

\end{abstract}


\section{Introduction}
\label{sec:intro}

During the evolution of massive stars (initial mass $\gtrsim$ 8 M$_{\odot}$), $^{12}$C fusion occurs in the stellar core and, during
more advanced phases, in convective shells. The lifetime of central carbon burning is governed by the balance between the energy generated by the fusion reaction and the energy lost by neutrinos
\citep[e.g.,][]{arnett:85,eleid:04}.
Carbon fusion has three relevant channels:\\
\newline
$^{12}$C+$^{12}$C $\rightarrow$ $^{24}$Mg* $\rightarrow$ $^{20}$Ne + $^{4}$He (Q = + 4.616 MeV)\\
$^{12}$C+$^{12}$C $\rightarrow$ $^{24}$Mg* $\rightarrow$ $^{23}$Na + p        (Q = + 2.238 MeV)\\
$^{12}$C+$^{12}$C $\rightarrow$ $^{24}$Mg* $\rightarrow$ $^{23}$Mg + n        (Q = - 2.605 MeV) \\
\newline
The first two channels have approximately the same probability, whereas the
probability of the neutron channel is about two orders of magnitude
lower \citep[][]{dayras:77}.
However, it is the uncertainty of the relative strength of these
channels at stellar energies (see next section for details),
where no direct measurements are available,
that is of particular relevance.

Despite considerable experimental efforts, the total $^{12}$C+$^{12}$C fusion
reaction rate remains uncertain at stellar temperatures.
Recent heavy ion fusion systematic studies have indicated that the fusion
cross section may be hindered at low energies \citep[][]{jiang}.
This would result in a lower rate than commonly used
\citep[][hereafter CF88]{caughlan:88}.
On the other hand, recent low-energy experiments \citep[][]{spillane} suggest an
enhancement of the rate due to the presence of resonant structure effects at
lower energies than previously considered \citep[][]{gasques05,jiang,gasques:07}.
This could result in a much higher $^{12}$C+$^{12}$C fusion rate close to the
Gamow peak temperature. Therefore, the present uncertainty of the
$^{12}$C+$^{12}$C rate still covers orders of magnitude at stellar temperatures.

Because both protons and $\alpha$ particles are products of $^{12}$C+$^{12}$C
fusion, C$-$burning is the first phase during massive star evolution where
proton- and $\alpha$-capture reactions can be efficiently activated at the
same time. Left over $^{22}$Ne, from the previous convective He$-$burning core,
efficiently captures $\alpha$ particles via the reaction
$^{22}$Ne($\alpha$,n)$^{25}$Mg, driving neutron capture nucleosynthesis
on the stellar material \citep[e.g,][]{raiteri:91b}.
The $^{22}$Ne($\alpha$,n)$^{25}$Mg reaction has been recognized as the dominant  source of neutrons for the $s$-process in massive stars.
Indeed, the reaction is active
in the convective core He$-$burning and subsequent C$-$burning phases, and
is responsible for the majority of the $s$-process species between iron and
strontium observed in the Solar System
\citep[][]{peters:68,lamb:77,couch:74,kaeppeler:89,prantzos:90,raiteri:93,the:07,pignatari:10}.
A large fraction of pre-supernova material is affected by
convective carbon shell nucleosynthesis. On the other hand,
material processed by C$-$burning in the core is usually
modified and processed by later evolutionary burning phases
\citep[e.g.,][]{woosley:95,limongi:00,rauscher:02,woosley:02}.
This scenario may be modified however by uncertainties in the
$^{12}$C+$^{12}$C reaction rate.

The impact of the $^{12}$C+$^{12}$C uncertainty in massive star simulations has
been studied previously, by e.g., \cite{gasques:07}, where the consequences
arising from a lower rate were considered, and by \cite{bennett:12}
(hereafter paper I), where a higher rate was used for stars of various initial masses at solar metallicity. The main results obtained in paper I
are the following:
$i$)~For all stellar masses considered ($M$ = 15-60 M$_{\odot}$) using the
upper limit $^{12}$C+$^{12}$C rate (CU rate, see also the next
sections) results in $^{13}$C($\alpha$,n)$^{16}$O becoming the predominant
 neutron source, compared to the established source $^{22}$Ne($\alpha$,n)$^{25}$Mg,
and increases the $s$-process stellar yields by up to $\sim$ 2 orders of magnitude.
$ii$)~When an intermediate rate (CI, paper I) is used, $s$-process rich material
 from the C core is only ejected for the 20 M$_{\odot}$ mass. For this one case,
 the stellar model shows similar results to that of the 20 M$_{\odot}$ star
using the CU rate, whereas all other models show results comparable to the
standard case using the CF88 $^{12}$C+$^{12}$C rate. $iii$)~For all of the
 masses considered, the lifetime of central C$-$burning
increases with increasing the $^{12}$C+$^{12}$C rate. The density and
temperature of C$-$burning decreases with increasing $^{12}$C+$^{12}$C rate.
No significant variation trends are observed for more
advanced phases. $iv$)~Point $i$) suggests that the CU rate is likely too high,
since it would be hard to reconcile the predicted yields with observations,
such as the solar $s$-process distribution.

In this paper, for one 25 M$_{\odot}$ star we consider the complete range of the present $^{12}$C+$^{12}$C uncertainty, in agreement with
\cite{gasques:07} and paper I. We discuss the importance of the rate uncertainty on the $s$-process and on explosive $p$-process yields, which are driven by photodisintegration on $s$-process rich material during core
collapse \citep[e.g.,][and reference therein]{arnould:03}.
We consider the implications for the $s$-process arising
from the large uncertainty regarding the relative strengths of the three carbon fusion channels, an effect which was not included in the two works
mentioned above. The impact of the neutron source
$^{13}$C($\alpha$,n)$^{16}$O is also considered for non-rotating and fast-rotating massive stars, with halo metallicities. Finally, the production of unstable long-lived species $^{26}$Al and $^{60}$Fe in the pre-explosive phase is revised within the $^{12}$C+$^{12}$C uncertainty
\citep[e.g.,][]{arnett:69,timmes:95,limongi:06}.\\
The paper is organized as follows:
the present uncertainty of the $^{12}$C+$^{12}$C reaction rate, and its
nucleosynthesis channels, is discussed in \S \ref{sec:c12c12status}.
Presented in \S \ref{sec:gnv} are the main features of the stellar models used
 in
this work. In \S \ref{sec:he core} and \S \ref{sec:c12c12sprocess} the
$s$-process in the convective He$-$burning core and in C$-$burning conditions is
described. The impact on nucleosynthesis of different $^{12}$C+$^{12}$C
channels is discussed in \S \ref{sec:c12c12a and c12c12p} and
\S \ref{sec:c12c12n}.
The $p$-process nucleosynthesis is analyzed in \S \ref{sec:pprocess}, and in
\S \ref{sec:low_metallcity} the importance of the $^{12}$C+$^{12}$C uncertainty is
explored at low metallicity, for non-rotating and fast-rotating stars.
Finally, conclusions are
presented in \S \ref{sec:concl}.

\section{($^{12}$C + $^{12}$C) reaction}
\label{sec:c12c12status}

The $^{12}$C+$^{12}$C reaction is one of the most studied heavy
ion fusion processes. Considerable effort has been expended over the last
decades on the measurement of the $^{12}$C+$^{12}$C fusion cross
section at very low energies
\citep[][]{Patterson,Mazarakis,High,Kettner,Becker,Eli}.

Despite these dedicated efforts, the low energy reaction rate still carries
considerable uncertainties due to pronounced resonance structures which are
thought to be associated with molecular $^{12}$C+$^{12}$C configurations in the
$^{24}$Mg$^*$ system \citep[][]{bromley}. While this observation is significant
for the understanding and interpretation of nuclear structure
configurations \citep[][]{freer}, it may also have a significant impact on
the actual reaction rate of $^{12}$C+$^{12}$C fusion in stellar
burning environments. The stellar reaction rate depends critically on the
S-factor in the stellar energy regime, which so
far has been inaccessible to laboratory studies.
The low, sub-picobarn cross sections and the substantial
beam induced background events on target impurities make
the identification and measurement of the reaction
products difficult \citep[][]{strieder}, and generally the expanding of the measurement of
the $^{12}$C+$^{12}$C fusion cross section to very low energies \citep[][]{Zickefoose}.
In particular, the resonance configurations are characterized
by pronounced structures associated with the $^{24}$Mg$^*$ compound
nucleus in the corresponding excitation range, and
should be strongly populated in the
fusion process except for the hindrance by Coulomb and orbital
momentum barrier.
The experimental data indicate that these
resonances have relatively narrow total widths, $\Gamma\approx$100 keV.
Since these resonance levels are highly unbound with
respect to $\alpha$ and proton decay, the $\alpha$ cluster and proton
single particle component in the wave function is relatively small,
corresponding to spectroscopic factors of $\le$ 0.01 for both
decay channels.
This is also reflected in the results of elastic
$^{20}$Ne($\alpha$,$\alpha$) \citep[][]{Abegg,Davis} and $^{23}$Na(p,p)
\citep[][]{Mitchell} scattering experiments, which indicate no
pronounced $\alpha$ or single particle structure in this particular
excitation range of $^{24}$Mg. Despite considerable theoretical
effort, the exact nature and structure of these molecular resonance
levels is not fully understood and low energy reaction cross
section extrapolations are mostly based on averaging the cross
section over the resonance structure \citep[][]{FCZ75}. A recent
potential model analysis of the fusion cross
section \citep[][]{gasques05} was based on the Sao Paulo model and resulted in a
theoretical prediction of the cross section, describing well
the experimental data and allowing extrapolation to lower energies.
The results were in good
agreement with the previous phenomenological approach
by \cite{FCZ75}.

More recently, however, it has been argued on the basis of heavy ion
fusion systematics that the low energy cross section of fusion
reactions declines faster with decreasing energy than projected by
the potential model. This has been modeled by introducing an
additional term in the potential which is related to the
incompressibility of nuclear matter \citep[][]{esbensen}. This term
translates into a hindrance effect for the fusion process
suggesting a significant reduction of the overall cross section
towards low energies. This effect has been confirmed by a number of low
energy fusion reactions with positive Q-values \citep[][]{janssens}, and
it is argued that the hindrance effect may also reduce the low
energy S-factor of the $^{12}$C+$^{12}$C fusion process
significantly \citep[][]{jiang}. The impact of such a reduction
on a number of stellar burning scenarios was
discussed in a previous study \citep[][]{gasques:07}.

On the other hand, recent low energy experiments \citep[][]{spillane}
indicate the existence of
strong molecular resonances at even lower energies
than previously considered \citep[][]{gasques05,jiang,gasques:07}. These
resonances may contribute significantly to the low energy cross
section and enhance the reaction rate substantially. One
pronounced $^{12}$C+$^{12}$C resonance structure has been
reported in both the proton and the $\alpha$ channel, at
$E_{cm}$=2.138~MeV \citep[][]{spillane}, and another resonance at
$E_{cm}\approx$1.5~MeV has been projected on the basis of a theoretical
evaluation of recent $^{12}$C+$^{12}$C scattering data, which seems
to indicate the existence of a pronounced resonance also at 1.5~MeV \citep[][]{Perez}.
The total strength of the resonance at 2.14~MeV
is projected to be $\omega\gamma\le$0.13~meV. This is remarkably large,
and translates into a pronounced $^{12}$C+$^{12}$C cluster
configuration with a spectroscopic factor of
C$^2$S$_{\bf ^{12}C}$ $\approx$ 1.\footnote{During a nuclear reaction, a new nucleus is formed with the nucleons occupying a given configuration (which defines also spin and parity). There are two ways of measuring the likelihood of a configuration, one where the isospin of the nucleons is neglected, and one where it is included.
C$^2$S$_{\bf ^{12}C}$ is the notation for the $^{12}$C+$^{12}$C spectroscopic factor, including the isospins of the nucleons. The fact that it is nearly 1 means that this is a very likely configuration.}
Adopting this value for the resonance
at 1.5~MeV, we obtain an upper limit for the total resonance
strength of $\omega\gamma\le$0.01~$\mu$eV. A thorough search was
performed to identify this resonance at low $^{12}$C+$^{12}$C in
the proton channel $^{12}$C($^{12}$C,p)$^{23}$Na.
Unfortunately the measurements were affected by beam induced proton background
from the resonant $^{12}$C(d,p)$^{13}$C reaction,
triggered by deuteron recoils from elastic
$^{12}$C beam scattering on low level deuteron impurities in the
$^{12}$C target \citep[][]{Zickefoose}. The yield of this two-step
background process presently sets the upper limit for the resonance
strength of $^{12}$C($^{12}$C,p)$^{23}$Na below E$_{cm}$=2.5~MeV. Recent
unpublished experimental data with
ultra-pure $^{12}$C targets and greatly reduced deuterium contamination,
did not show any indications for lower energy resonances \citep[][]{Zickefoose2}.
While this background also affects the proposed
strong $^{12}$C($^{12}$C,p)$^{23}$Na resonance at
E$_{cm}$=2.138 MeV~\citep[][]{spillane}, the possibility of a
$^{12}$C+$^{12}$C cluster resonance reflected in
the $^{12}$C($^{12}$C,$\alpha$)$^{20}$Ne reaction channel cannot be excluded,
since the $\alpha$ channel was not observed in this particular experiment.


Complimentary information may come from $\alpha$-scattering,
$\alpha$-capture and $\alpha$-emission data because a
$^{12}$C cluster configuration may also show an enhanced $\alpha$-cluster
signature. Indeed, the $^{20}$Ne($\alpha$,$\alpha_0$)
elastic scattering data by \cite{Abegg} indicate a number of strong
natural parity states in the excitation range of interest. No
radiative capture data are available for this energy range
\citep[][]{En91}, but several levels have been observed in the
$^{23}$Na(p,$\alpha$)$^{20}$Ne reaction. Overall, the
complimentary information is sparse, since not many experiments
have systematically probed the level structure of $^{24}$Mg above 15~MeV.

Evaluating the available information, most notable are the two
states at 15.44 MeV (J$^{\pi}=0^+$, T=2), and 16.07~MeV
(J$^{\pi}=6^+$, T=0), which match the energies of the two postulated
low energy resonances in the $^{12}$C+$^{12}$C channel at
$E_{cm}$=1.5~MeV \citep[][]{Perez} and $E_{cm}$=2.14~MeV \citep[][]{spillane}.
The 15.44~MeV level has been observed
in a number of scattering and reaction studies, and was identified
as a T=2 isospin state. The total width, as well as the proton and
$\alpha$ partial widths, are known or can be deduced from the
available experimental information such as $^{23}$Na($p,\alpha$)
and $^{23}$Na(p,p) \citep[][]{McD78, En91}.

The critical parameters that need to be defined
in order to determine the resonance strengths in the fusion
process are the $^{12}$C partial widths for these levels.
The level at 16.07~MeV has only been observed in high
spin transfer reactions \citep[][]{FCR74}, and the proton and $\alpha$
partial widths can not be directly evaluated.
If these two levels correspond to the observed $^{12}$C+$^{12}$C
resonance states, it would require a pronounced $^{12}$C+$^{12}$C
cluster configuration for both levels. Some information can be
extracted on the basis of this assumption. The population of a
pure T=2 level is isospin forbidden; it is only possible with
strong isospin mixing. In the case of the 15.44~MeV state, the
existence of isospin mixing is confirmed by the observation of the
$\alpha$ emission \citep[][]{McD78} and the $\alpha$ decay of this
level \citep[][]{McG70}. The isospin mixing can actually be deduced
from the strength of the $\alpha$ decay branch to be $\approx$0.01,
adopting a pronounced $^{20}$Ne+$^4$He cluster configuration. This
result is relatively insensitive to the choice of the radius of
the system. On the other hand, a pronounced $^{12}$C+$^{12}$C
cluster structure is associated with a large interaction radius of
$R$=8.5~fm.
The $^{12}$C partial width of the 16.07~MeV 6$^+$ state is limited
by the $\ell$=6 orbital momentum barrier. This level has not been
observed in any of the proton or $\alpha$ induced scattering or
reaction processes. The available information about the proton and
$\alpha$ partial widths rely on the recent measurement of the
resonance strengths for both branches in the  $^{12}$C+$^{12}$C
fusion reaction \citep[][]{spillane}. The measured resonance strengths
are consistent with the large interaction radius of $R$=8.5~fm,
anticipated for a cluster configuration of this level.

On the basis of these considerations we have re-evaluated the reaction
rate for the $^{12}$C+$^{12}$C fusion reaction as well as the
rates for the two most important reactions channels
$^{12}$C($^{12}$C,$\alpha$)$^{20}$Ne and
$^{12}$C($^{12}$C,$p$)$^{23}$Na.
While these two channels are responsible for the energy production
by carbon fusion and for the production of seed material for the next
Ne-burning stage, the third channel $^{12}$C($^{12}$C,$n$)$^{23}$Mg
may contribute as neutron source to $s$-process nucleosynthesis at higher
temperature environments. This possibility will be discussed in \S \ref{sec:c12c12n}.
Fig. \ref{fig:c12c12_rate_and_ratio} shows the total $^{12}$C+$^{12}$C rate
with the evaluated uncertainties. Notice that in the figure
the neutron channel is neglected; the reaction is endothermic with a negative
Q-value of -2.6 MeV and contributes less than 4\% at higher energies
to the total cross section.
The ``lower limit'' (CL) of the rate
corresponds to the one suggested on the basis of an extrapolation
of the averaged data, using a potential model approach which includes a
hindrance term for low energy fusion processes \citep[][]{gasques:07}.
The recommended rate is based on the classical extrapolation of
the averaged S-factor data using a standard potential model
\citep[][]{gasques05} plus the contribution of a single resonance
observed at $E_{cm}$=2.13~MeV, and is quite similar to the standard
CF88 rate. The ``upper limit'' (CU) includes an additional term resulting from
a possible strong
$^{12}$C+$^{12}$C cluster resonance at $E_{cm}$=1.5~MeV. The resonance parameters
have been estimated as outlined
above. The total rates are listed in
Table \ref{tab: c12c12rate_and_channels}.

\section{Stellar Models}
\label{sec:gnv}

In order to assess the importance of the $^{12}$C+$^{12}$C rate for the
$s$-process in massive stars, five 25 M$_{\odot}$ stellar models were calculated
using different $^{12}$C+$^{12}$C rates.
Apart from the $^{12}$C+$^{12}$C rate, all of the models were calculated with identical
 input physics, and were therefore identical at the end of He$-$burning.
Calculations were performed using the Geneva stellar evolution code (GENEC),
 described in \citet{hirschi:04} and \citet{eggenberger:08}. We recall here the main input physics used.

The initial metal content was $Z$ = 0.01,
where the solar abundances were given by
\cite{anders:89}
 for the total elemental abundances, and by
\cite{lodders:03} for the relative isotopic abundances. The corresponding OPAL opacity tables are taken from \citet{rogers:96}, except at low temperatures, where instead opacities from \citet{ferguson:05} were used.
The convective boundaries were defined according to the Schwarzschild criterion. Overshooting was applied for core H$-$ and core He$-$burning only, using an overshooting parameter, $\alpha= 0.2$\,H$_{\rm P}$, as in \cite{maeder:92}.
Effects from rotation and magnetic fields were not included in the calculation,
 but this does not detract from the main focus of the present investigation.
For $\log T_{\rm eff} > 3.9$, O--type star mass loss rates were adopted from \citet{vink:01}, and from \citet{dejager:88} otherwise. With a few exceptions,
 reaction rates were taken from the NACRE compilation \citep{angulo:99}. For
temperatures below 0.1~GK, the $^{14}$N(p ,$\gamma$)$^{15}$O rate is taken from
\citet{mukhamedzhanov:03}. Above 0.1~GK, the lower limit NACRE rate was used.
This combined rate is very similar to the more recent LUNA rate \citep{imbriani:05} which is used in the post-processing at relevant temperatures. The $3\alpha$ rate
adopted is from \cite{fynbo:05}, and the $^{12}$C($\alpha, \gamma$)$^{16}$O rate from \cite{kunz:02}. For the $^{22}$Ne($\alpha$, n)$^{25}$Mg reaction we
used the rate of \citet{jaeger:01} for temperatures of 1~GK and below. The
NACRE rate was used for higher temperatures. The $^{22}$Ne($\alpha$, n)$^{25}$Mg rate competes with $^{22}$Ne($\alpha, \gamma$)$^{26}$Mg, where the NACRE rate is used.



\begin{figure}
\centering
\resizebox{12cm}{!}{\rotatebox{0}{\includegraphics{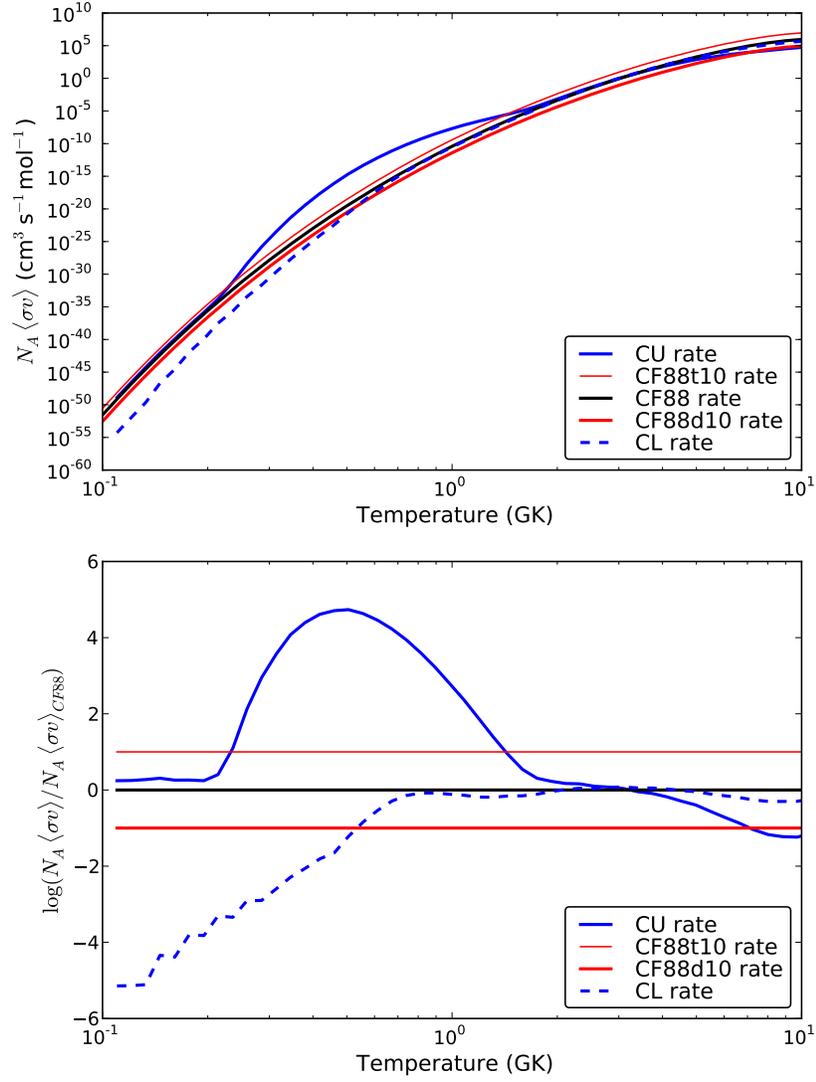}}}
\caption{$Top$ $panel$ $-$ The $^{12}$C+$^{12}$C reaction rate used in this study. The CU rate and CF88 rate
are the same used in paper I \citep[][]{bennett:12}.
$Bottom$ $panel$ $-$ The rates are shown normalized to the standard case (CF88).}
\label{fig:c12c12_rate_and_ratio}
\end{figure}

\begin{figure}
\centering
\resizebox{6.5cm}{!}{\rotatebox{0}{\includegraphics{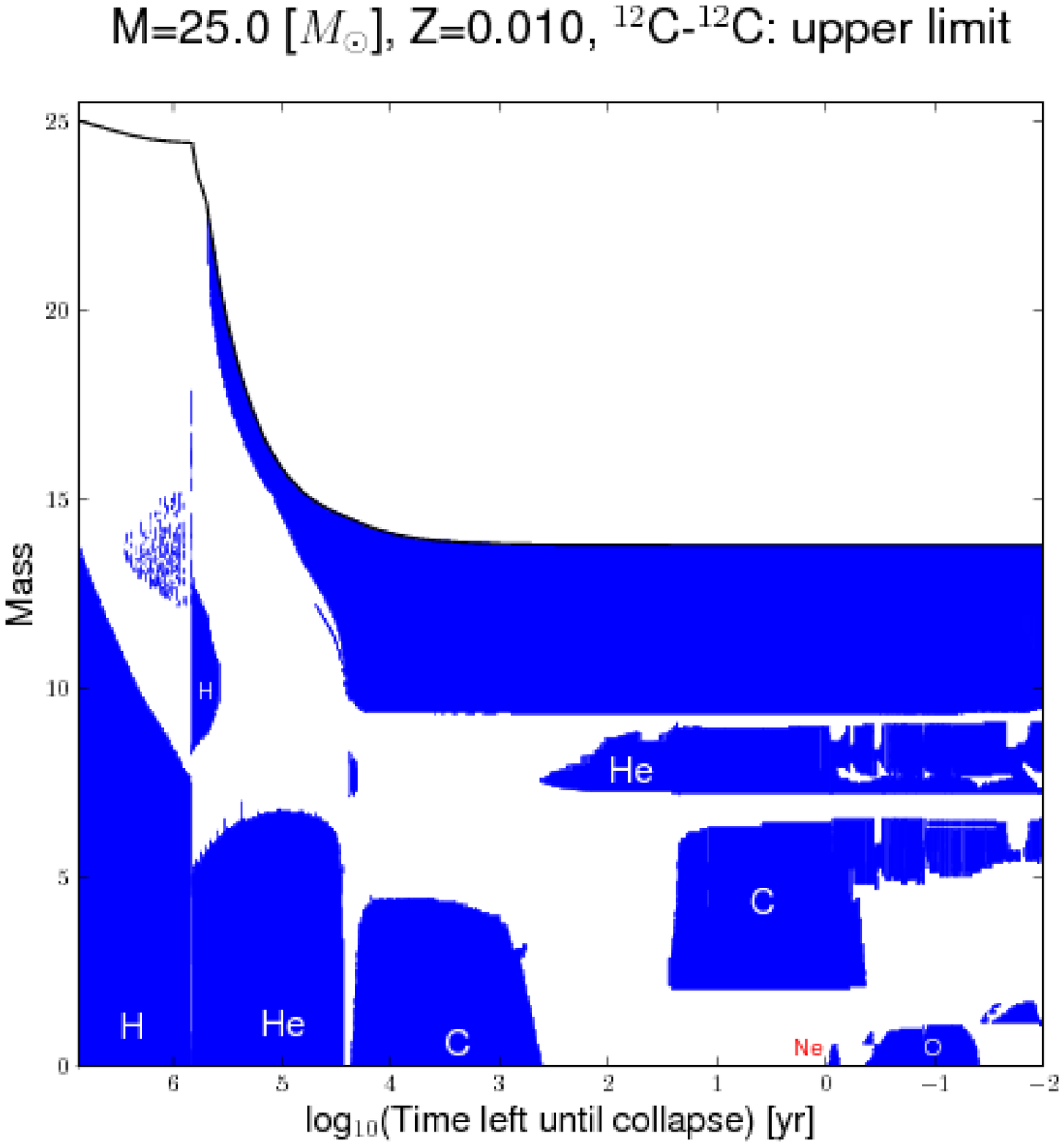}}}
\resizebox{6.5cm}{!}{\rotatebox{0}{\includegraphics{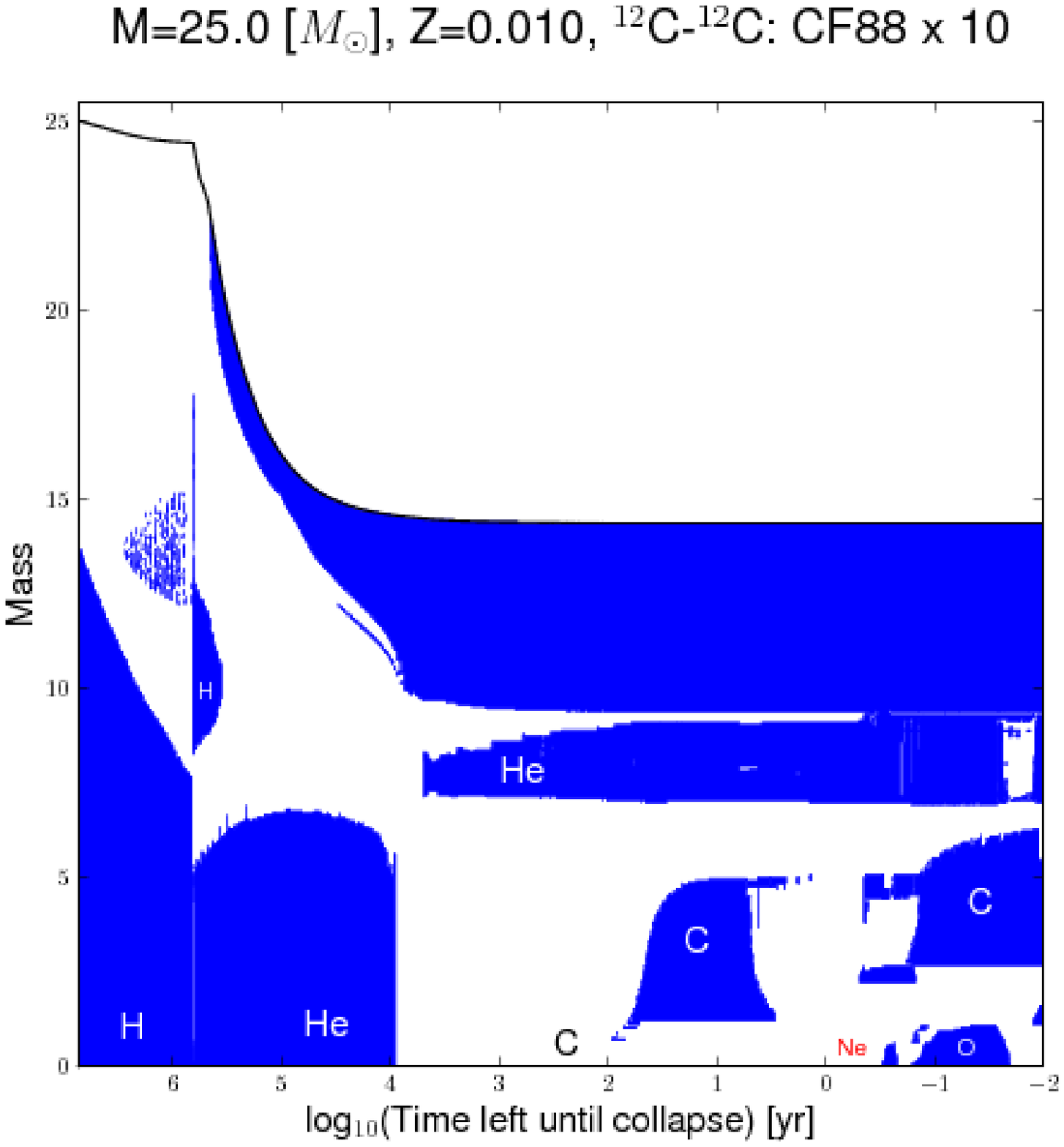}}}
\centering
\resizebox{6.5cm}{!}{\rotatebox{0}{\includegraphics{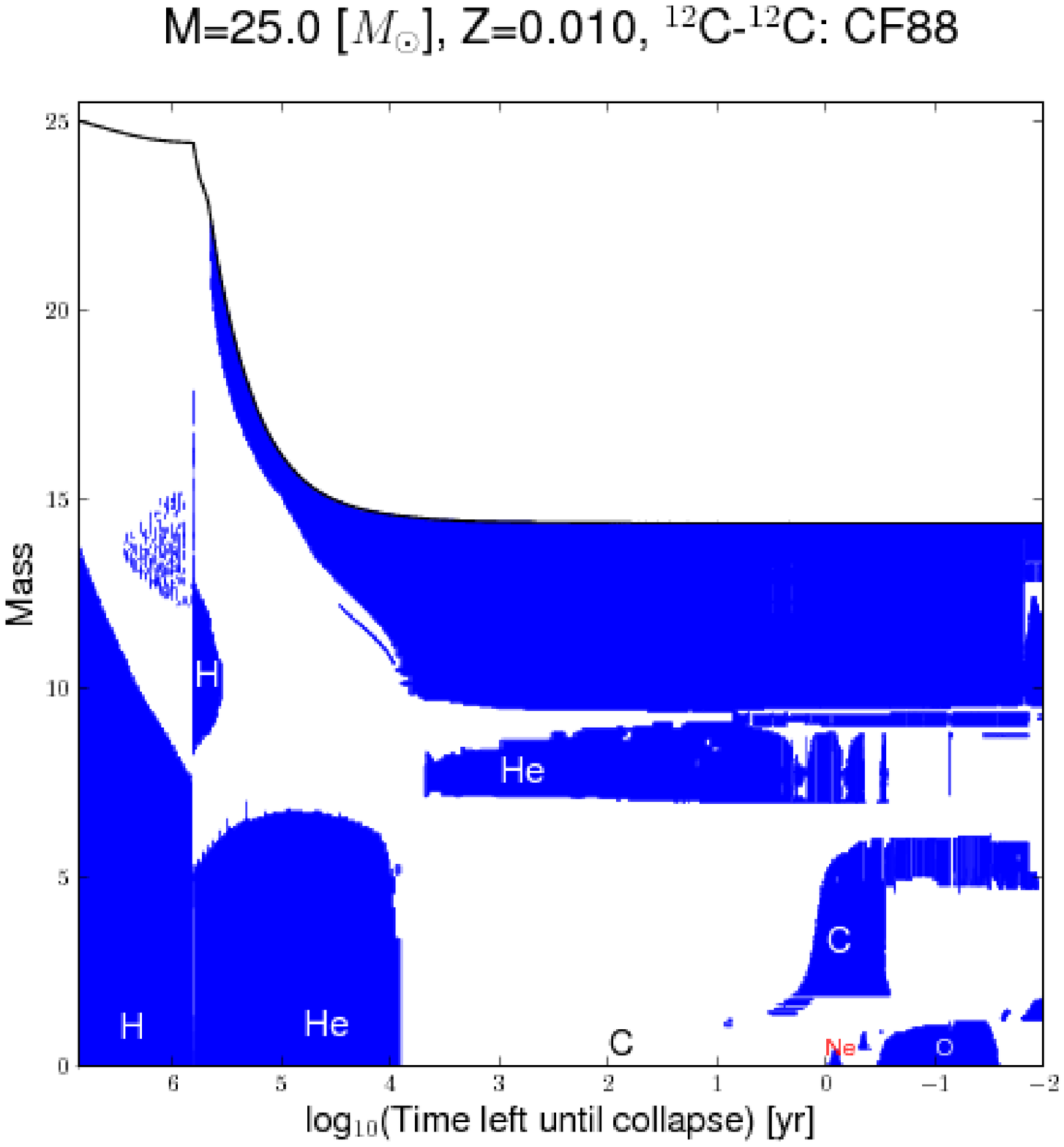}}}
\hfill

\resizebox{6.5cm}{!}{\rotatebox{0}{\includegraphics{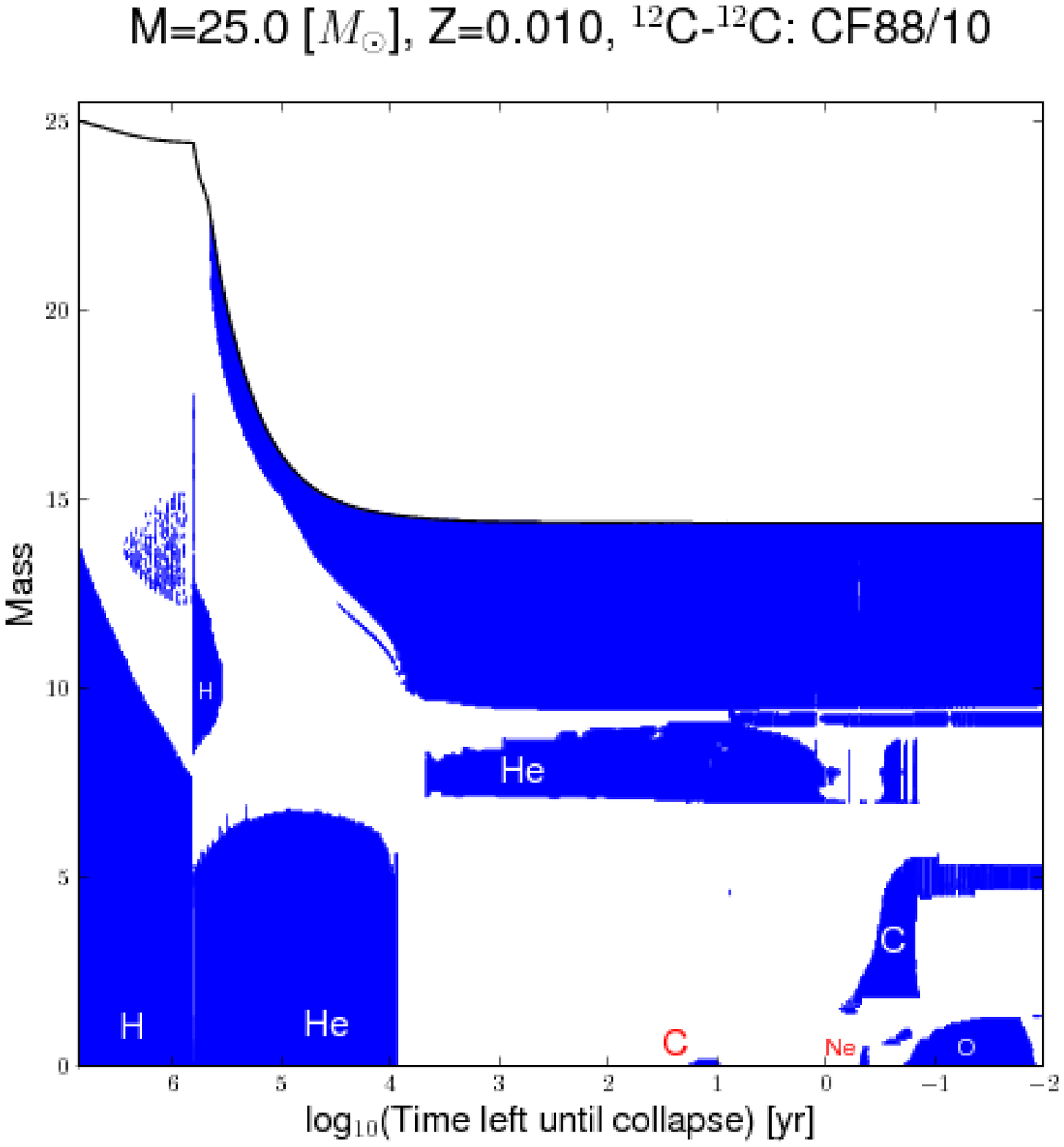}}}
\resizebox{6.5cm}{!}{\rotatebox{0}{\includegraphics{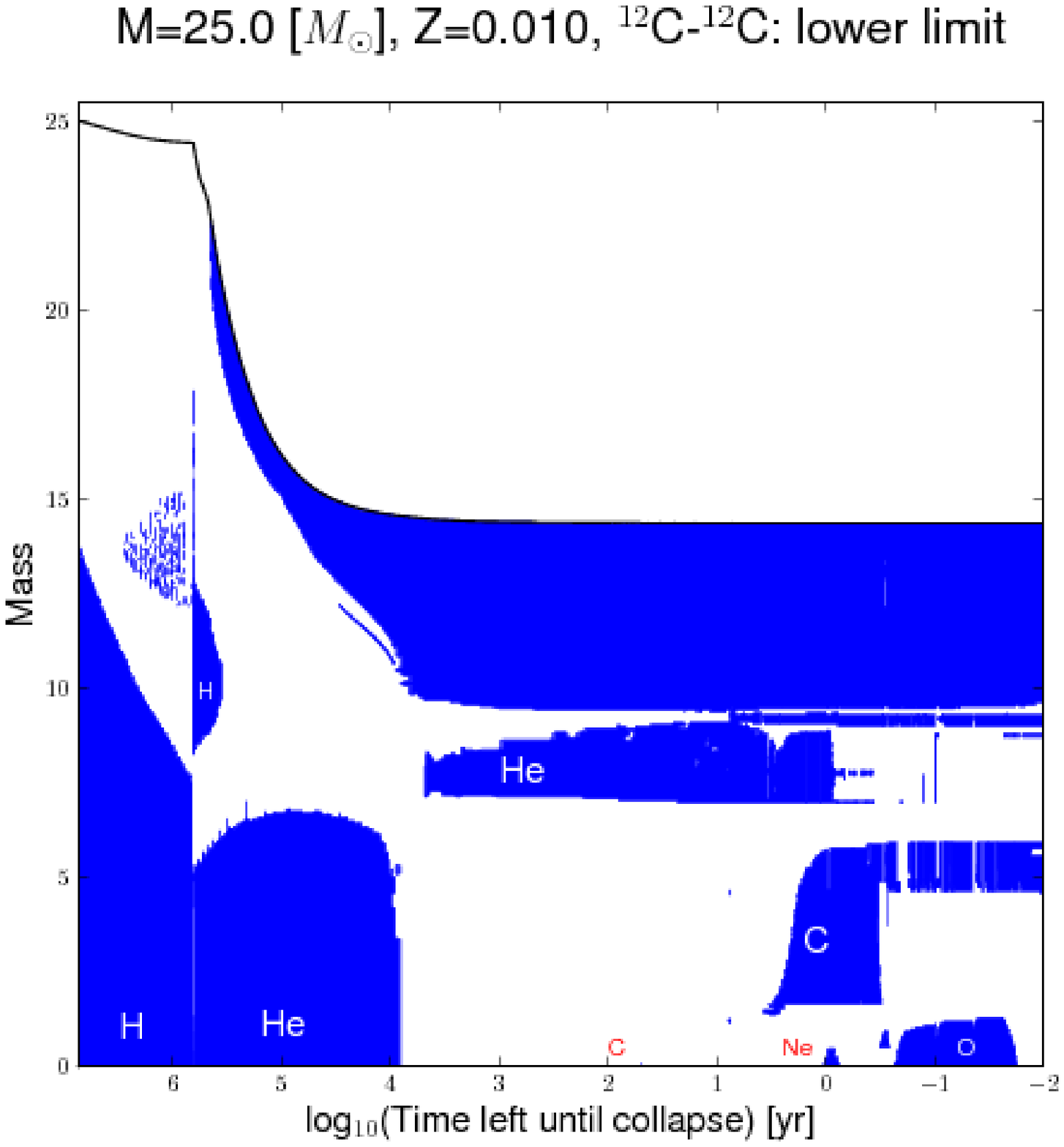}}}
\caption{The Kippenhahn diagram is provided for five stellar models of a
25 M$_{\odot}$ star, and $Z$ = 0.01, calculated
using different $^{12}$C + $^{12}$C rates (see the text for more details):
the upper limit rate (CU, Upper left Panel),
the \cite{caughlan:88} rate (CF88, Central Panel), multiplied and divided
by a factor of ten (CF88t10 and CF88d10, Upper Right Panel and Lower Left Panel, rispectively),
and the lower limit rate (CL, Lower Right Panel).
}
\label{k_c12c12}
\end{figure}

In Fig. \ref{k_c12c12}, we
present the Kippenhahn evolution diagram for the five stellar models, where the
$^{12}$C+$^{12}$C rate used in the calculations are, respectively:
$i$) The upper limit provided in \S \ref{sec:c12c12status} (M25CU model, Panel $a$),
$ii$) The CF88 rate multiplied by a factor of 10 (M25CF88t10 model, Panel $b$),
$iii$) The CF88 rate (M25CF88 model, Panel $c$), $iv$) The CF88 rate divided
by a factor of 10 (M25CF88d10 model, Panel $d$), $v$) The lower limit provided in
\S \ref{sec:c12c12status} (M25CL model, Panel $e$). The rates used for the different cases are shown
in Fig. \ref{fig:c12c12_rate_and_ratio}.

\begin{figure}
\centering
\resizebox{12cm}{!}{\rotatebox{0}{\includegraphics{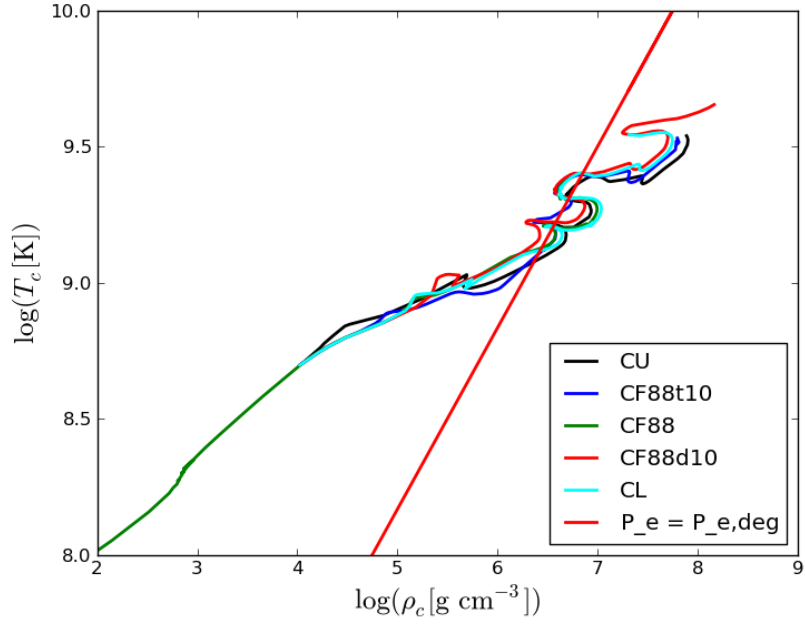}}}
\caption{The evolution of central temperature versus central density for the models considered in this work. The red straight line identifies the limit between a non-degenerate and a degenerate
electron gas, $P_{\rm gas}$ = $P_{\rm e,deg}$.}
\label{fig:tc_rhoc}
\end{figure}

All models were evolved beyond the end of central O$-$burning.
The central temperature and density evolution are shown in Fig. \ref{fig:tc_rhoc}.
As mentioned, the evolutionary curve in Fig. \ref{fig:tc_rhoc} is the same for all the models until central He exhaustion.
Models calculated using the higher $^{12}$C+$^{12}$C rates generally show the signature of central C ignition at lower temperature and density conditions (upward kink in the curve in the range log($\rho_{\rm c}$)= 4-6\,g\,cm$^{-3}$).
After central C$-$burning, no significant variations may be noticed in the $T_{\rm c}$-$\rho_{\rm c}$ diagram in the following evolutionary phases (see below for more
details), as already noticed in paper I.
The main parameters of the stellar
models presented in Fig. \ref{k_c12c12} and \ref{fig:tc_rhoc}
are summarized in Tables \ref{tab:modeldata_1} and
\ref{tab:modeldata_2}.
Similar data are provided for a grid of masses and at solar metallicity
in paper I, that is similar to the case discussed here at
lower metallicity.
However, the range of $^{12}$C+$^{12}$C rate explored in this work
is larger than in paper I, including the possibility of a rate
lower than CF88. At $T$ $\sim$ 0.7-0.8 GK the CL rate varies from the CF88 rate by
less than a factor of 10. However there is
a factor of 10 difference between CF88d10 and CF88 over the same
temperature range (Fig. \ref{fig:c12c12_rate_and_ratio}). Therefore because the CL rate rapidly increases with
temperature, CF88d10 is actually the lowest carbon
fusion rate we consider during central C$-$burning conditions.
Notice that the CF88d10 and CF88t10 rates are given by applying a
temperature invariant correction factor to the CF88 rate over the entire
temperature range. Though such constant variation is unlikely, the present
models calculated with these rates are still useful as a guide, and can provide
qualitative
insights about the impact of the $^{12}$C+$^{12}$C rate on stellar evolution,
including later stages.

\begin{figure}
\centering
\resizebox{12cm}{!}{\rotatebox{0}{\includegraphics{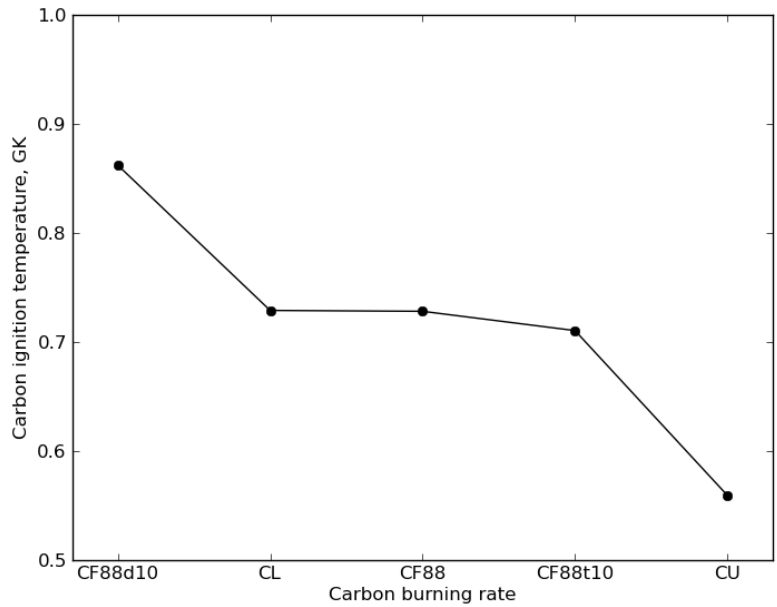}}}
\caption{The central carbon ignition temperature is shown for each stellar model,
according to the $^{12}$C + $^{12}$C rate used. Notice that the case CF88d10 shows a central C ignition at higher temperature than the case CL.
This is expected, since the difference between the standard rate and the CL rate drops quickly below a factor of 10 with increasing temperature
(see Fig. \ref{fig:c12c12_rate_and_ratio}).
}
\label{fig:ignition_temp}
\end{figure}

\begin{figure}
\centering
\resizebox{12cm}{!}{\rotatebox{0}{\includegraphics{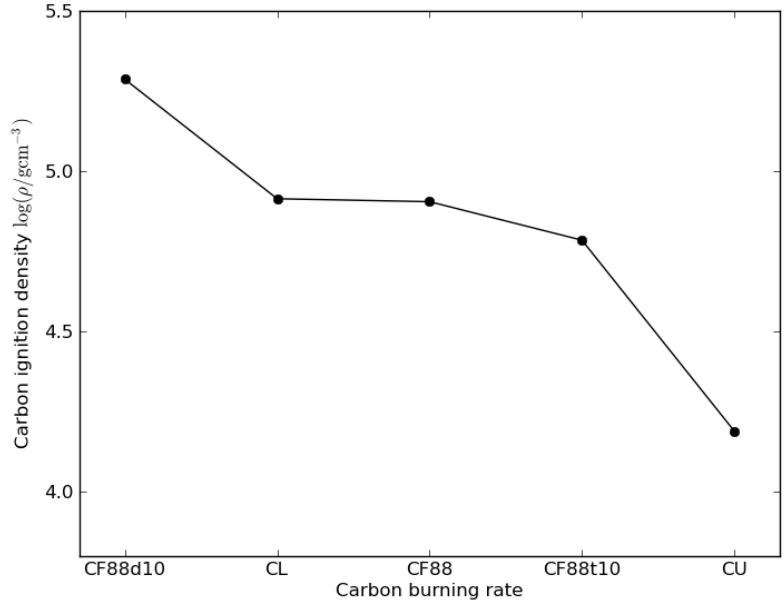}}}
\caption{As Fig. \ref{fig:ignition_temp}, but for the central density.}
\label{fig:ignition_rho}
\end{figure}

\begin{figure}
\centering
\resizebox{12cm}{!}{\rotatebox{0}{\includegraphics{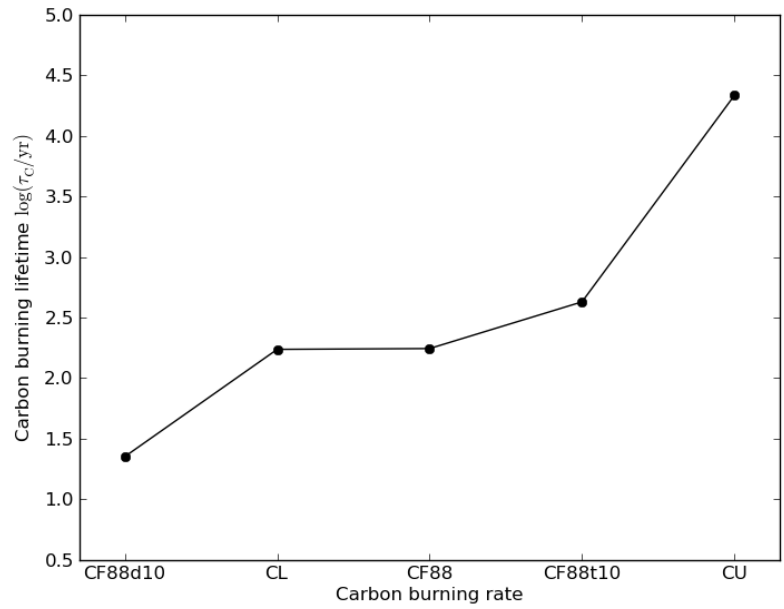}}}
\caption{As Fig. \ref{fig:ignition_temp}, but for for core C-burning lifetimes.}
\label{fig:lifetimes}
\end{figure}

The time elapsed between He exhaustion and core C$-$burning activation
decreases with increasing $^{12}$C+$^{12}$C rate.
However, the early C ignition is compensated by a longer core burning phase
(see Fig. \ref{fig:lifetimes} and Table \ref{tab:modeldata_2}).
In Figgs. \ref{fig:ignition_temp}, \ref{fig:ignition_rho}, and \ref{fig:lifetimes} the central temperature and density at C ignition
and C$-$burning lifetimes\footnote{The central C-burning lifetime is calculated as follow: the C fusion is assumed to start once central
$^{12}$C from the He-core ashes is decreased by 3\%, and is assumed to finish once central $^{12}$C mass fraction is less than 0.001.} are compared.
The trend observed in paper I is
confirmed also for $^{12}$C+$^{12}$C rates lower
than CF88: the temperature and density of central C ignition decreases with
increasing carbon fusion rate, whereas the
C$-$burning lifetime shows the opposite behavior.
The M25CL and M25CF88 models show similar conditions for central C burning. Indeed,
at the point where central carbon ignition occurs in these models, the CL rate
and the CF88 rate are quite similar, despite the fact that for lower
temperatures the CL rate drops quickly compared to the standard rate.
On the other hand, the M25CF88d10 and M25CF88t10 models show significantly different
behaviors; reducing the $^{12}$C+$^{12}$C rate has a larger impact, compared to the CF88 case.
Finally, because of the high upper limit, the M25CU model
(which is about 50,000 larger than the CF88 rate, Fig. \ref{fig:c12c12_rate_and_ratio})
shows a large effect in the stellar conditions
(see Table \ref{tab:modeldata_1} and \ref{tab:modeldata_2}),
in agreement with calculations presented in paper I.

Notice that the behavior of C-burning temperature with the
$^{12}$C+$^{12}$C rate can be simply derived analytically, if we consider
the temperature dependence of the C-burning rate, where
$\lambda$$_{^{12}{\rm C}^{12}{\rm C}}$ $\propto$ T9$^{29}$
\citep[see][where T9 is temperature in GK units]{woosley:02}. Indeed,
if we simply consider that the amount of energy required for the
stellar structure is the same, regardless of the C-burning rate,
we obtain for the upper limit a burning temperature of T9 = 0.50,
and for the model M25CF88d10 T9 = 0.79. These values are only about
10\% lower than the activation
temperatures showed in Table \ref{tab:modeldata_1}.
Such small differences are
likely due to minor effects considered in the full stellar model, as
for instance the neutrino feedback.

Fig. \ref{k_c12c12} shows that carbon core burning is radiative
 for the models M25CL, M25CF88 and M25CF88t10. For the lowest rate, CF88d10, core
C$-$burning induces a tiny convective core. The CU case, on the other hand,
shows an extended convective core (up to 4.12~M$_{\odot}$ in size).
Generally, the standard 25~M$_\odot$ model is close to the lower mass limit for
radiative core carbon burning, which is around 22~M$_\odot$ \citep{hirschi:04}.
Core C$-$burning becomes convective once the energy produced by $^{12}$C+$^{12}$C is higher than the energy lost
by neutrinos \citep[][]{eleid:04}. In the CF88d10 case, the lower rate leads to
 a delayed C ignition at higher temperatures, where the energy generation
slightly overtakes the neutrino losses.
In the M25CU model, the formation of an extended convective core is due to the
early activation of C$-$burning, where the lower temperature makes
the neutrino energy loss much less efficient \citep[][see also discussion in paper I]{itoh:96}.

After central carbon exhaustion,  C$-$burning develops in outer
convective shells. Models M25CF88d10, M25CL, M25CF88 and M25CU show one
extended convective shell, whereas M25CF88t10 develops 2 convective shells. In this last
case, the first convective shell overlaps with the second and
final convective shell by about 2/3 of its total extent. In a similar way,
the M25CU model shows a similar overlap between the central
convective C core and the subsequent convective C shell. Such overlap has a
strong impact on the composition of the final stellar ejecta.
Looking at Table \ref{tab:modeldata_2}, we can see that contrary to carbon burning,
all stellar models are quite similar during the subsequent burning stages
(neon and oxygen burning stages, see also paper I).
%

As mentioned before, the $^{12}$C+$^{12}$C rate is the only reaction
relevant for energy generation tested in this work. Notice however that
the uncertainty of other reactions such as
$^{12}$C($\alpha$,$\gamma$)$^{16}$O
\citep[e.g.,][]{imbriani:01,eleid:04,woosley:02,eleid:09}
or $^{16}$O+$^{16}$O may affect stellar structure.
In particular, the
$^{12}$C($\alpha$,$\gamma$)$^{16}$O affects the amount of $^{12}$C
left after He-burning and therefore the ensuing C-burning
and Ne-burning stages. In principle, using a different
$^{12}$C($\alpha$,$\gamma$)$^{16}$O rate
could affect the results discussed in the following sections.
However, after having explored the impact of the $^{12}$C + $^{12}$C rate
for stars of different masses (and therefore with different $^{12}$C and $^{16}$O
composition at the onset of central C ignition) in paper I, we believe that
our present results are robust and cannot be significantly changed by
using different $^{12}$C($\alpha$,$\gamma$)$^{16}$O rates.

Stellar physics uncertainties \citep[mixing-length theory approximations,
prescriptions adopted for convective boundary mixing,
mass loss, impact of rotation and magnetic field,
see e.g.,][]{limongi:00, woosley:02, meynet:06} may also affect stellar models.
Furthermore, multi-dimensional simulations for advanced stages of massive stars
showed several features that cannot be properly reproduced by basic
one-dimensional models \citep[e.g.,][]{arnett:11}.
However, as also discussed for paper I, all those uncertainties do not modify
the general conclusions of this work on the impact of the $^{12}$C+$^{12}$C rate
uncertainty, for both the stellar structure and the nucleosynthesis.

\section{He core - weak $s$-process}
\label{sec:he core}

In massive stars central He$-$burning evolves in a convective core.
The nuclear reactions driving the energy generation are the 3$-\alpha$ reaction, converting
three $^4$He nuclei into $^{12}$C, and the $^{12}$C($\alpha$,$\gamma$)$^{16}$O reaction.
The most abundant species at the end of core He burning are
$^{12}$C and $^{16}$O \citep[e.g.,][]{arnett:85}.
The $^{14}$N left from the CNO cycle during the previous H$-$burning is fully converted
to $^{22}$Ne at the beginning of core He$-$burning, via the reaction chain
$^{14}$N($\alpha$,$\gamma$)$^{18}$F($\beta^+$$\nu$)$^{18}$O($\alpha$,$\gamma$)$^{22}$Ne.
Therefore the amount of $^{22}$Ne produced depends on the initial abundance
of the CNO species in the star \citep[e.g.,][]{prantzos:90}.
The main source of neutrons for the $s$-process is $^{22}$Ne,
via the reaction $^{22}$Ne($\alpha$,n)$^{25}$Mg.
This occurs at the end of the He core when the He abundance drops below 10 \%,
in the last $\sim$ 10$^4$ yr, the central temperature becomes larger than
2.5$\times$10$^8$ K and $\alpha$-capture on $^{22}$Ne is efficiently activated
\citep[e.g.,][]{raiteri:91a,the:07,pignatari:10}.
As mentioned in \S \ref{sec:gnv}, all models treated share the same
evolution and nucleosynthesis history until central He exhaustion.

The complete calculations for the $s$-process
were performed using the parallel version of the multizone driver of the
post-processing code PPN, developed by the NuGrid research platform (http://www.astro.keele.ac.uk/nugrid) and described by \cite{herwig:08} (MPPNP).
Reference sources for basic charged particle reactions,
3$-\alpha$ and $^{12}$C($\alpha$,$\gamma$)$^{16}$O have been selected
in agreement with the stellar code \citep[][respectively]{fynbo:05,kunz:02}.
As outlined in \S \ref{sec:gnv}, the $^{22}$Ne($\alpha$,n)$^{25}$Mg and $^{22}$Ne($\alpha$,$\gamma$)$^{26}$Mg rates
are from \cite{jaeger:01} and \cite{angulo:99}, respectively.
Neutron capture reaction rates for stable isotopes, and also for unstable isotopes if available, are provided by the KADoNIS websource
\citep[][]{dillmann:06}.
For neutron capture rates not included in KADoNIS, we refer to the Basel REACLIB database, revision 20090121 \citep[][]{rauscher:00}.
The $\beta-$decay rates are provided by \cite{oda:94} or \cite{fuller:85} for light species,
and by \cite{aikawa:05} for species heavier than iron.
Exceptions are $^{26}$Al and $^{85}$Kr $\beta-$decay rates, where the isomeric state and the
ground state must be considered as separate species at He$-$burning temperatures
and terrestrial rates must be used
\citep[e.g.,][and reference therein]{pignatari:10}.

\begin{figure}
\centering
\resizebox{8cm}{!}{\rotatebox{0}{\includegraphics{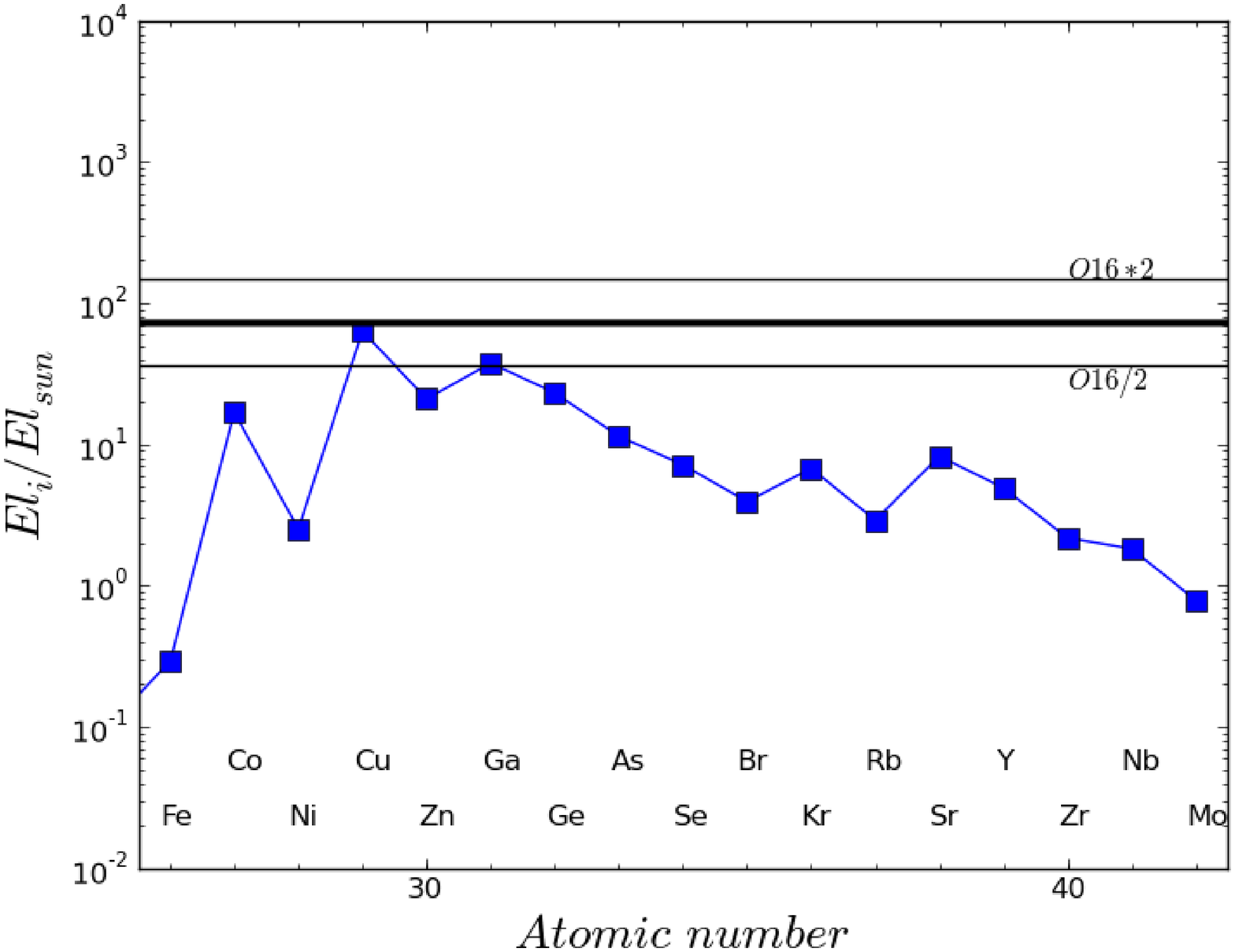}}}
\resizebox{8cm}{!}{\rotatebox{0}{\includegraphics{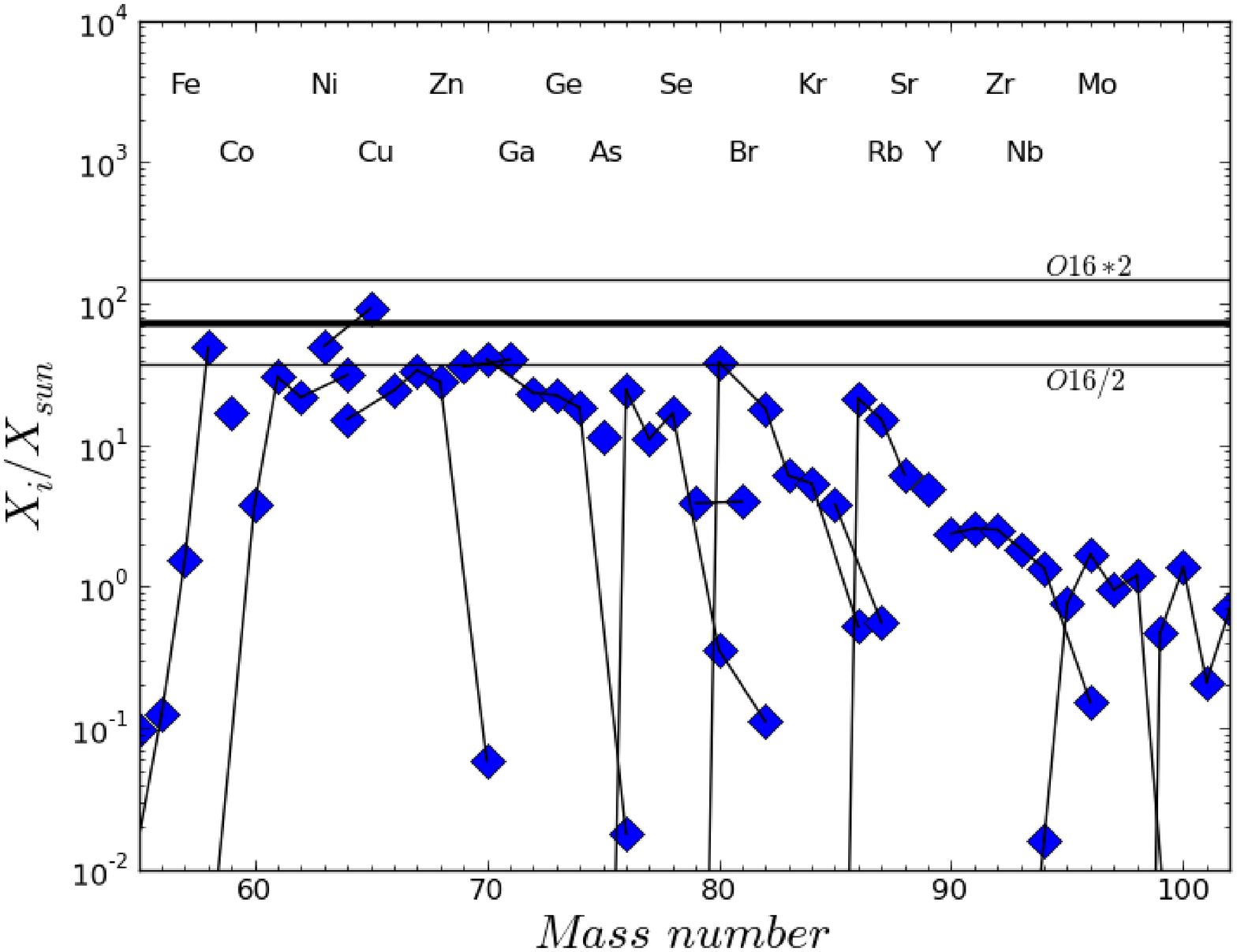}}}
\caption{Left Panel $-$ Overabundances for elements between Fe and Mo at the end of the core He exhaustion
for the 25 M$_{\odot}$ star, half solar metallicity.
Right Panel $-$ For the same model and for the same elements the distribution of stable isotopes is given, including the decay of
unstable species. In both panels the overabundance of $^{16}$O is provided as reference, together with its value multiplied and divided
by a factor of two (indicated with the labels $O16*2$ and $O16/2$ in the plots).}
\label{fig:hecore}
\end{figure}

The isotopic and elemental distributions between Fe and Mo
at the end of core helium burning
are presented in
Fig. \ref{fig:hecore}.
For elements between iron and strontium, most of the $s$-process abundances in the solar system have been produced by massive stars
\citep[weak $s$-process component, e.g.,][]{kaeppeler:82,kaeppeler:94,rauscher:02,the:07,pignatari:10}.

The overabundance of $^{16}$O is reported in Figure \ref{fig:hecore},
as a reference. $^{16}$O is a primary isotope,
therefore the production of it does not change with the initial
metallicity of the star.
The $^{16}$O observed in the solar system today is mainly produced in
massive stars, in the same region of the star where the $s$-process
yields are synthesized. Unlike primary isotopes, however,
$s$-process yields in massive stars show a direct dependence on the
initial stellar metal content, which is closer to a secondary-like
nucleosynthesis
\citep[see e.g.,][for a definition of secondary nucleosynthesis and for the analysis of the metallicity dependence for
$s$ process in massive stars]{tinsley:80,raiteri:92}.
%
%
According to \cite{tinsley:80}, isotopes that are fully produced by the $s$-process in massive stars are expected to show an overabundance a factor of two
higher than $^{16}$O at solar metallicity.
Actually, the weak $s$-process is not a pure secondary process
 \citep[e.g.,][]{raiteri:92,baraffe:92,pignatari:08,nomoto:06}.
However, we may still use the $^{16}$O overabundance as a guidance for the $s$-process efficiency.
From Figure \ref{fig:hecore} (Left Panel), Cu and Ga are the elements with the
highest $s$-process production.
In the Right Panel the isotopes show an odd-even pattern typical of the $s$-process, due to their neutron capture cross sections behavior \citep[see e.g.,][]{kaeppeler:82}.
Among the others, the isotopes $^{58}$Fe and $^{63,65}$Cu show the highest overabundances, just above the iron seeds.
As expected, the $s$-process efficiency drops beyond the Sr-Y-Zr neutron
magic peak (N=50), due to the
total amount of neutrons produced.
In Table \ref{table:hecore} we report the abundances of the main
He core products $^{12}$C and $^{16}$O, of the main neutron source
$^{22}$Ne, of the $s$-process seed $^{56}$Fe, of the $s$-only
isotopes between Fe and Sr and of neutron magic $^{88}$Sr at
the He exhaustion.
Notice that the amount of $^{22}$Ne available before the $s$-process starts is about 1\% by mass (given mostly by the initial CNO)
and that 36 \% of it is still available for more
$s$-processing during the following C$-$burning phase.

The abundance of $^{54}$Fe is also reported in Table \ref{table:hecore},
together with the neutron exposure.
The neutron exposure $\tau$$_{\rm n}$ is calculated according to the
formula proposed by \cite{woosley:95} and it is given by the $^{54}$Fe
depletion via neutron capture (notice that $^{54}$Fe is destroyed
by neutron capture):\\

$\tau$$_{\rm n}$ = $-$ (MACS$_{\rm ^{54}Fe(n,\gamma)}$)$^{-1}$$\times$ln($X$(54)/$X$(54)$_{\rm ini}$)
[mbarn$^{-1}$], \\
\newline
where the MACS$_{\rm ^{54}Fe(n,\gamma)}$ is the Maxwellian averaged neutron capture cross section on $^{54}$Fe
\citep[for simplicity hereinafter in this formula we choose for every temperature
the rate at 30 keV by][30.03 mbarn$^{-1}$]{coquard:06},
$X$(54) is the present $^{54}$Fe mass fraction
and $X$(54)$_{\rm ini}$ its initial abundance
($X$(54)$_{\rm ini}$ = 5.339$\times$10$^{-5}$). The total neutron exposure at the end of the He core
is reported in Table \ref{table:hecore}.
Final mass fraction abundances of $^{12}$C and $^{16}$O are 0.27
and 0.71 respectively.

\section{Carbon burning nucleosynthesis and $s$-process}
\label{sec:c12c12sprocess}

As discussed in \S \ref{sec:gnv}, increasing the $^{12}$C + $^{12}$C rate results
 in an early central C ignition.
This is because increasing the rate affects the equilibrium between the energy
generation due to C fusion reaction and the energy lost by neutrinos.
Therefore, the total lifetime of the central C$-$burning is affected and
more massive stars may develop a convective C core (see paper I).
On the other hand, a lower $^{12}$C + $^{12}$C rate delays the C$-$burning
activation to higher temperature and density
conditions \citep[see e.g.,][]{gasques:07}. In the following section, we analyze
the impact of these features on $s$-process nucleosynthesis.

\subsection{The $s$-process nucleosynthesis in C$-$burning conditions,
and the cs component}
\label{subsec:sprocess_cs}

\begin{figure}
\centering
\resizebox{6.5cm}{!}{\rotatebox{0}{\includegraphics{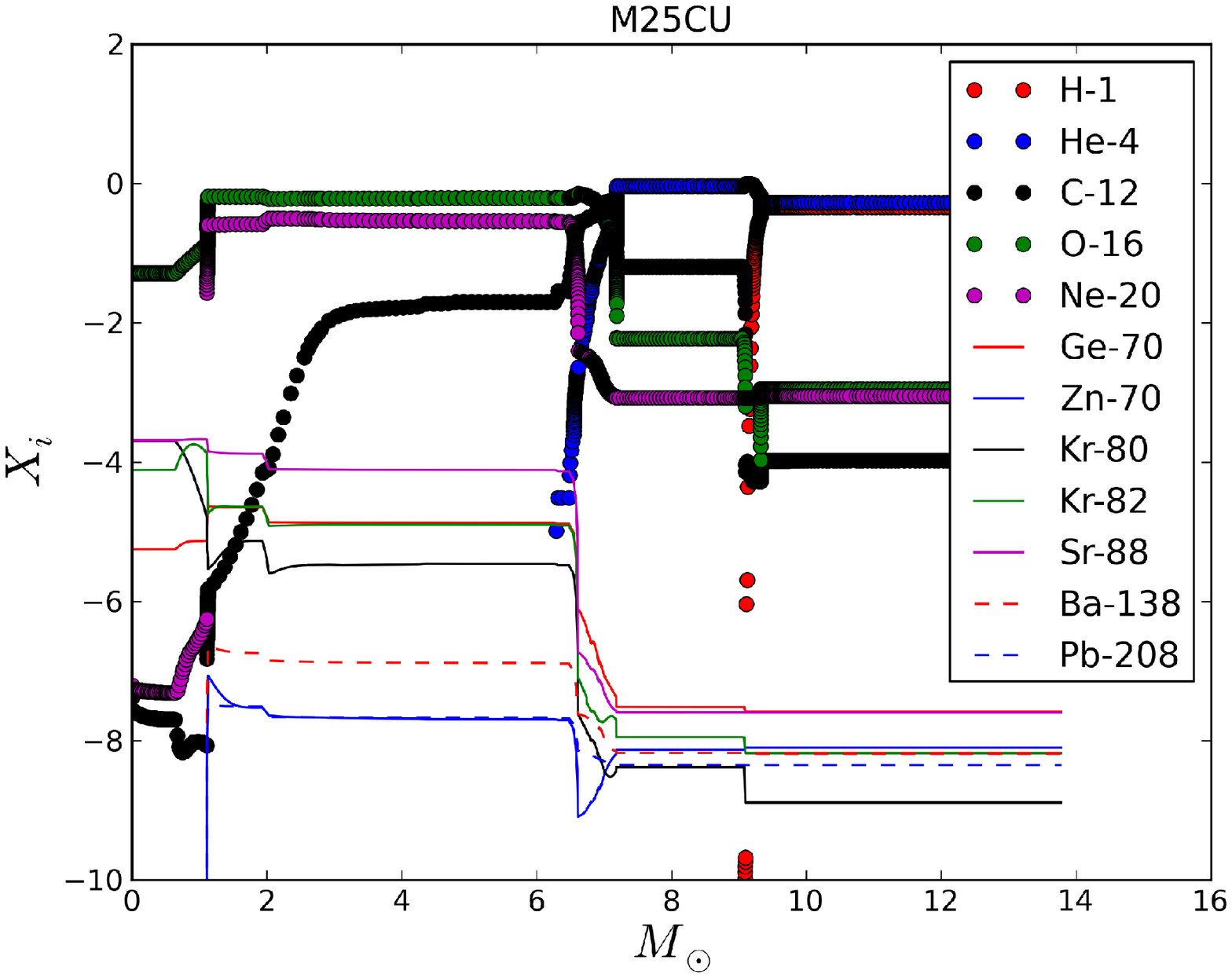}}}
\resizebox{6.5cm}{!}{\rotatebox{0}{\includegraphics{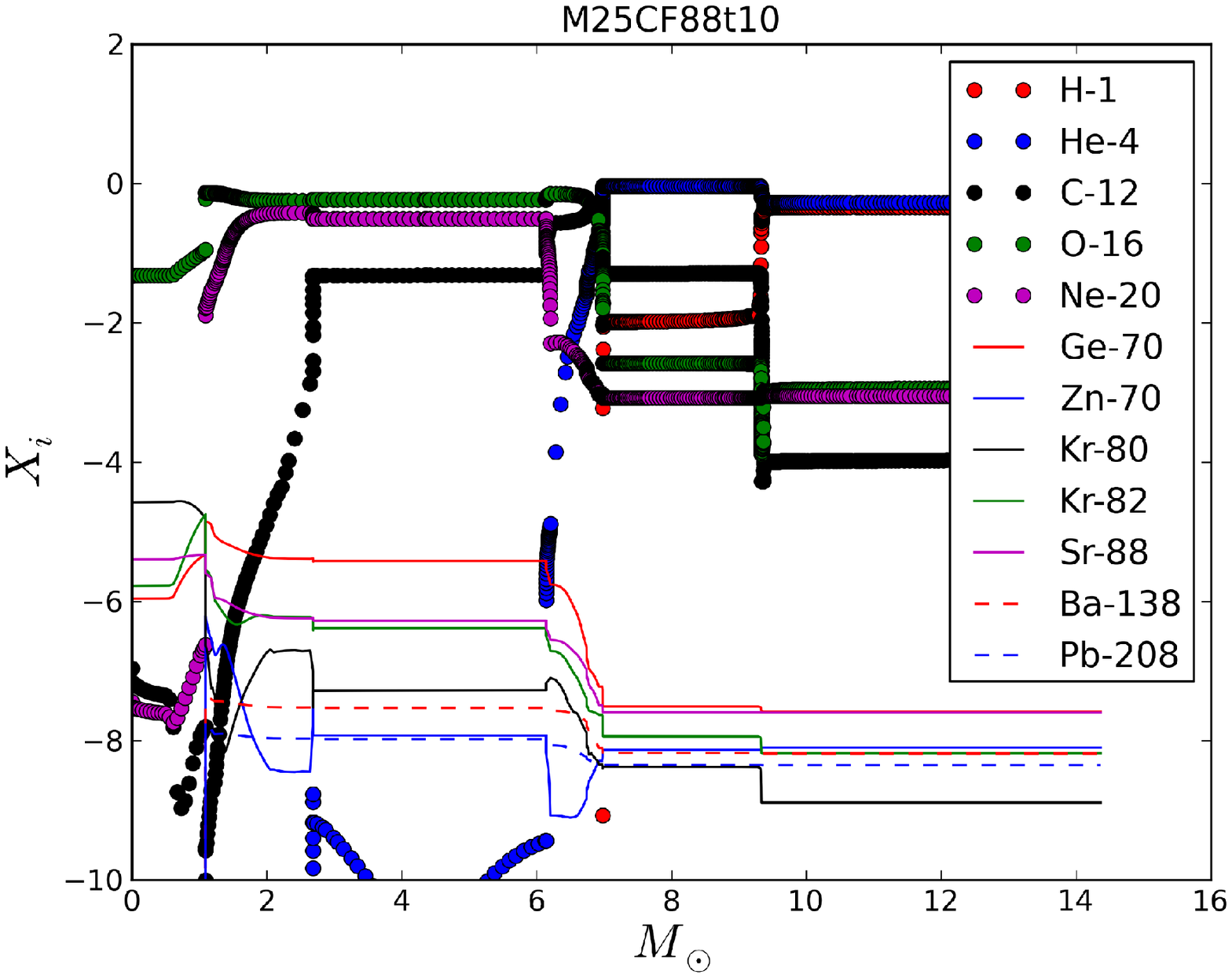}}}
\centering
\resizebox{6.5cm}{!}{\rotatebox{0}{\includegraphics{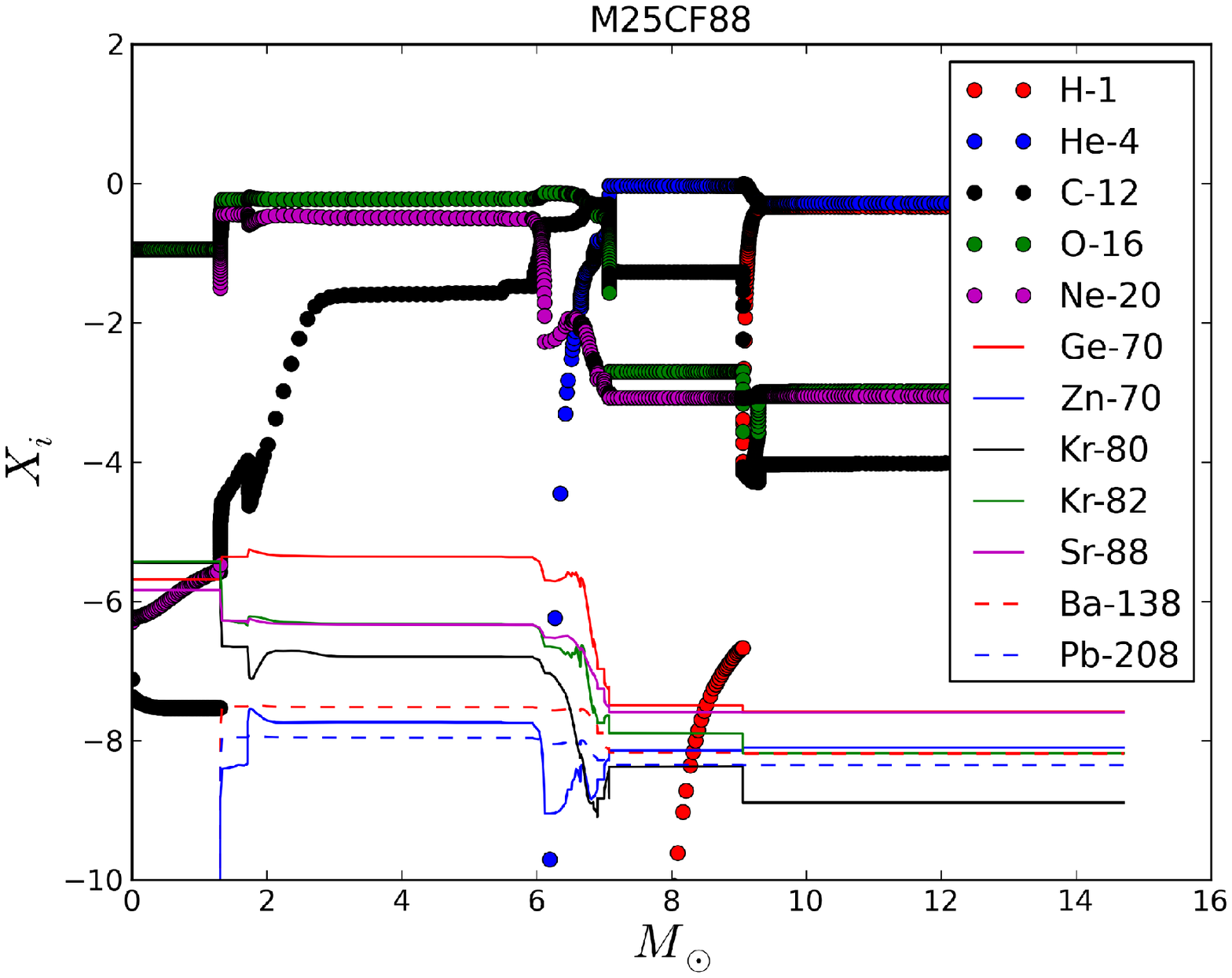}}}
\hfill

\resizebox{6.5cm}{!}{\rotatebox{0}{\includegraphics{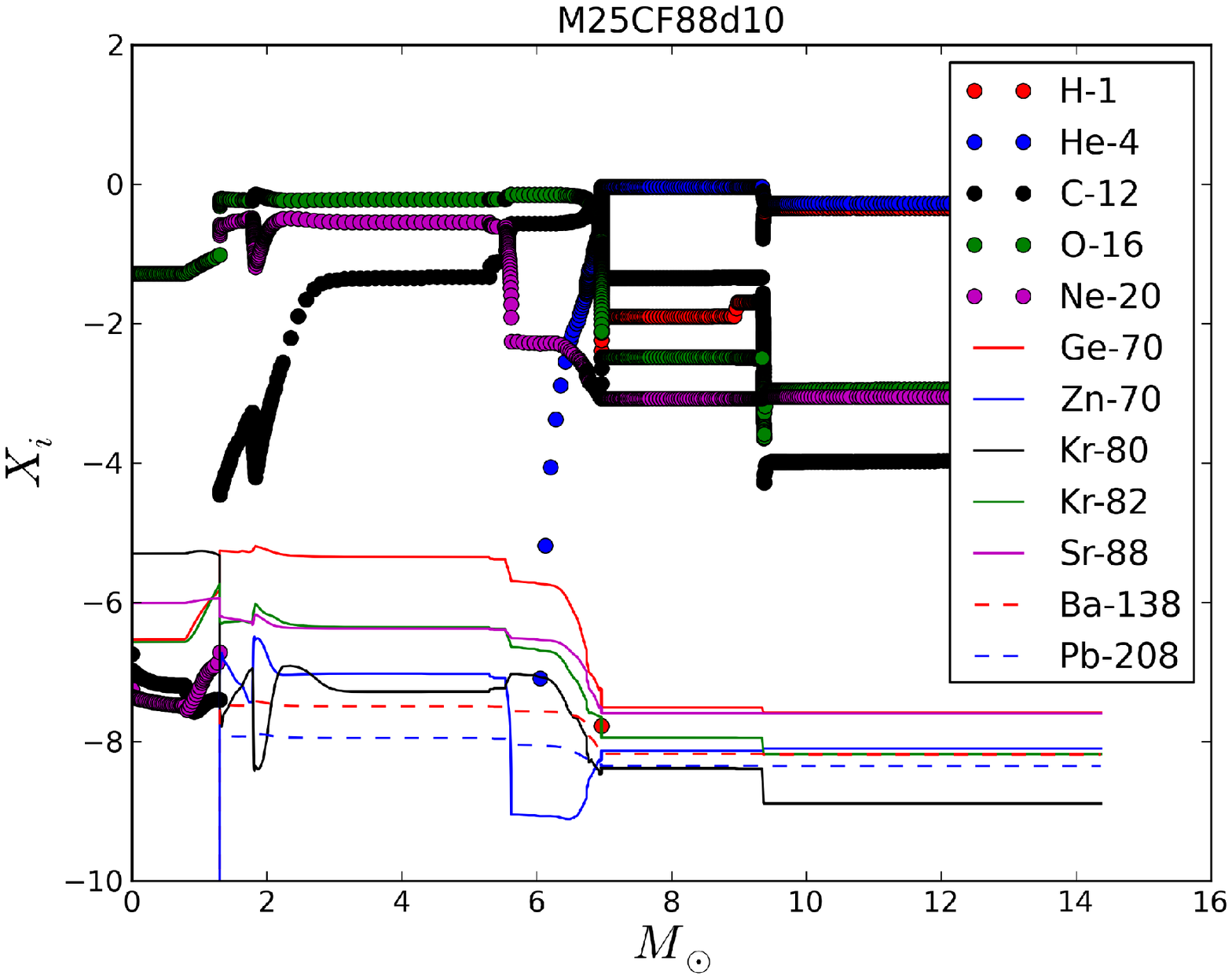}}}
\resizebox{6.5cm}{!}{\rotatebox{0}{\includegraphics{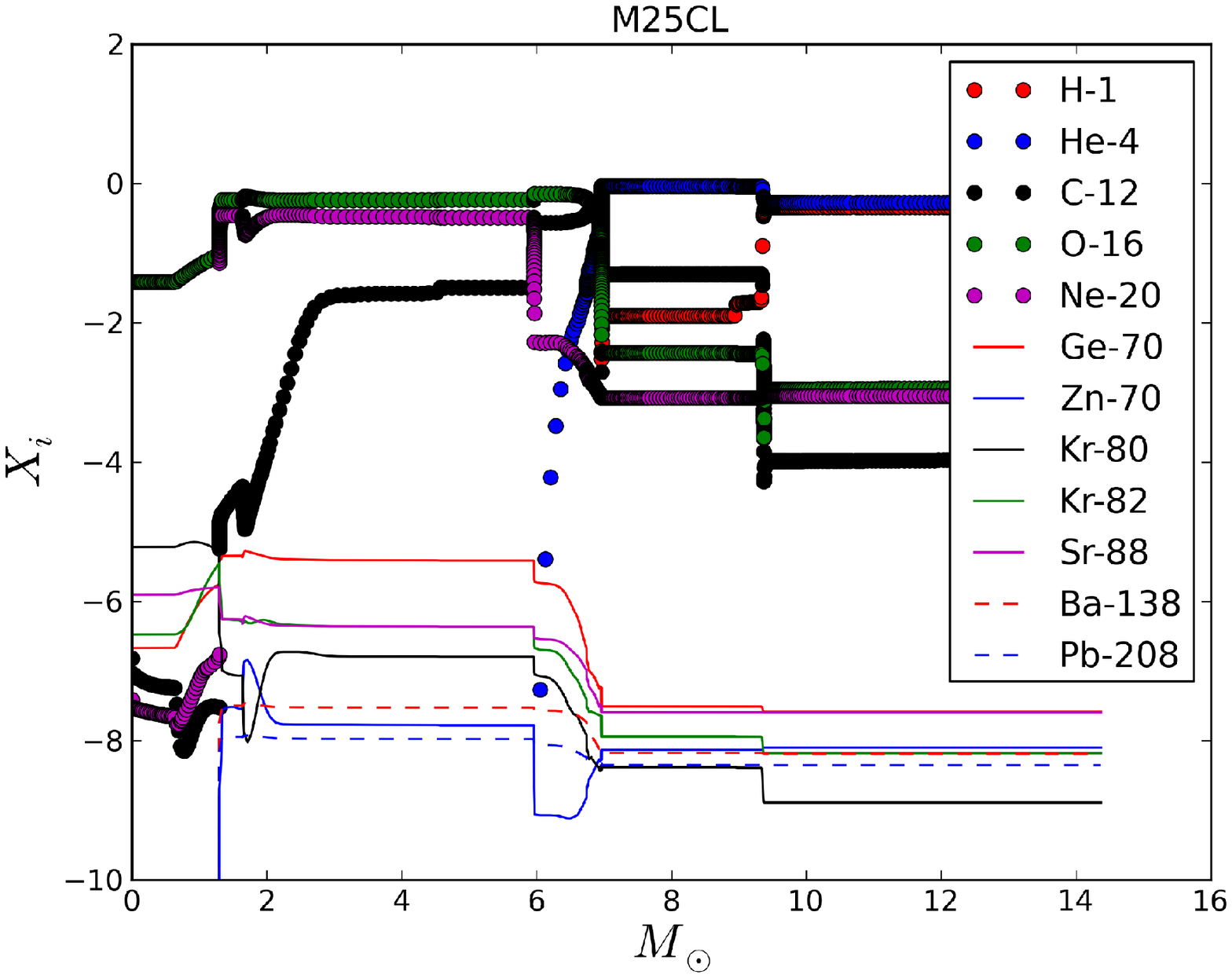}}}
  \caption{Abundance distribution (in mass fraction) for a sample of selected species close to the end of central O burning is given as a function of the mass coordinate (solar masses unit) for the models considered. The model is identified by the label on top of each panel.
}
\label{cshell:element_distribution}
\end{figure}

In Fig. \ref{cshell:element_distribution}, we present the abundances
for a sample of species at the end of central O$-$burning as a function of the
mass coordinate, for the models CU, CF88t10, CF88, CF88d10 and CL.
Together with light isotopes indicative of different burning regions
($^1$H, $^4$He, $^{12}$C, $^{16}$O and $^{20}$Ne),
we report the abundance profile of
the $s$-only isotopes $^{70}$Ge, $^{80,82}$Kr, $r$-only isotope $^{70}$Zn
(which can be also be produced by direct neutron capture via the
$^{69}$Zn branching and is therefore a good tracer for neutron
densities higher than $\sim$ 10$^{11}$ cm$^{-3}$),
and the isotopes $^{88}$Sr, $^{138}$Ba and $^{208}$Pb.
These last three nuclides belong to the three neutron magic peaks
along the neutron capture path (N=50, 82, 126 respectively),
and are mostly of $s$-process origin. However, the bulk of their $s$-process abundance
in the Solar System distribution was produced by low mass AGB stars,
which are responsible
for the main and strong $s$-process components \citep[see][and reference therein]{gallino:98}.
By comparing the abundances of these isotopes in the stellar envelope (which
roughly coincides with the initial abundance) with the abundances
in the O-rich zone, the $s$-process efficiency producing these species can be
deduced. In all cases the production of $^{88}$Sr is significantly higher than
$^{138}$Ba and $^{208}$Pb,
as expected from the neutron exposure regime typical for $s$-process
in massive stars \citep[e.g.,][]{raiteri:91b}.
The further $s$-process contribution during C$-$burning over the He core ashes
 may be observed in Fig. \ref{cshell:element_distribution},
by comparing the abundances in the He core window with the deeper profiles in
the convective C shell region.
Production of $^{70}$Zn, as well as the isotopic ratio
$^{80}$Kr/$^{82}$Kr, provides an indication of the neutron densities reached
during the convective He$-$burning core, in the He core window and in the
C shell material. This is due to the effect of neutron capture branching points
at $^{69}$Zn and at $^{79}$Se
\citep[e.g.,][and reference therein]{pignatari:10}. In particular, $^{70}$Zn and
 $^{80}$Se both increase with increasing the neutron density.
In Table \ref{tab:sprocess_cshell_end} we report the mass-fraction
abundances for a sample of species in the C shell
(including the ones listed above)
and for the cases reported in Fig. \ref{cshell:element_distribution}.

The abundances of the main products of C-burning,
$^{20}$Ne and $^{23}$Na, do not show a linear correlation with the C-burning
rate. In particular, the $^{20}$Ne/$^{23}$Na isotopic ratio ranges
from about 12 (model M25CU)
to 50 (model M25CF88d10). The maximum abundance of $^{20}$Ne is obtained
in M25CF88t10, whereas the largest $^{23}$Na amount is in M25CU,
since in this last case the abundances are dominated by the C-burning
at lower central temperatures \citep[][]{arnett:72}.
The distribution of light species in the C-burning region is important
for the $s$-process, since 90\% or more of the neutrons produced are captured
by light neutron poisons \citep[e.g.,][]{pignatari:10}.
Therefore, the abundances of light nuclides needs
to be considered to preserve the consistency of nucleosynthesis
calculations.
In particular, at metallicities close to solar the main neutron poison is
$^{25}$Mg, with an important contribution from
$^{24,26}$Mg, $^{20,22}$Ne, $^{23}$Na.

From Table \ref{tab:sprocess_cshell_end}, we see that for model CF88d10
the C shell material is exposed to the highest
neutron density. This is because of the late
C$-$burning activation at higher temperature and
density \citep[see later in this section, and][]{gasques:07}.
On the other hand, the $s$-process is more efficient in the CU  model;
compared to the standard case, the $^{88}$Sr abundance is boosted by more than
2 orders of magnitude.
Such a result has already been mentioned \cite[]{bennett:10}, and is discussed
 in more detail in paper I.

Interestingly, all models show an overproduction of $^{88}$Sr in the central
 zone, at mass coordinate
lower than 2 M$_{\odot}$, compared to more external C$-$burning regions.
Indeed, during both Ne$-$burning and early O$-$burning conditions, $^{88}$Sr is
 still produced, or at least not efficiently depleted,
in conditions of partial photodisentegration. Also,
typical $p$-rich species,
 e.g., $^{78}$Kr or $^{84}$Sr, are produced in this phase with different
efficiencies. Such nuclear species will be further created or destroyed
by photodisintegrations in later O$-$burning and Si$-$burning evolutionary
stages \citep[e.g.,][]{thielemann:85}. However, as pointed out
 by \cite{rauscher:02}, such pre-explosive abundances could be
mixed by the merging of different shells into more external regions of the star,
where they can survive and provide a significant contribution
to $p$-process species yields, together with the explosive $p$-process abundances.
Such contributions can also be relevant for non-$p$-rich isotopes,
like $^{88}$Sr.
Along similar lines, \cite{arnett:11} showed how multi-dimensional effects
may affect the stellar structure before the supernova explosion,
allowing a more efficient exchange of material between different burning zones
deep in the star compared to standard one-dimensional models.
Therefore, caution must be maintained concerning nucleosynthesis abundances in
the later evolutionary stages of massive stars.

The CF88d10 model is characterized by a lower production of neutrons
compared to the CF88 model ($^{88}$Sr is lower by 11 \%),
but it has a higher neutron density (higher $^{70}$Zn amount, lower
$^{80}$Kr/$^{82}$Kr ratio, see Table \ref{tab:sprocess_cshell_end}).
This effect was discussed by \cite{gasques:07}.
The CL and CF88 models show similar results, because of the similarities
between the two stellar models (\S \ref{sec:gnv}).
The CF88t10 model shows an increase of $^{88}$Sr by only a factor of 1.2 in the
final convective C shell compared to CF88
(Table \ref{tab:sprocess_cshell_end}), confirming earlier results presented
by \cite{bennett:10}.
In the CU case $^{88}$Sr increases by a factor of 166 relative to the
standard case.
Major differences are observed for the whole $s$-process distribution. The
reason for such a large overproduction in the CU case
is the strong $s$-process efficiency reached in the convective
C$-$burning core.
For CU, in the C core the dominant neutron source is
$^{13}$C($\alpha$,n)$^{16}$O, not $^{22}$Ne($\alpha$,n)$^{25}$Mg.
When the burning temperature is lower than $\sim$ 0.70$-$0.75 GK,
$^{13}$C is efficiently produced via
the $^{12}$C(p,$\gamma$)$^{13}$N($\beta^+$$\nu$)$^{13}$C channel.
The ashes of the C core are later mixed in the
convective C shell and ejected in the interstellar medium by the final SN
explosion (Fig. \ref{k_c12c12}).
For the CU $^{12}$C+$^{12}$C rate, the overlap between the convective
core and the final shell acts to
drastically increase the $s$-process efficiency, as explained in paper I.

The activation of the $^{13}$C($\alpha$,n)$^{16}$O reaction
in the C$-$burning core was previously discussed by \cite{chieffi:98}.
However in stellar models using the standard
$^{12}$C+$^{12}$C rate, the central C-burning is not convective. Even in
 the eventuality that a small convective core is developed,
the enriched material is further processed by the more advanced burning phases
and eventually collapses to form the neutron star.
On the other hand, when
the temperature is higher than $\sim$ 0.70$-$0.75 GK,
$^{13}$N($\gamma$,p)$^{12}$C reaches equilibrium with its
direct reaction \citep[][]{clayton:68}.
In these conditions, the only way to produce $^{13}$C is via
$^{12}$C(n,$\gamma$)$^{13}$C, and so the $^{13}$C($\alpha$,n)$^{16}$O reaction
only recycles neutrons
captured by $^{12}$C. The neutrons have to be produced by other reactions, e.g.,
by $^{22}$Ne($\alpha$,n)$^{25}$Mg.

\begin{figure}
\centering
\resizebox{8cm}{!}{\rotatebox{0}{\includegraphics{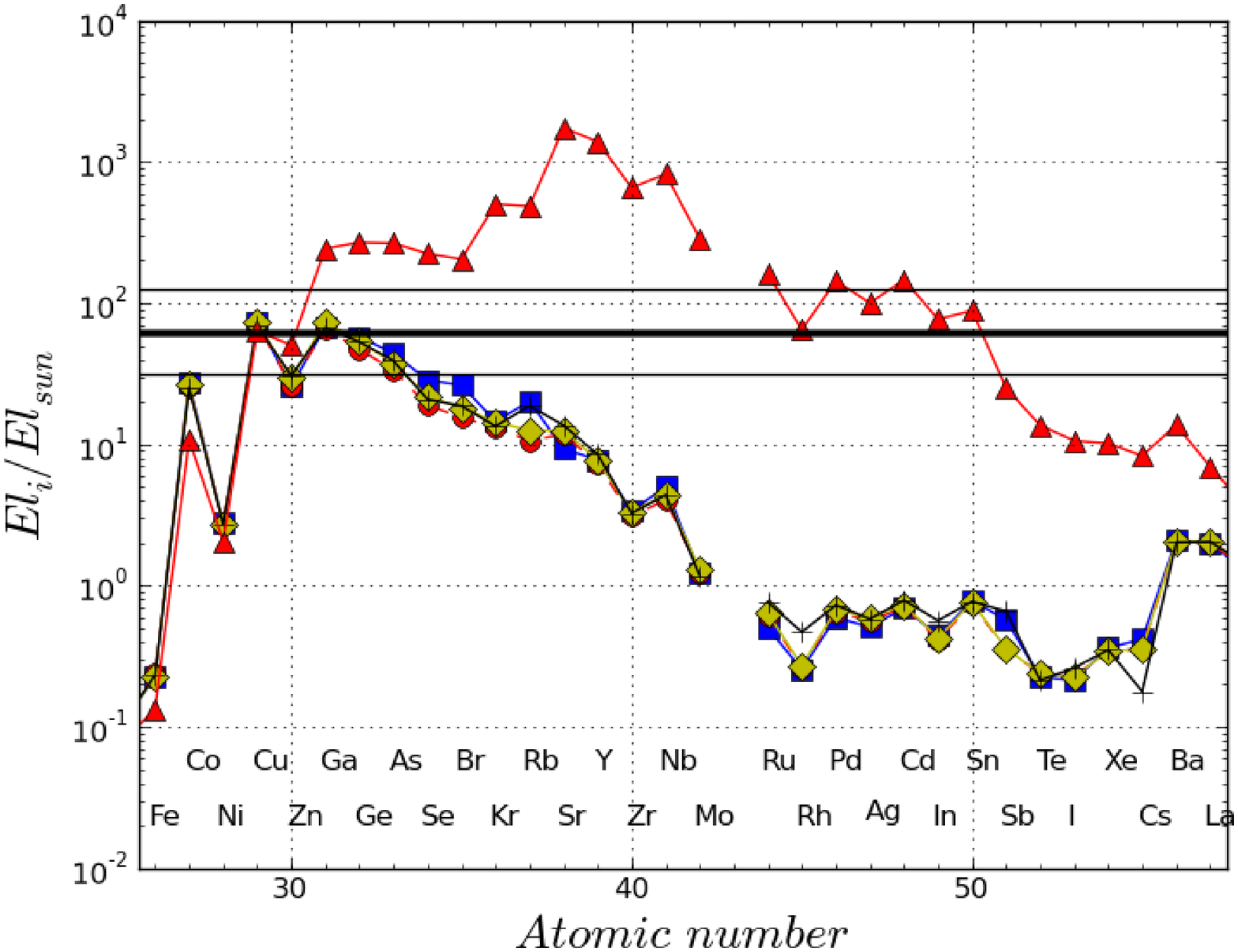}}}
\resizebox{8cm}{!}{\rotatebox{0}{\includegraphics{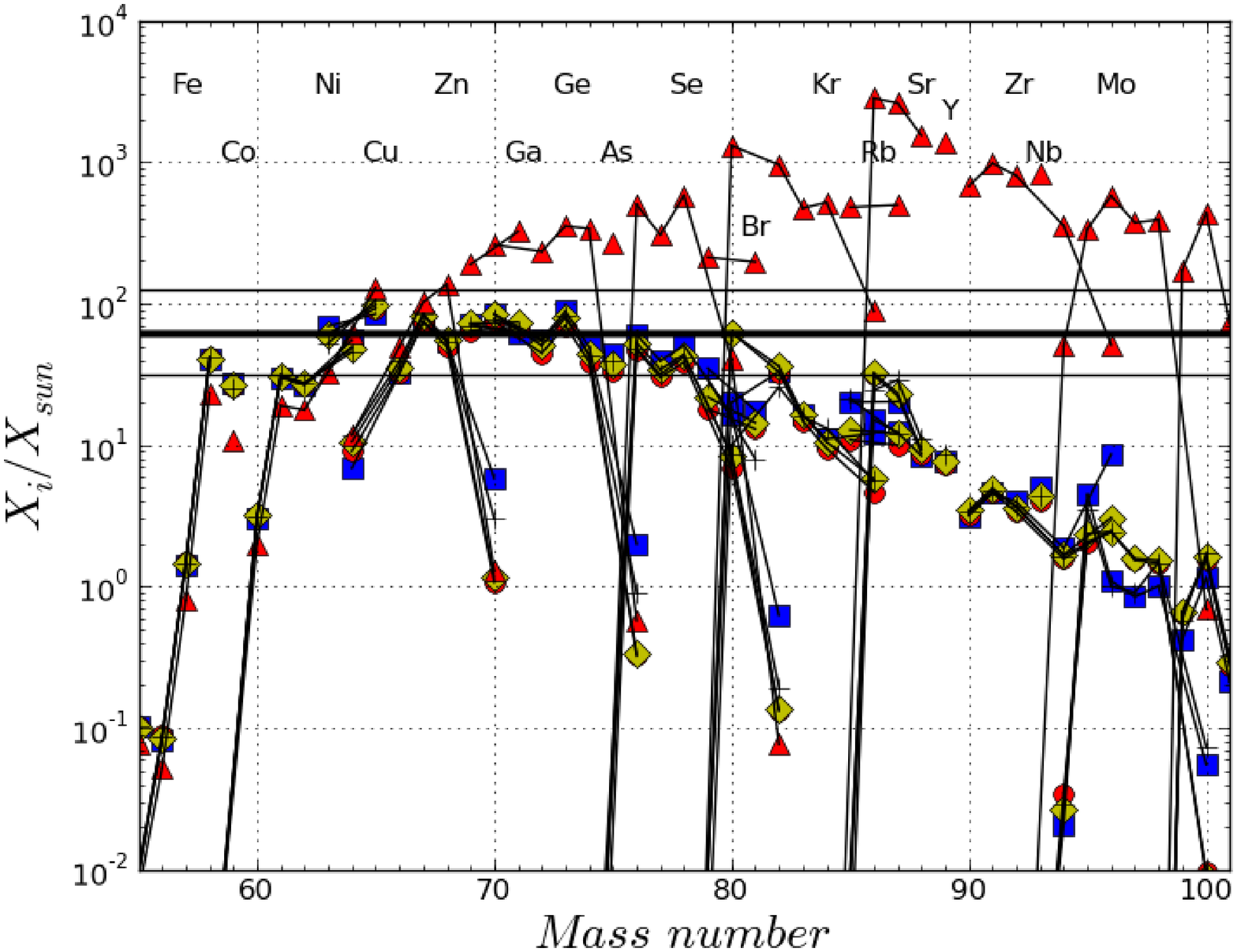}}}
  \caption{Left Panel $-$ The final element production factors between Fe (Z=26) and La (Z=57) in the C shell region for CF88d10 (blue squares),
CL (red circles), CF88 (yellow diamonds), CF88t10 (black crosses) and CU (red triangles).
Right Panel $-$ The isotopic distribution zoomed in the mass region
between $^{56}$Fe and $^{100}$Mo,
for the same stellar models presented in the Left Panel.
Isotopes of a given element are connected with lines.
}
\label{fig:s_distribution_cshell}
\end{figure}

In Fig. \ref{fig:s_distribution_cshell} (Left Panel), the final $s$-process
elemental distribution between Fe and La is shown for all models
considered. The first four cases
present similar $s$-process abundance distributions.
In Table \ref{tab:sprocess_cshell_end_element_overabb_small}
the the overproduction factors for elements between C and Mo are
reported for a detailed comparison.
The CU model produces an $s$-process distribution which is peaked in the
region of Sr-Y-Zr-Nb (with overproduction factor larger than 1000
for Sr and Y).
The neutron-magic peak at N=50 is bypassed by the neutron capture path
efficiently and a relevant production of $s$ elements
(with overproduction factor larger than 100) is observed up to Sn (Z=50).
Negligible abundances are observed above Ba-La (Z=56,57),
compared to the Sr-Y-Zr-Nb bulk.
In Fig. \ref{fig:s_distribution_cshell} (Right Panel), the respective isotopic distributions between Fe and Mo are presented.
We recall again that the different distributions shown in
Fig. \ref{fig:s_distribution_cshell} share the same neutron capture process history
until the end of the core He$-$burning. After core He exhaustion,
the heavy isotope distribution in the C shell ejecta is driven
by different neutron exposure histories.

In the CF88 model, any $^{22}$Ne left in the He-core ashes is burnt with a peak neutron
density of $\sim$ 10$^{11}$ cm$^{-3}$ \citep[e.g.,][]{raiteri:91b}
and a neutron exposure of $\tau$ $\sim$ 0.05 mbarn$^{-1}$.
The final convective C$-$burning shell episode is switched off by
the onset of central O$-$burning and is not active in the last
evolutionary phase before the SN explosion.
The final distribution in Fig. \ref{fig:s_distribution_cshell}
is peaked in the Cu-Ga-Ge region.
In the CU model, the dominant neutron exposure contribution is
provided during the convective C$-$burning core, driven by
the $^{13}$C($\alpha$,n)$^{16}$O reaction.
During the final convective C shell, $^{22}$Ne drives the neutron flux.
The component built in the C core and peaked in the Sr-Y-Zr-Nb region is
dominant in the final $s$-process rich ejecta
because of the larger neutron exposure and the large overlapping factor
between core and shell ($\sim$ 2/3 of the mass
of the convective C shell is made of C$-$burning core ashes).
Similar results were found in paper I for a large range of masses using
the CU rate, and for the specific case of the 20 M$_{\odot}$ star
model and a rate intermediate between CF88 and CU (CI rate, paper I).
In cases where the convective C core overlaps with the final convective C shell,
the material from the core dominates the final $s$-process ejecta.
The resulting $s$-process yields are anomalous compared to the classical
weak $s$-process component.

If all, or most, of the $s$-process rich material lost to the
interstellar medium by core collapse supernovae is similar to the CU ejecta,
then the predicted abundances cannot be reconciled with the solar system
distribution. In other words, if all
massive stars were to eject material exposed to $^{13}$C($\alpha$,n)$^{16}$O
in the central C$-$burning conditions observed for CU, an anomalous abundance
distribution would be observed.
This may indicate that the upper limit $^{12}$C+$^{12}$C rate is overestimated,
a conclusion already reached in paper I.
We will discuss in \S \ref{sec:c12c12a and c12c12p}
the impact of the $^{12}$C($^{12}$C,$\alpha$)$^{20}$Ne and
$^{12}$C($^{12}$C,p)$^{23}$Na strength uncertainty on these results.
Tackling the discussion from a more speculative point of view, two $s$-process
components may be identified: the standard
weak $s$-process component, dominated by $^{22}$Ne($\alpha$,n)$^{25}$Mg, and
a stronger $s$-process component coming
from $^{13}$C($\alpha$,n)$^{16}$O, triggered when the temperature
is low enough to prevent the
$^{13}$N($\gamma$,p)$^{12}$C reaction from
becoming efficient. We refer to the latter component as
the `cold` C$-$burning component,
or `cs component` hereafter.

\subsection{Possible implications of the cs component for
the Lighter Element Primary Process}
\label{subsec:csprocess_LEPP}

\cite{travaglio:04} performed Galactic chemical evolution (GCE)
calculations, which included $s$-process contributions from massive and
low mass AGB stars
and the $r$-process signature evaluated from the $r$-process residual and
from spectroscopic observations at low metallicity.
A new unknown component was identified in the solar system distribution
(termed the Lighter Element Primary Process, or LEPP),
spanning the region from Sr-Y-Zr to $\sim$ Te-Xe. The LEPP component comprises
10$-$20 \% of the total elemental abundances in the region, including a similar
contribution to the $s$-only species.
This component has been connected to the anomalous abundances observed
in a group of very low metallicity stars that are enriched in the
Sr-Y-Zr mass region compared to the $r$ process
\citep[see also recent observations by][]{hansen:11,chiappini:11}.
Besides the $s$-process, various scenarios involving primary explosive
nucleosynthesis in massive stars has been invoked
\citep[e.g.,][]{hoffman:96,froehlich:06,qian:07,montes:07,pignatari:08,farouqi:09,farouqi:10,roberts:10,arcones:11}.

Is it possible that the cs component
corresponds or partially explains the LEPP component? In other words,
if only a limited fraction of $s$-process rich material with the distinct
$^{13}$C($\alpha$,n)$^{16}$O signature is ejected from massive stars,
could this be a feasible scenario to explain LEPP in the solar system?
According to paper I, it is possible that
the cs component is only ejected for a limited range of stellar masses.
By using a $^{12}$C+$^{12}$C rate given by the geometric average between CU and CF88 (CI),
it was found that only a 20 M$_{\odot}$ star ejects material
carrying the cs component signature, all other models displayed the
standard weak $s$-process signature.
From Table \ref{tab:sprocess_cshell_end_element_overabb_small}, oxygen
overproduction is $\sim$ 60. Therefore, at solar metallicity an element is
fully produced by a secondary $s$-process if its overabundance
is $\sim$ 120, and 20 \% of its production is explained if the
overproduction
is $\sim$ 24 (see discussion in \S \ref{sec:he core}).
The cs component in CU is peaked at Sr-Y-Zr, where the overproduction of Sr
is  $\sim$ 1740
(Table \ref{tab:sprocess_cshell_end_element_overabb_small}). This means that if
only $f$ = 1$-$2 \% of the $s$-process rich material
ejected by massive stars carries a cs-like component, $\sim$ 20 \% of the
solar Sr could be explained.
The factor $f$ could correspond to the contribution from a narrow mass range of stars, for instance.
Notice that the current astrophysics scenario for elements typically produced by the
weak $s$-process would not be compromised, since with
$f$ = 0.02 the cs-component contribution to Cu or Ga would be marginal
($\sim$ 1 \% and $\sim$ 5 \% of their solar abundances, respectively).
Only Sr, Y, Zr and Nb would receive a significant contribution by the cs-process in this case (see
Table \ref{tab:sprocess_cshell_end_element_overabb_small} for comparison).

Such considerations provide a warning to bear in mind in our discussion and more
generally for the present understanding of the $s$-process
in massive stars. If the $^{12}$C+$^{12}$C rate is higher
than presently used, and the cs-process does exist,
only 1$-$2 \% of the $s$-process rich material ejected from massive stars
carrying such a component would be enough to explain
the LEPP signature at the Sr peak.
Therefore, it is extremely important to measure the $^{12}$C+$^{12}$C rate down
to stellar energies, to constrain C$-$burning conditions and $s$-process
yields from massive stars.

\subsection{Pre-explosive production of $^{26}$Al and $^{60}$Fe.}
\label{subsec:al26fe60}

In Table \ref{tab:sprocess_cshell_end} we report our
$^{26}$Al and $^{60}$Fe predictions for the different models.
In particular, using a weaker $^{12}$C+$^{12}$C rate,
\cite{gasques:07} noticed a stronger production in the total
stellar yields of these
long-lived radio nuclides compared to standard models.
This is characteristic of the higher proton and neutron density, which affects
the convective C shell abundances.
We confirm this trend for $^{60}$Fe pre-explosive production. Indeed, because
of the long half life at stellar C$-$burning temperatures
\citep[few tens of years, e.g.][]{langanke:00}, $^{60}$Fe can be produced and
maintained in stellar material even if the convective C-burning shell ends before the
SN explosion, which is the case for all of our models except CF88t10.
In Table \ref{tab:sprocess_cshell_end}, CF88d10 shows the highest $^{60}$Fe abundance
compared to the other models.
On the other hand, CU has the lowest abundance (about a factor of 50 less than
in CF88d10).
During C$-$burning the $^{58}$Fe abundance (the main seed of $^{60}$Fe) is
mostly depleted compared to other models. The $^{60}$Fe is not efficiently
produced in the C core because of the lower neutron density compared to the
convective C shell, resulting in a weaker neutron capture channel at the
$^{59}$Fe branching.
Consequently the pre-explosive contribution to the final $^{60}$Fe ejecta tends to be
inversely proportional to the $^{12}$C+$^{12}$C rate, increasing as the rate
decreases.

The pre-explosive $^{26}$Al abundance in the C shell region
is not significantly enhanced in the CF88d10 model. In fact the $^{26}$Al half life
at C$-$burning temperatures is quite low \citep[few tens of minutes, e.g.][]{oda:94}, allowing it to almost
completely decay in stellar material before the SN explosion.
The model showing the highest pre-explosive $^{26}$Al abundance is CF88t10,
the only one in the present set of models having a convective C shell lasting
until the latest evolutionary phases before the onset of core collapse.
This stellar structure feature is in common with the stellar models used by \cite{gasques:07}.

To first order, the pre-explosive $^{60}$Fe abundance correlates
with the neutron density in the final convective C shell, and is anti-correlated
with the $s$-process efficiency of the CU model, where most of the neutrons
are produced during earlier central C$-$burning.
Conversely the dependence of pre-explosive $^{26}$Al abundance
on the $^{12}$C+$^{12}$C rate is model dependent.
For a more comprehensive discussion about the production of $^{26}$Al and
$^{60}$Fe in massive stars, we refer to e.g.,
\cite{timmes:95a,limongi:06}, where the explosive component is also
taken into account.

\section{$^{12}$C($^{12}$C,$\alpha$)$^{20}$Ne and $^{12}$C($^{12}$C,p)$^{23}$Na channels}
\label{sec:c12c12a and c12c12p}

\begin{figure}
\centering
\resizebox{10cm}{!}{\rotatebox{0}{\includegraphics{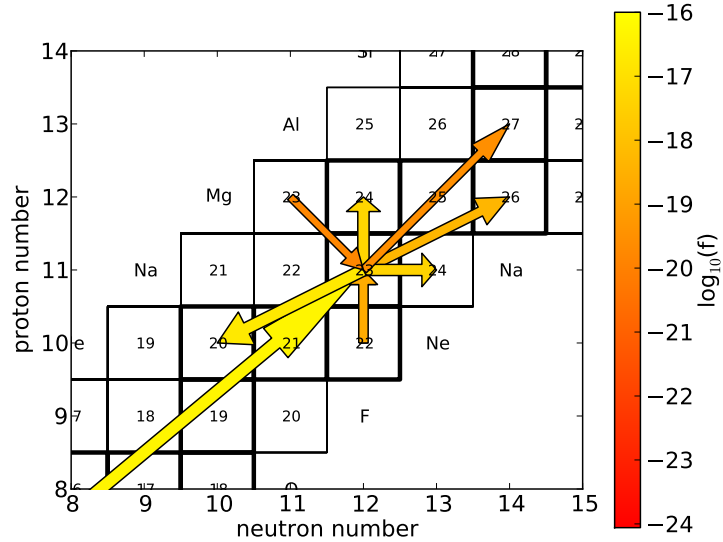}}}
\caption{The nucleosynthesis fluxes ([$\delta$$Y_{\rm i}$/$\delta$t]$_{\rm j}$, i.e., the variation of the abundance $Y_{\rm i}$ =
$X_{\rm i}$/$A_{\rm i}$ due to the reaction j)
producing and depleting $^{23}$Na are shown in central C-burning conditions.
The arrow size and color correspond to the flux strength.}
\label{fig:flux_na23_st}
\end{figure}

The $^{12}$C+$^{12}$C reaction has two main channels with comparable
strength, producing respectively one $\alpha$ particle and one $^{20}$Ne
nucleus, or one proton and one $^{23}$Na nucleus.
However, the predicted C$-$burning yields consist of more than
20 \% $^{20}$Ne, and only few per cent of
$^{23}$Na. This is due to the efficient activation of the
$^{23}$Na(p,$\alpha$)$^{20}$Ne, which processes $^{23}$Na in $^{20}$Ne
\citep[e.g.,][]{arnett:85,chieffi:98}.
As an example, in Fig. \ref{fig:flux_na23_st},
we show the isotopic chart with the main reaction fluxes responsible
for productive and destructive nucleosynthesis of $^{23}$Na,
at typical shell C$-$burning conditions.
Most of the $^{23}$Na produced via $^{12}$C($^{12}$C,p)$^{23}$Na
is depleted by the $^{23}$Na(p,$\alpha$)$^{20}$Ne reaction.
Therefore, the final $^{23}$Na abundance is mostly
given by the equilibrium between those two reactions.
The other significant flux in Fig. \ref{fig:flux_na23_st} is due to the $^{23}$Na(p,$\gamma$)$^{24}$Mg reaction which
competes with the (p,$\alpha$) channel (with a typical flux ratio of (p,$\alpha$)/(p,$\gamma$) $\sim$ 4, e.g., NACRE).
Notice that most of the sodium observed today in the Solar System is produced in massive stars in the conditions described.

At stellar temperatures, the uncertainty in the relative strength between the $^{24}$Mg$^{*}$ $\alpha$-decay and proton-decay channels has a relevant impact on nucleosynthesis calculations.
In the present work, we use the ratio $R_{\alpha}$/$R_p$ = 0.65/0.35, keeping in mind that we also properly consider
the small contribution from the neutron channel (see \S \ref{sec:c12c12n} for a more detailed discussion about the neutron channel).
However, this ratio becomes more uncertain at temperature conditions closer to the Gamow peak energies and
typical C core conditions.

\begin{figure}
\centering
\resizebox{12cm}{!}{\rotatebox{0}{\includegraphics{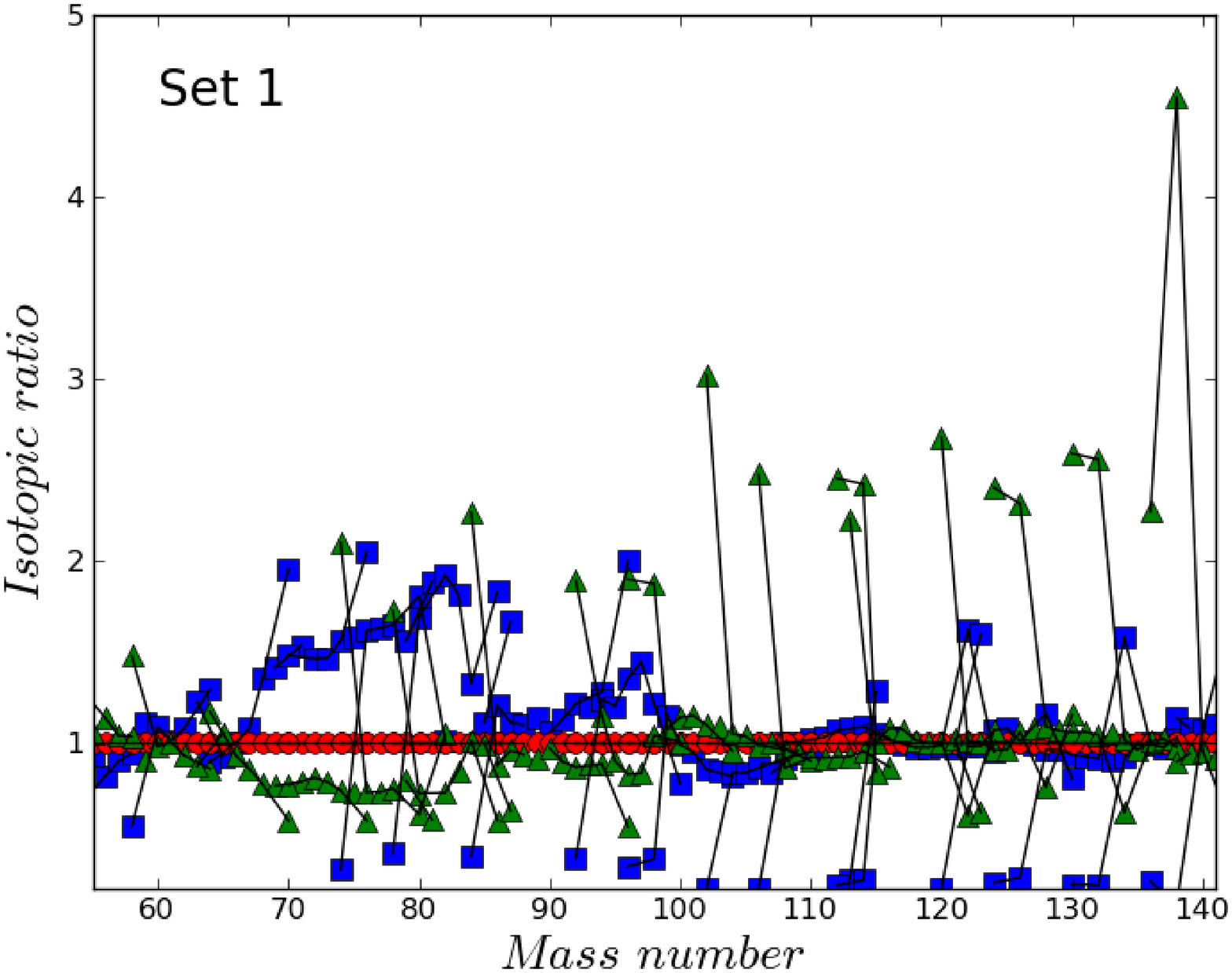}}}
\resizebox{12cm}{!}{\rotatebox{0}{\includegraphics{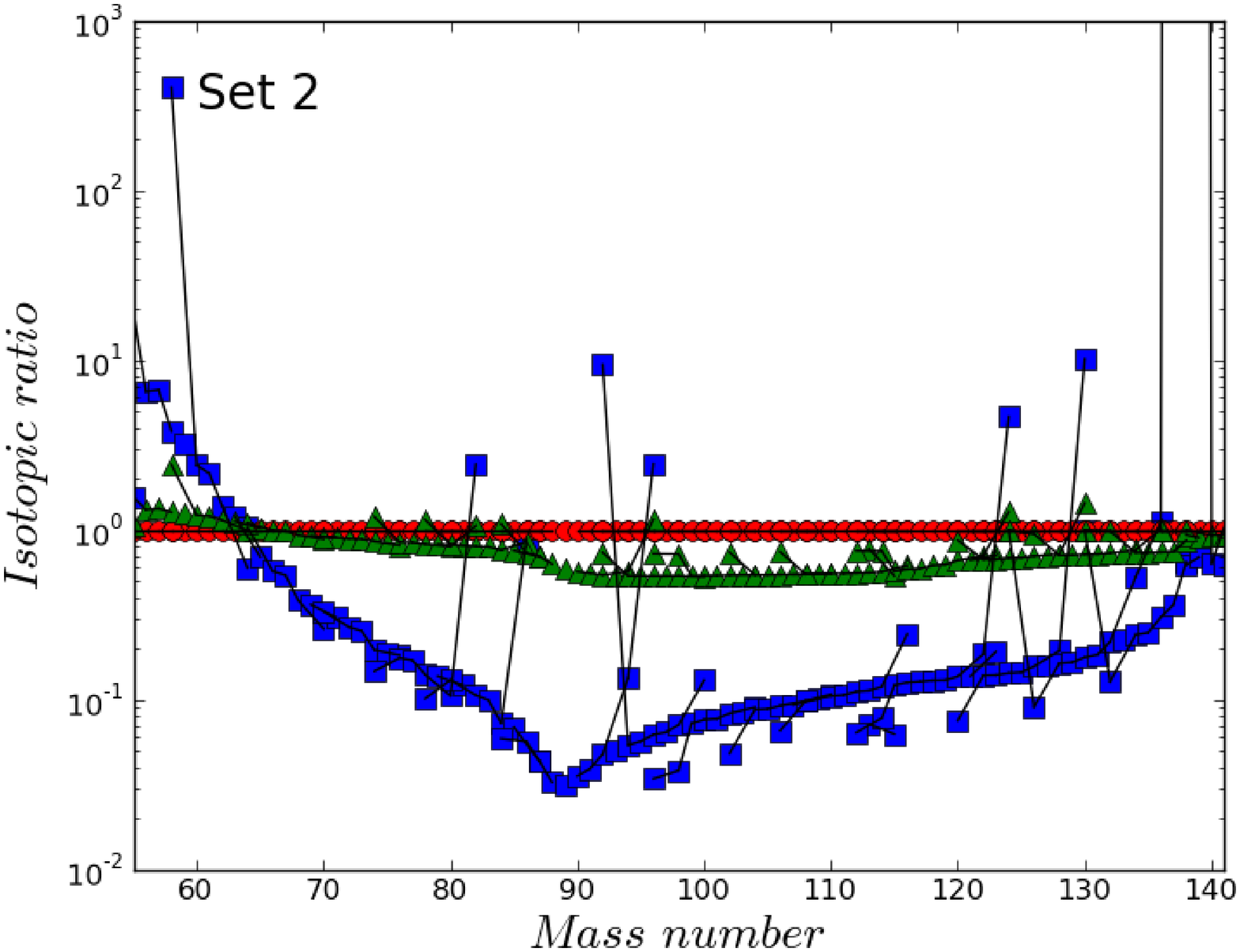}}}
  \caption{Left Panel $-$ The isotopic abundances at the end of C-burning (when the mass fraction of $^{12}$C left is less than 2\%) for the cases using the $^{12}$C+$^{12}$C channel ratio $R_{\alpha}$/$R_p$ = 0.05/0.95 (green triangles)
and $R_{\alpha}$/$R_p$ = 0.95/0.05 (full blue squares), normalized to the isotopic distribution obtained using the standard
$R_{\alpha}$/$R_p$ = 0.65/0.35. Species belonging to the same element are connected with lines.
The temperature and density used for the calculations are
$T$ = 1.0 GK and $\rho$ = 10$^5$ cm$^{-3}$, the $^{12}$C+$^{12}$C rate used is CF88 ($set1$).
Right Panel $-$ As in the left panel, but for $T$ = 0.65 GK and $\rho$ = 10$^4$ cm$^{-3}$
and $^{12}$C+$^{12}$C CU rate ($set2$).
Note that the relative abundance distributions are shown with a different scale in
the y-axis compared to the left panel.
}
\label{fig:c12c12ap_range}
\end{figure}

In order to test the impact of this uncertainty, we have performed two sets of calculations using a simple single-zone post-processing method:
using the CF88 $^{12}$C+$^{12}$C rate ($set1$, CF88 rate, $T$ = 1.0 GK, $\rho$ = 10$^5$ g cm$^{-3}$) for typical shell C$-$burning conditions and the CU rate for
conditions dominated by core C$-$burning ($set2$, CU rate, $T$ - 0.65 GK, $\rho$ = 10$^4$ g cm$^{-3}$), the ratio  $R_{\alpha}$/$R_p$ has been modified within the range 0.95/0.05 $\lesssim$ $R_{\alpha}$/$R_p$ $\lesssim$ 0.05/0.95.
In Fig. \ref{fig:c12c12ap_range},
the C$-$burning yields from the most extreme cases are compared with the yields obtained using the
standard ratio 0.65/0.35. Results for $set1$ and $set2$ are presented in the left and right panel, respectively.
For a more direct comparison, in Tab. \ref{tab:c12c12_ap_summary} are reported the abundances for selected species corresponding to the cases presented in Fig. \ref{fig:c12c12ap_range}.

\begin{figure}
\centering
\resizebox{10cm}{!}{\rotatebox{0}{\includegraphics{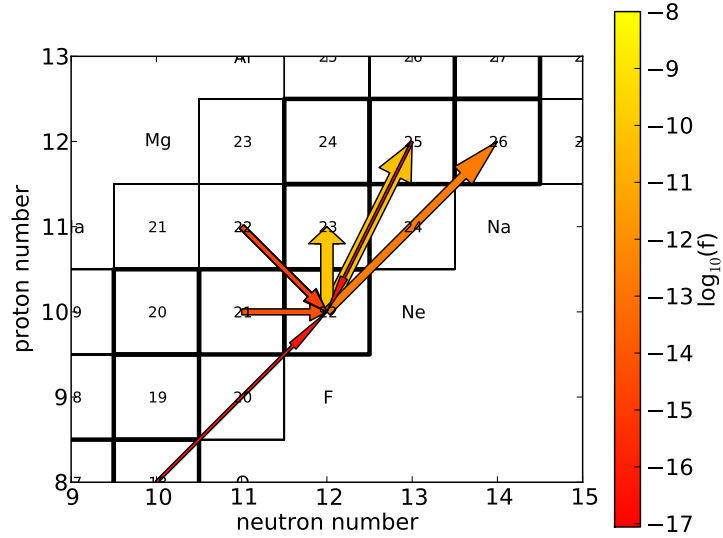}}}
\caption{The nucleosynthesis fluxes producing and depleting $^{22}$Ne are shown in shell C-burning conditions.
The arrow size and color correspond to the flux strength.}
\label{fig:flux_ne22_st}
\end{figure}

Concerning $set1$, despite the large difference in the $R_{\alpha}$/$R_p$ ratio, the variation is below a factor of two for most species.
A larger departure from the standard isotopic distribution is related
to $p$-process species (e.g., $^{92}$Mo), which are only depleted
by neutron captures and that have in all cases negligible abundances.
Increasing the number of $\alpha$ particles compared to protons has the effect of increasing the number of neutrons available, therefore increasing the $s$-process efficiency. In particular, the $s$-process species between Fe and Zr are boosted (see Table \ref{tab:c12c12_ap_summary}).
On the other hand, increasing the number of protons reduces the $s$-process efficiency,
even if protons are mostly converted to $\alpha$ particles via $^{23}$Na(p,$\alpha$)$^{20}$Ne.
The main reason for this behavior is that in C$-$burning conditions the $^{22}$Ne($\alpha$,n)$^{25}$Mg nucleosynthesis channel
competes with $^{22}$Ne(p,$\gamma$)$^{23}$Na. This is shown in Fig. \ref{fig:flux_ne22_st} for the standard case $set1$,
where the fluxes producing and depleting $^{22}$Ne are presented.
Therefore, variation of the $R_{\alpha}$/$R_p$ ratio also affects $^{22}$Ne nucleosynthesis. Notice that during C$-$burning the
$^{22}$Ne($\alpha$,$\gamma$)$^{26}$Mg rate is marginal compared to the
($\alpha$,n) rate.

In Fig. \ref{fig:c12c12ap_range}, right panel, $set2$ calculations show
a larger variation than $set1$. In particular, a stronger
$^{12}$C+$^{12}$C proton channel does not significantly modify the
$s$-process yields and the results discussed in
\S \ref{sec:c12c12sprocess} are confirmed. Conversely, a stronger
$\alpha$ channel reduces the $s$-process yields by
more than a factor of 10 at the Sr-Y-Zr peak.
For instance, $^{88}$Sr (neutron magic number, N=50) in the a95p05*
case (where $\alpha$-channel probability = 95\%, proton-channel probability = 5\%) is reduced by a factor of $\sim$ 30
compared to the st* case ($\alpha$-channel probability = 65\%, proton-channel probability = 45\%, see Tab. \ref{tab:c12c12_ap_summary}).
This effect is the opposite to that in $set1$, where a stronger
$\alpha$ channel leads to a more efficient $s$-process production.
The main neutron source in $set2$
(and the $s$-process ejecta when using the CU $^{12}$C+$^{12}$C rate) is the $^{13}$C($\alpha$,n)$^{16}$O, not $^{22}$Ne($\alpha$,n)$^{25}$Mg.
In the temperature conditions of $set2$, the photodisintegration timescale of $^{13}$N via $^{13}$N($\gamma$,p)$^{12}$C is comparable with the $\beta$-decay of $^{13}$N to $^{13}$C (see \S \ref{sec:c12c12sprocess}). As a result, the available $^{13}$C abundance depends on the amount of protons available to activate the $^{12}$C(p,$\gamma$)$^{13}$N reaction.
As such, increasing the $R_{\alpha}$/$R_p$ ratio with respect to the
standard value reduces the proton density
and consequently the $^{13}$C neutron source.
Calculations in $set2$ also show significant variations for light species; a stronger $\alpha$-channel leads to a significant decrease of
$^{16}$O and $^{23}$Na, and an increase of $^{20}$Ne (see details in Tab. \ref{tab:c12c12_ap_summary}).
Conclusions obtained in \S \ref{sec:gnv} need to be viewed cautiously
since the CU model, as well as the other four models, was calculated
using a constant value of $R_{\alpha}$/$R_p$ = 0.65/0.45.
If the $^{12}$C+$^{12}$C rate is indeed larger than CF88,
then the impact of the  $R_{\alpha}$/$R_p$ ratio uncertainty on stellar
calculations becomes more important and may be not negligible with
regard to the main features of subsequent evolutionary stages.

In summary, the total
$^{12}$C+$^{12}$C fusion rate is not the only uncertainty to
consider when comparing different
nucleosynthesis calculations. In this section, we showed that the
$R_{\alpha}$/$R_p$ ratio must also be considered and more
experimental efforts are required to constrain the relative
strength of the primary decay channels.

\section{$^{12}$C($^{12}$C,n)$^{23}$Mg channel}
\label{sec:c12c12n}

Besides the two main nucleosynthesis channels,
feeding $^{20}$Ne and $^{23}$Na, respectively, the
$^{12}$C+$^{12}$C reaction can also result in the production
of neutrons via the $^{12}$C($^{12}$C,n)$^{23}$Mg
channel. The isotope $^{23}$Mg is unstable, decaying rapidly
($\tau_{1/2}$ = 11.3 s) into $^{23}$Na. The reaction
is endothermic with a Q-value of Q=-2.6MeV and will contribute only
at higher temperatures.
Yet, even at high energies the contribution of the reaction branch
is small. An early analysis by \cite{dayras:77} (D77 hereafter)
indicates 2\% at 3 MeV gradually increasing up to 10\%.
This is confirmed by more recent measurement
(X. Tang, private communications).
For the present calculations,
the $^{12}$C($^{12}$C,n)$^{23}$Mg rate of D77
has been used.
In simulations where the total $^{12}$C+$^{12}$C fusion rate
corresponds to the CF88 rate, the neutron flux produced by the
$^{12}$C($^{12}$C,n)$^{23}$Mg branch is weaker than the
$^{22}$Ne($\alpha$,n)$^{25}$Mg,
at least for solar-like metallicities.
Notice, however, that the $^{22}$Ne($\alpha$,n)$^{25}$Mg neutron source
is secondary and its efficiency depends on the initial metal content,
whereas $^{12}$C($^{12}$C,n)$^{23}$Mg is a primary source.
In comparison with the third potential neutron source
$^{13}$C($\alpha$,n)$^{16}$O,
the $^{12}$C($^{12}$C,n)$^{23}$Mg probability is marginal for models
using the CU rate as will be discussed below.

The experimental uncertainty in the $^{12}$C($^{12}$C,n)$^{23}$Mg data
is approximately 50\% at energies
above 4 MeV, towards lower energies it increases up to 80\%.
The cross section is expected to decline rapidly since the neutron
channel closes at 2.6 MeV, yet the existence of molecular resonance
structures near the threshold cannot be excluded, because such features are visible
in the proton and $\alpha$-channels at the corresponding energy range.
For low temperature environments typical for C shell burning near 1 GK we
therefore consider an uncertainty range of approximately a factor of 10;
for higher temperatures as found at the bottom of the C shell
during the later evolutionary stages prior to core collapse we reduce the
uncertainty range to a factor of five since the Gamow window extends into the
excitation range well covered by experiment. We do not consider the reaction
for typical core C-burning conditions, since
the neutron channel of C-burning is endoergic and therefore suppressed
towards the low temperatures associated with the core burning environment .

\begin{figure}
\centering
\resizebox{10cm}{!}{\rotatebox{0}{\includegraphics{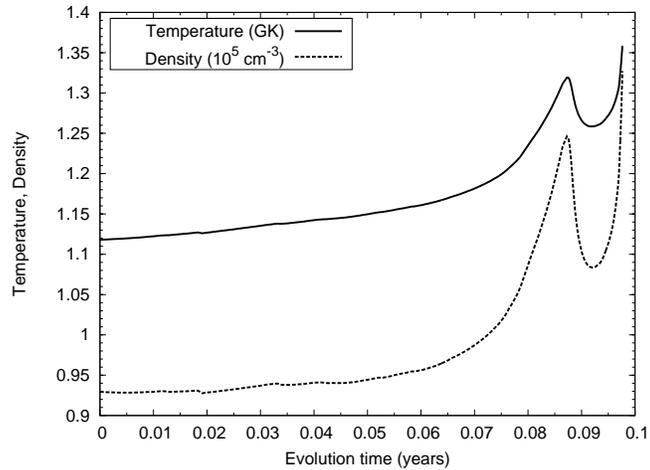}}}
\caption{Temperature and density profiles at the bottom of the final convective carbon shell for the CF88t10 stellar model. Evolution time = 0 corresponds to the start of shell C-burning.
}
\label{fig:temp_density_profile_for_c12c12n}
\end{figure}

In order to test the impact of the neutron channel, and taking into account its uncertainty, we have performed two sets of
calculations using different $^{12}$C($^{12}$C,n)$^{23}$Mg rates (D77, and D77 multiplied by a factor of 2, 5 and 10,
which are called D77t2, D77t5 and D77t10):
$set3$, where typical C-shell conditions with constant temperature and density are considered
(CF88 $^{12}$C+$^{12}$C rate, $T_{\rm set3}$ = 1.1 GK and
$\rho$ = 10$^{5}$ g cm$^{-3}$, the $^{12}$C($^{12}$C,n)$^{23}$Mg probability is $\sim$ 0.12 \%
of the total rate), and $set4$, where realistic temperature and density profiles are extracted
at the bottom of the final convective carbon shell
of the CF88t10 model (using CF88t10 rate).
Notice that cases with rates lower than D77 are not considered here,
since the the $^{12}$C($^{12}$C,n)$^{23}$Mg contribution
becomes negligible.
The temperature and density profiles used for $set4$ calculations are reported in
Fig. \ref{fig:temp_density_profile_for_c12c12n}
and are characterized by an increase in the final temperature and density, up to
$T$ = 1.36 GK and $\rho$ = 1.33$\times$10$^5$ g cm$^{-3}$.
During core collapse, and shortly before the supernova explosion,
thermal instabilities occur in O$-$burning layers
and temperature and density
may rapidly increase in the outer stellar material
\citep[][]{arnett:74,arnett:78}.
We choose the CF88t10 model because it shows
a convective C shell lasting until the SN explosion.
In all of the other models considered in this work, the convective phase terminates at an earlier point in the evolution.
In these cases, stellar structure modifications triggered by O ignition
in the core prohibits convective shell C$-$burning from continuing
\citep[e.g.,][]{hirschi:04} and
the yields are not affected by the temperature increase shown in
Fig. \ref{fig:temp_density_profile_for_c12c12n}.
In the CF88t10 model, the final temperature and density rise causes
a boost in the neutron density, resulting in distinctive effects in the
$s$-process distribution. For instance, all the isotopes considered as
$r$-only between iron and strontium may be signicantly produced
($^{70}$Zn, $^{76}$Ge, $^{80,82}$Se),
$^{75}$As overproduction might be comparable to neighbor $s$-only
isotopes (e.g., $^{76}$Se), and, in general,
$s$-process branchings are significantly affected
\citep[e.g.,][]{pignatari:10}.

During this final phase, the $^{12}$C($^{12}$C,n)$^{23}$Mg channel may also become more efficient,
since its strength compared
to the $\alpha$ and the proton channel increases with temperature
\citep[][]{dayras:77}.
The main purpose of the two sets presented in this section is to
test the impact of the $^{12}$C($^{12}$C,n)$^{23}$Mg channel
in typical C shell conditions ($set3$) and in the final
temperature-rise phase ($set4$).

\begin{figure}
\centering
\resizebox{12cm}{!}{\rotatebox{0}{\includegraphics{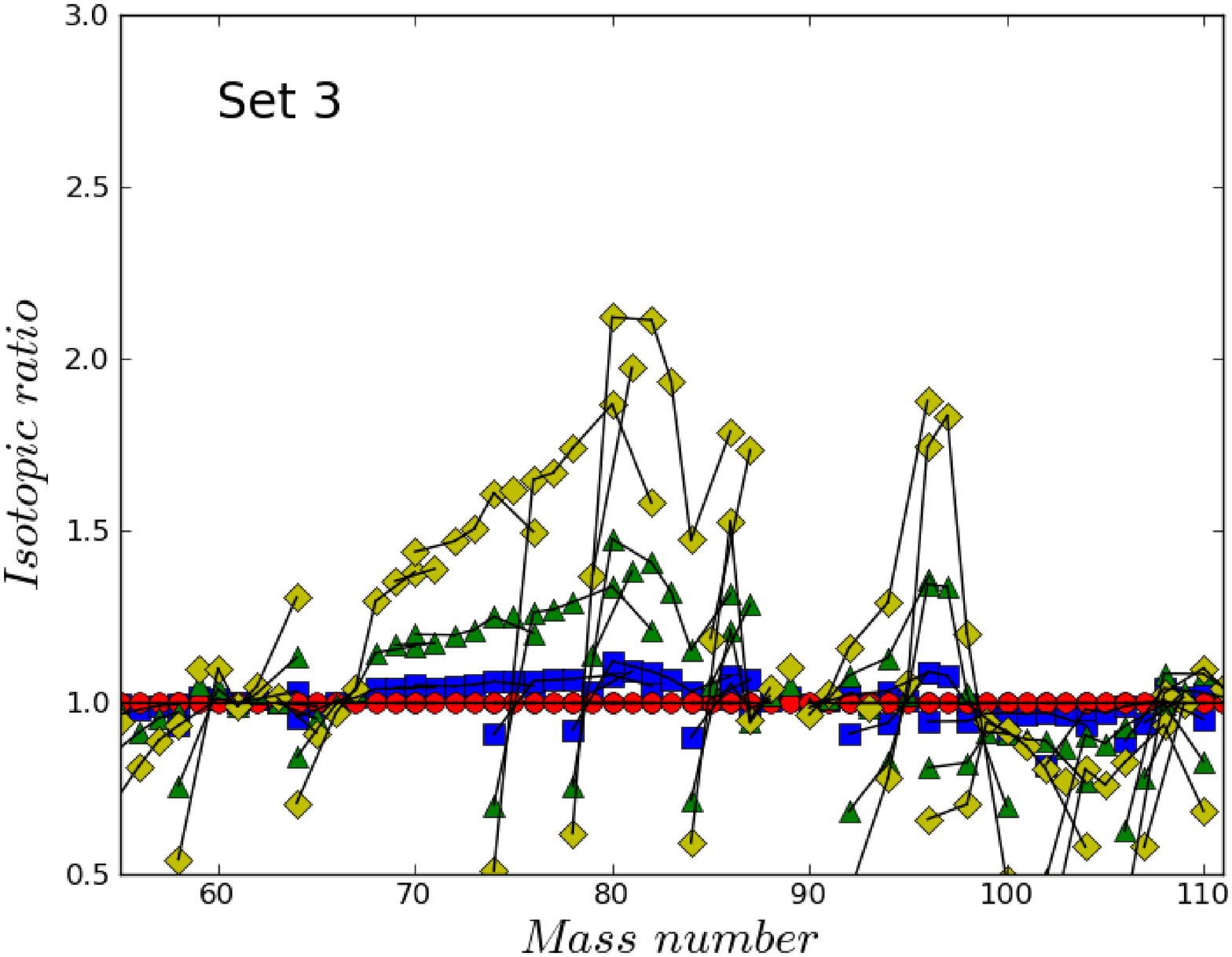}}}
\resizebox{12cm}{!}{\rotatebox{0}{\includegraphics{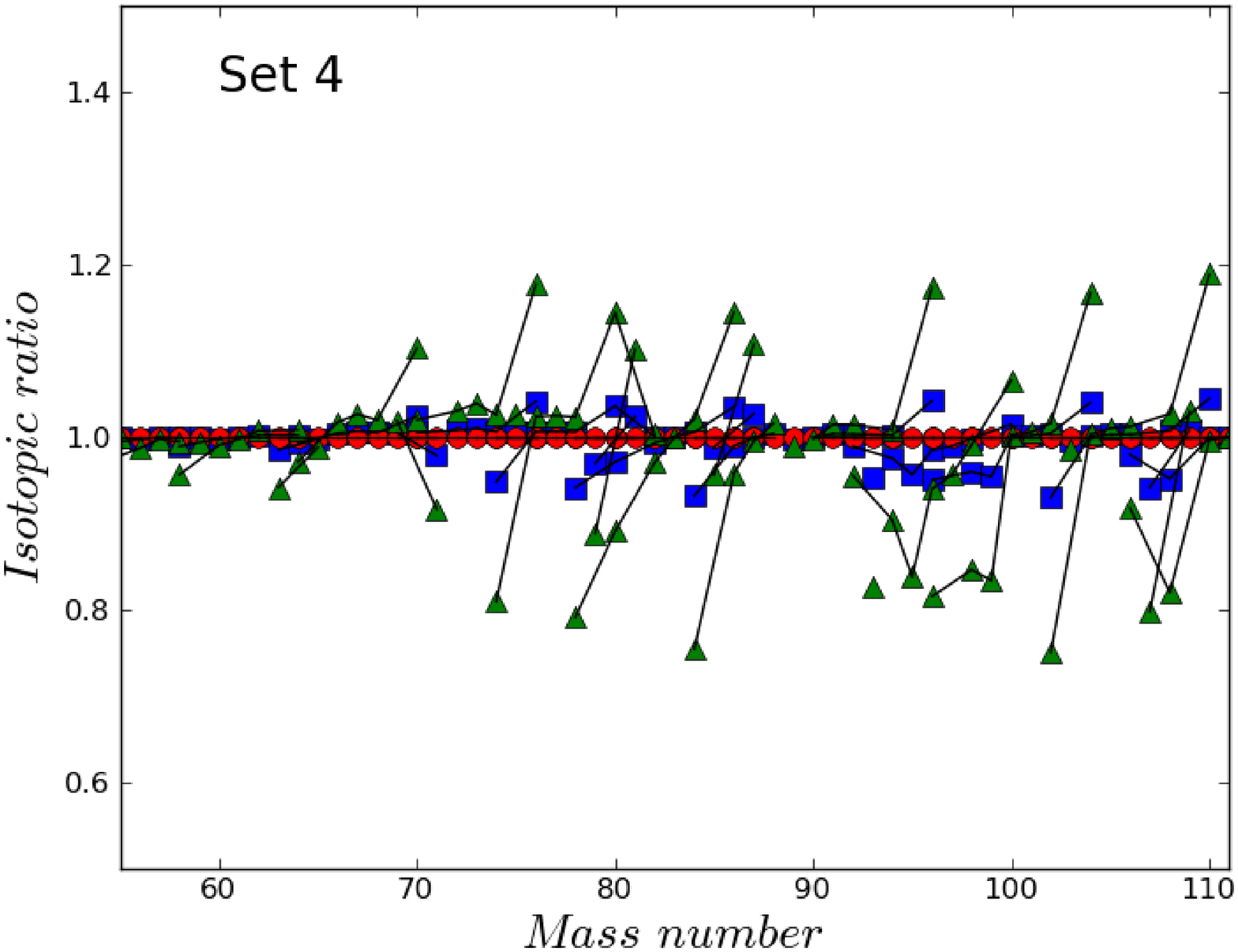}}}
  \caption{Left Panel $-$ The abundances at the end of shell C-burning (when the mass fraction of
$^{12}$C left is less than 2\%) for the $set3$ calculations, using the $^{12}$C($^{12}$C,n)$^{23}$Mg rate by \cite{dayras:77}
multiplied by a factor of two (blue squares), by a factor of five (green triangles) and by a factor of ten (yellow diamonds), normalized to the abundance distribution obtained using the standard $^{12}$C($^{12}$C,n)$^{23}$Mg rate.
Species belonging to the same element are connected with lines. The temperature and density of these runs are $T$ = 1.1 GK and $\rho$ = 10$^5$ g cm$^{-3}$.
Right Panel $-$ As in the left panel, but only for the
$^{12}$C($^{12}$C,n)$^{23}$Mg rate multiplied by 2 (blue squares) and by a factor of 5 (green triangles). The C shell trajectory
used in the calculations is given in Fig. \ref{fig:temp_density_profile_for_c12c12n}.
}
\label{fig:c12c12n_range}
\end{figure}

The relative isotopic distributions for the D77t2, D77t5 and D77t10 cases are shown for $set3$ in Fig. \ref{fig:c12c12n_range}.
The right panel shows the same calculations for $set4$. Abundances for selected species are reported in Tab. \ref{tab:c12c12_n_t9p1} and \ref{tab:c12c12_n_trajCF88p10} for the most extreme cases,
compared to D77.
In both sets of calculations presented in Fig. \ref{fig:c12c12n_range}, the D77t2
case shows small departures from the standard case within 10 \%.
On the other hand, in the D77t5 and D77t10 cases the $s$-only species between iron and strontium ($^{70}$Ge, $^{76}$Se, $^{80,82}$Kr and
$^{86,87}$Sr) show an increase in their average mass fraction
of 1.30 and 1.77, respectively (Tab. \ref{tab:c12c12_n_t9p1}).
This means that the average neutron exposure has increased (see also $^{88}$Sr in Tab. \ref{tab:c12c12_n_t9p1}).
Interestingly, an increase in the availability of neutrons due to a more efficient $^{12}$C($^{12}$C,n)$^{23}$Mg channel
causes a propagation effect over the $s$-process species up to Sr, with significant differences along the neutron capture path.
Illustrating this point, the variation of $^{80,82}$Kr is larger than lighter or heavier $s$-only species (e.g., in the D77t10 case
compared to the D77 case, the $^{82}$Kr abundance increases by a factor of 2.1, compared to 1.4 for $^{70}$Ge, Tab. \ref{tab:c12c12_n_t9p1}).
The $r$-only isotopes like $^{70}$Zn or $^{76}$Ge are enhanced for the cases D77t5 and D77t10 compared
to the standard case, with a production comparable to the $s$-process species
(see Tab. \ref{tab:c12c12_n_t9p1}).

The evolution history of the stellar structure has a relevant impact on the C shell nucleosynthesis, affecting
for example the amount of $^{12}$C available when C ignites in a convective shell
\citep[e.g.,][]{imbriani:01,eleid:09}.
The $set4$ calculations presented in Fig. \ref{fig:c12c12n_range}
use the initial abundances and the temperature-density
evolution of the last convective C shell in the CF88t10 model.
Compared to $set3$, $set4$ models show less $^{12}$C fuel available
(the $^{12}$C abundance is $\sim$ 0.075,
compared to 0.27 in $set3$).
This is because of the large overlap with the previous convective shell, where most of the carbon fuel was consumed.

Therefore, the relatively low amount of $^{12}$C significantly reduces the impact of a higher $^{12}$C($^{12}$C,n)$^{23}$Mg rate
compared to $set3$, and the average mass fraction is only marginally modified.
On the other hand, in the D77t5 case, most of the $r$-only isotopes are increased by a factor of $\sim$ 1.2 compared to the standard case,
and, at the same time, the $p$-only isotopes are reduced by a similar factor.
In fact, such modifications are triggered by local neutron captures due to the final short neutron burst, where $r$-only isotopes are fed from local abundant
$s$-process isotopes and the $p$-rich isotopes are more depleted via neutron capture. The impact of a final neutron burst,
arising from a higher $^{12}$C($^{12}$C,n)$^{23}$Mg rate, may be more severe than the case considered in $set4$
since its relevance is directly related to the amount of
carbon fuel that is still available when the temperature starts to rise.
As such, the estimation regarding the impact of
the $^{12}$C($^{12}$C,n)$^{23}$Mg uncertainty on C shell yields may
change according to the
$^{12}$C+$^{12}$C rate used and to the stellar model
considered.

Summing up, present $s$-process abundance predictions
may be affected locally (at the different branching points producing more $r$-only species, less $^{80}$Kr, etc), and/or more generally
over the entire $s$-process distribution, by the uncertainty associated with the
  $^{12}$C+$^{12}$C neutron channel. Fig. \ref{fig:c12c12n_range} shows the two different effects that can be obtained,
depending on the stellar model.
The $^{12}$C($^{12}$C,n)$^{23}$Mg probability needs to be known at stellar carbon
 burning conditions with a precision significantly better than a factor of 5.
We have shown that a factor of 2 precision would have an impact on the
 $s$-process predictions within an acceptable 10 \%.



\section{Implications for the $p$-process nucleosynthesis}
\label{sec:pprocess}


In this section we explore the impact of the $^{12}$C+$^{12}$C rate uncertainty
on $p$-process nucleosynthesis.
In the solar abundance distribution there are 35 $p$-rich stable nuclei,
from $^{74}$Se to $^{196}$Hg, that are called
$p$-only isotopes since they cannot be produced by either
the $s$-process or the $r$-process.
In the solar system the $p$-only nuclei are about
two orders of magnitude less abundant
when compared to the other stable isotopes
of the same element fed by the $s$- and $r$-process.
Exceptions are the $p$-nuclei $^{92,94}$Mo and $^{96,98}$Ru, which represent respectively 14.84, 9.25
\% and 5.52, 1.88 \% of the total abundance of the respective element.

Different astrophysical sources have been proposed to either reproduce or
contribute to the solar $p$-process distribution
\citep[see e.g.,][]{arnould:03,froehlich:06,travaglio:11}.
At present, the most well-established astrophysical site for $p$-process nucleosynthesis is
the O/Ne-rich layers in massive stars, before and after
the SN explosion \citep[e.g.,][]{arnould:76,woosley:78}.
The pre-explosive $p$-process component is mostly re-processed
by the explosive nucleosynthesis triggered by the SN shock wave.
Its effective impact on the total $p$-process yields depends
on the explosion mechanism and
on the stellar structure behavior in the last days before core collapse.
For instance, extensive convection and
mixing between different burning shells
\citep[e.g., convective O shell mixing with convective C shell,][]{rauscher:02}
may trigger the enrichment of SN O-rich ejecta by pre-explosive $p$-process nuclei
\citep[see also][]{arnett:11}.

Explosive $p$-process nucleosynthesis in standard SNe type-II explosions
shows a weak dependence on the initial mass of the star \citep[e.g.,][]{rayet:95}.
The initial abundances for the $p$-process are provided by
the $s$-process distribution, built in the previous evolutionary phases.
This implies that the $p$-process efficiency depends
on the initial metallicity of the star, showing a secondary nature similar to the
weak $s$-process \citep[][]{rayet:95, arnould:03}.

\begin{figure}
\centering
\resizebox{8cm}{!}{\rotatebox{0}{\includegraphics{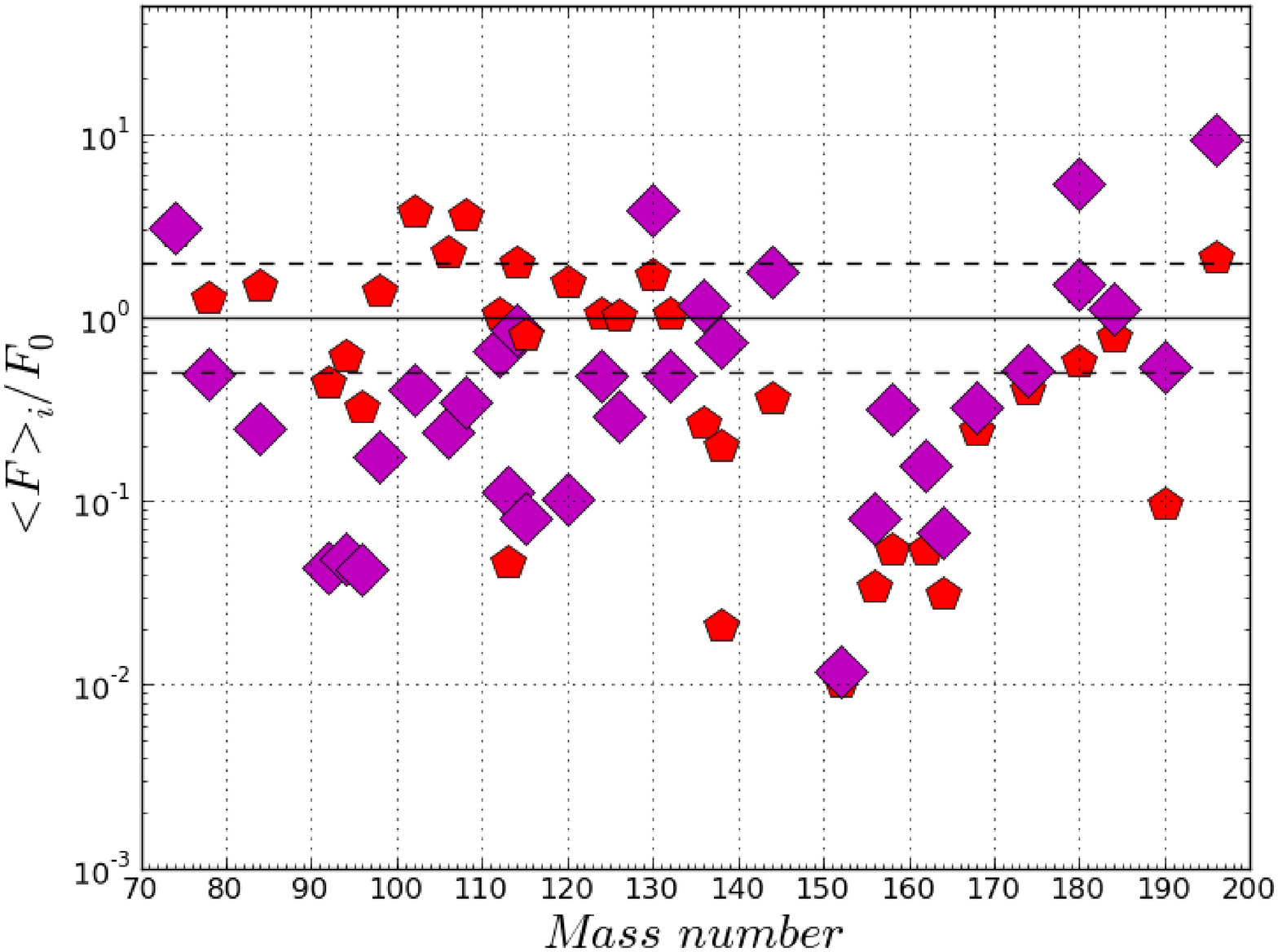}}}
\resizebox{8cm}{!}{\rotatebox{0}{\includegraphics{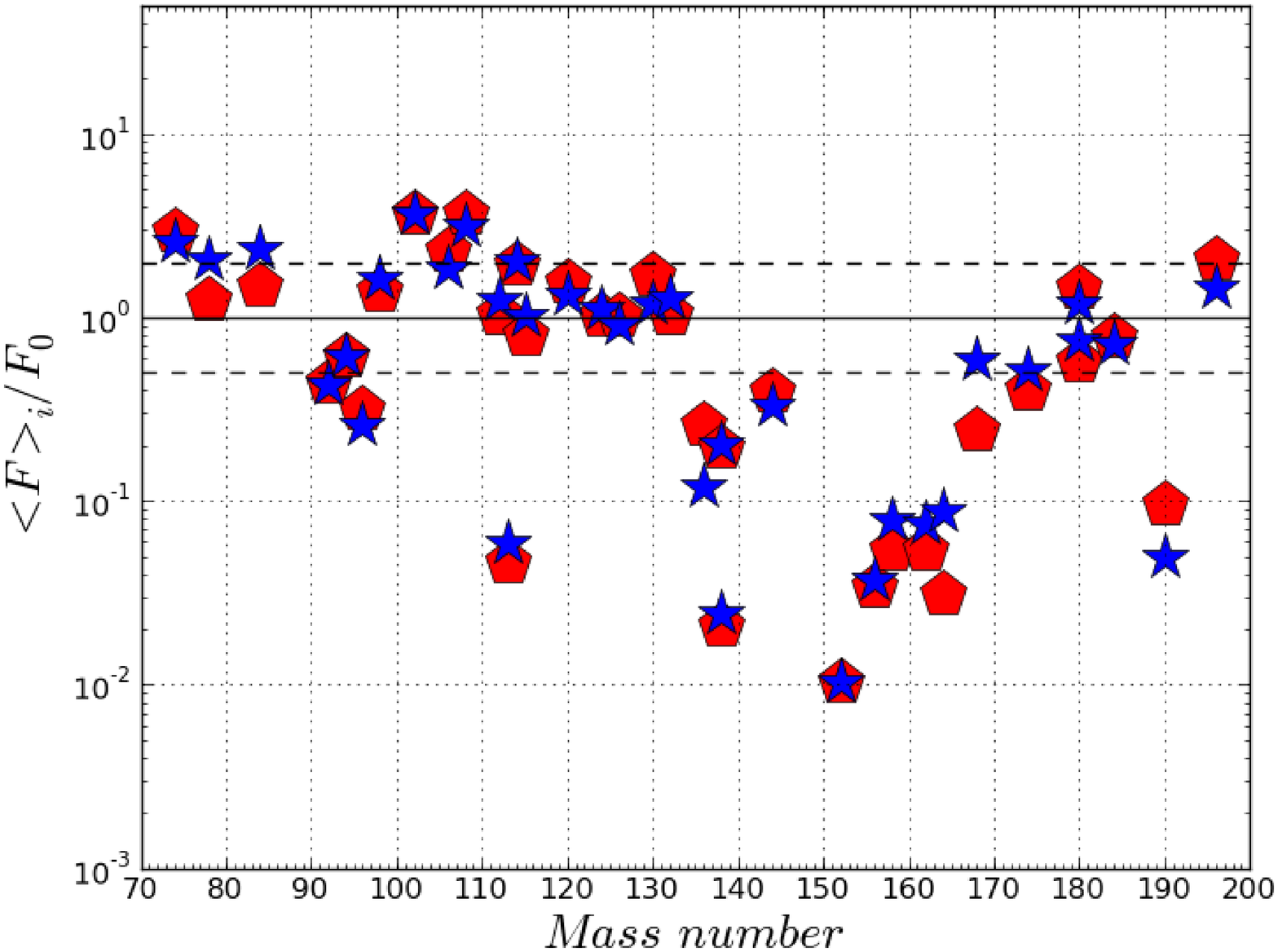}}}
\caption{Left Panel $-$ Average overabundance distribution of the $p$-process nuclei over the 14
trajectories by \cite{rapp:06}, for the 25 M$_{\odot}$ star, Z = 0.01,
using as seeds the abundances from the CF88t10 model (pCF88t10, purple diamonds) and CU
(pCU, red pentagons).
The average overproduction factor $F_o$ is 42.3 and 347.7, respectively.
Right Panel $-$ The pCU distribution is compared with the same case,
calculated using the post-processing code by \cite{rapp:06}.
}
\label{fig:pprocess_mean_25_z1m2_all}
\end{figure}

In order to investigate the impact of the $^{12}$C+$^{12}$C rate on the
$p$-process abundances, we have performed a set of one-zone explosive nucleosynthesis simulations. We have used the same explosive trajectories used by \cite{rapp:06}.
Initial abundances are given by the pre-explosive composition calculated in the
models presented in this paper.
In Fig. \ref{fig:pprocess_mean_25_z1m2_all} (left panel), we compare the
$p$-process abundance distribution obtained using the abundance seeds from the
CF88t10 and CU models
(pCF88t10 and pCU,
respectively). The other models discussed in the
previous sections are similar to pCF88t10
(\S \ref{sec:gnv}).
The abundance distribution of pCF88t10 is consistent with previous analyses
of the $p$-process nucleosynthesis in explosive O/Ne layers in massive stars
\citep[e.g.,][]{prantzos:90a, rayet:95}.
About 60~\% of the $p$-process isotopes are produced in comparable
amounts \citep[e.g.,][]{rayet:90,rayet:95}.
A sample of isotopes are underproduced compared to the average production
factor $F_o$.  In all present models, the nuclei $^{92,94}$Mo and $^{96,98}$Ru are
systematically underproduced by
an order of magnitude or more in all nucleosynthesis scenarios using realistic
massive stars conditions \citep[][]{arnould:03}.
This problem cannot be resolved by nuclear physics uncertainties
\citep[][]{rauscher:06,rapp:06}.
A possible solution was proposed by \cite{arnould:92}, showing that
artificially increasing
$s$-process abundances for A $\geq$ 90 would solve the Mo-Ru puzzle.
It was argued by \cite{costa:00} that using a larger $^{22}$Ne($\alpha$,n)$^{25}$Mg rate (within the upper limit of \cite{angulo:99}) could
increase $s$-process yields in massive stars at the Sr peak and beyond.
However, not only is this finding unconfirmed by other results \citep[][]{heger:02}, such a high rate would also result in disagreement between present weak and main $s$-process component predictions in the solar system abundance
distribution. Furthermore, the capability of the present $s$-process stellar AGB models
to reproduce the $s$-process signature measured in carbonaceous
presolar grains would be called into question \citep[see also discussion in][]{pignatari:10}.

Alternatively, other processes in massive stars different from the
classical $p$-process have been proposed to contribute to the missing
Mo-Ru $p$ abundances, e.g., in $\alpha$-rich freezout conditions
during the SNII explosion \citep[e.g.,][]{hoffman:96} or the
$\nu$$p$-process in proton-rich neutrino-wind conditions
\citep[][]{froehlich:06}.
The $p$-process underproduces $^{113}$In and $^{115}$Sn within present nuclear uncertainties, despite the potential contribution
from the $r$-process
\citep[][and references therein]{rapp:06,dillmann:08a}.
On the other hand, $^{152}$Gd and $^{164}$Er receive a significant contribution from the $s$-process
\citep[e.g.][]{arlandini:99,abbondanno:04,best:01}.
Finally, $^{138}$La might receive a significant contribution from neutrino
capture on $^{138}$Ba \citep[][]{goriely:01}, which is not considered in these calculations.

For the pCF88t10 case, we obtain an overproduction of
$^{130}$Ba, $^{180}$Ta and $^{196}$Hg.
The overproduction is acceptable within the nuclear uncertainties for
 $^{130}$Ba.
The $p$-process production of $^{180}$Ta is affected by the pre-explosive
$s$-process  abundance distribution.
The pCF88t10 yields presented in Fig. \ref{fig:pprocess_mean_25_z1m2_all}
are characterized by a general underproduction compared
to $^{16}$O.
According to \cite{rayet:95}, the total production factors for $p$-nuclei and $^{16}$O in the ejected SN material
should be comparable, assuming that massive stars are their main astrophysical site.
For instance, \cite{rayet:95} estimated such underproduction
to be at least a factor of two for a 25 M$_{\odot}$ star, considering
uncertainties associated with both the stellar model and nuclear input, such
as the
$^{12}$C($\alpha$,$\gamma$)$^{16}$O rate.
In the pCF88t10 case, the average overproduction factor is $F_{0}$ = 42.3.

Considering a mass correction factor of $f_m$ = 0.2, the p/O yield
ratio is $\sim$ 0.12. This takes into account an $^{16}$O
overproduction factor compared to the solar overabundance ($\sim$ 70)
and the amount of p-rich and O-rich mass ejected by a standard 25
M$_{\odot}$ core collapse SN model
\citep[see for instance Table 3 of][where $f_m$ ranges
from 0.19 to 0.23 depending on the different model
prescriptions]{rayet:95}.

The average overproduction factor of the pCF88t10 case ($F_{0}$ = 42.3) is
consistent with the value provided by \cite{rayet:95} for a 25 M$_{\odot}$ model
 at solar metallicity and for the IMF weighted $p$-process
distribution ($F_{0}$ = 130 and 100, respectively), considering that the classic $p$-process in massive stars is a secondary process.
Indeed, its efficiency critically depends on the $s$-process abundance seeds
from the previous stellar phases \citep[][]{arnould:03}.
It is also important to notice that according to \cite{tinsley:80},
 at solar metallicity secondary species (in this case, the $p$-process yields)
should be produced twice as much as primary species (e.g., $^{16}$O).
Consequently the underproduction of the $p$-process in the SNII scenario is
likely to be more severe than mentioned by \cite{rayet:95}; the yield ratio
required to reproduce the $p$-process inventory should be $p$/O = 2 at
solar metallicity,
twice as much as previously estimated.

In Fig. \ref{fig:pprocess_mean_25_z1m2_all}
(left panel), the pCU case shows significant
differences compared to the pCF88t10 case.
First, $F_{0}$ is more than a factor of eight higher than in the standard case
($F_{0}$ = 348).
Second, a good fit is obtained for $^{92,94}$Mo and $^{98}$Ru within a factor of
two uncertainty, and only a factor of three underproduction for $^{96}$Ru.
The higher $s$-process efficiency of the cs component in the CU model at the
Sr peak and beyond provides the required seed distribution to feed this mass
region.
It is important to mention that for the pCU case the lighter $p$-process
species are mostly produced by the coldest (and more external)
trajectories, whereas in standard $p$-process conditions they are
mostly produced by the hottest (and deepest) particles
due to photodisintegration flows destroying heavier species.
Therefore, in this case, local photodisintegration
channels are mainly responsible for $p$-nuclei yields,
and previous impact studies of nuclear
uncertanties \citep[e.g.,][]{rauscher:06,rapp:06} should be reconsidered.
The underproduction of $^{113}$In, $^{138}$La, $^{152}$Gd and $^{164}$Er
is confirmed also for the pCU case. The problem of
$^{130}$Ba, $^{180}$Ta and $^{196}$Hg overproduction, seen in pCF88t10,
is also solved
due to the increase of $F_{0}$.
However $^{190}$Pt is now underproduced.

As a verification test, we compare the pCU $p$-process abundances with the results obtained by the code used
in \cite{rapp:06}, using the same initial distribution and the same trajectories (Fig. \ref{fig:pprocess_mean_25_z1m2_all}).
Considering that the two codes used different nuclear networks,
the final compositions are consistent, with maximum yield variation within a
factor of two.
We also tested how the present results may be affected by using different trajectories.
Only marginal variations in the final $p$-process distribution were
obtained by either changing the freezout temperature in the explosive
particles, using different power law prescriptions, or assuming an
exponential decrease. Therefore, the
present conclusions are not compromized.

From the present calculations, for the higher (CU) $^{12}$C+$^{12}$C rate, the
overlap between
the convective C core and the convective C shell provides a higher
$s$-process efficiency at the Sr peak and beyond, forming the cs component
(\S \ref{sec:c12c12sprocess}).
Such a distribution feeds enhanced $p$-process yields, also reproducing the
solar $p$-process abundances in the Mo-Ru region within a factor 2-3
(Fig. \ref{fig:pprocess_mean_25_z1m2_all}).
Therefore the cs component obtained from the CU model provides a set of
$s$-process seeds that may efficiently feed the final abundances of the
puzzling Mo and Ru $p$-only species,
and provide a satisfactory distribution of heavier $p$-nuclei.

We have seen that the cs component is difficult to reconcile with the
$s$-process abundances observed in the solar system,
unless only a few per cent of the total $s$-process rich mass ejected by
massive stars can carry it. By extention the $p$-process rich distribution
obtained from CU (cp component, hereafter)
has to follow the same restrictions discussed in \S \ref{sec:c12c12sprocess}.
Indeed, we have shown that if about 2 \%
of the $s$-process rich mass ejected in to the ISM has the
enhanced component,
then this could provide a possible scenario to solve the LEPP puzzle in
the solar system distribution.
Assuming that 2 \% of the ejected $p$-process rich mass carries the cp
component, (with $F_{0}$ = 348 in the pCU case), only about 2 \% of the
average $p$-process abundance
in the solar system distribution could be reproduced by the cp component
(348$\times$$f_m$$\times$0.02/70, where $f_m$=0.2 is the mass correction
factor described above and 70 the $^{16}$O overproduction).
The $F_{0}$ from the pCF88t10 case, assuming that the remaining ejected mass
shows a standard $s$-process and $p$-process signature, would correspond
to $\sim$ 12 \% of the average $p$-process solar distribution.
The low $p$-process efficiency in massive stars simulations cannot
therefore be solved by considering the cp component in the $p$-process
inventory, and it is even worse than previously estimated
if we also consider the secondary nature of the classic $p$-process.
An amount of only a few per cent of solar Mo-Ru $p$-nuclei could
be produced.
If such speculations will be confirmed,
another astrophysical source for the cosmic abundances of
the $p$-process nuclei is needed.
In this direction, thermonuclear supernovae (SNIa) might efficiently produce
$p$-process abundances in their
ejecta \citep[][]{howard:91,howard:93,kusakabe:11,travaglio:11}.
In particular, \cite{travaglio:11} provides SNIa simulations where a
relevant fraction of the $p$-process abundances in the Solar System
inventory are reproduced.

These final considerations however are based on only one
stellar mass, and the trajectories used for the $p$-process
calculations are based on one-zone supernova explosion
trajectories \citep[][]{rapp:06}.
Recent calculations have shown that
fall-back in core collapse SN could actually drastically reduce
the amount of O-rich mass ejected for stars with masses heavier
than $\sim$ 20 M$_{\odot}$ \citep[][]{fryer:09}. The impact of this on
final  $p$-process yields still needs to be estimated
in detail, and for different masses.
One possible implication is that $p$-process yields
may depend on which SN engine is considered.
Nevertheless, the intriguing scenario where part of the missing
LEPP component may be related to the missing $p$-process abundances
of Mo and Ru, and more in general with a new component for the
$p$-process inventory, can be constrained or ruled out by an
experimental measurement of the $^{12}$C+$^{12}$C rate
at stellar temperatures.


\section{Implications for the $s$-process nucleosynthesis at low metallicity}
\label{sec:low_metallcity}

In previous sections the impact of the $^{12}$C+$^{12}$C rate uncertainty
on $s$- and $p$-process nucleosynthesis
near solar metallicity has been analyzed.
In this section we discuss the implications for the $s$-process at low
metallicity.
In \S \ref{sec:c12c12sprocess}, we showed that the cs-process may at
 least partially contribute to the LEPP component in the Solar System.
\cite{travaglio:04} proposed that the solar LEPP component is due to a primary
process already active in the early universe, leaving a signature that has
been observed in the Galactic halo for Sr, Y and Zr
\citep[e.g.][]{truran:02}.
Whether the `solar` LEPP and the `stellar`
LEPP \citep[see definitions and discussion in][]{montes:07}
are either the same process or due to a combination of different processes
is
still matter of debate. More observations involving Sr-Y-Zr and other elements
in the LEPP mass region (from the Sr peak up to Te-Xe)
are required in order to shed more light into the origin of
LEPP \citep[][]{farouqi:09,chiappini:11,hansen:11}.
Following the analysis of the previous sections, one possible question is:
can the cs-process efficiently contribute
to the so-called stellar LEPP at low metallicity?

Another efficient $s$-process component has been proposed, occurring in fast
rotating stars of low metallicity \citep[][]{pignatari:08a,frischknecht:12}.
This component is triggered by the production of primary $^{22}$Ne
and the consequent activation of $^{22}$Ne($\alpha$,n)$^{25}$Mg.
Since, in principle, the cs process may also be produced and ejected by fast rotating massive
stars, what is the impact on the $s$-process ejecta of these two combined
components?
Stellar evolution models including the effects of rotation are
successful in reproducing several observations, like
nitrogen chemical enrichment in the early Galaxy
\citep[][]{heger:00,meynet:00} or the ratio of Wolf-Rayet to O-type
stars \citep[][and reference therein]{vazquez:07,frischknecht:10}.
There are, however, some unresolved issues in rotating models. A more efficient braking mechanism is necessary to explain the rotation rates of white dwarfs and millisecond pulsars. Although magnetic fields help, it is not clear yet whether their effect is sufficient
\citep{suijs:08}. Rotation induced shear mixing at the envelope-core interface in AGB stars may be too strong and thus disable the $s$ process via radiative $^{13}$C burning in the interpulse phase of low-mass AGB stars \citep{herwig:02,siess:04}.
Keeping this in mind, the cs-process behavior with metallicity does not
depend on rotation induced mixing, and the $s$-process component in fast
rotators mentioned above is a realistic scenario to consider
within the uncertainties of the present models.

\begin{figure}
\centering
\resizebox{8cm}{!}{\rotatebox{0}{\includegraphics{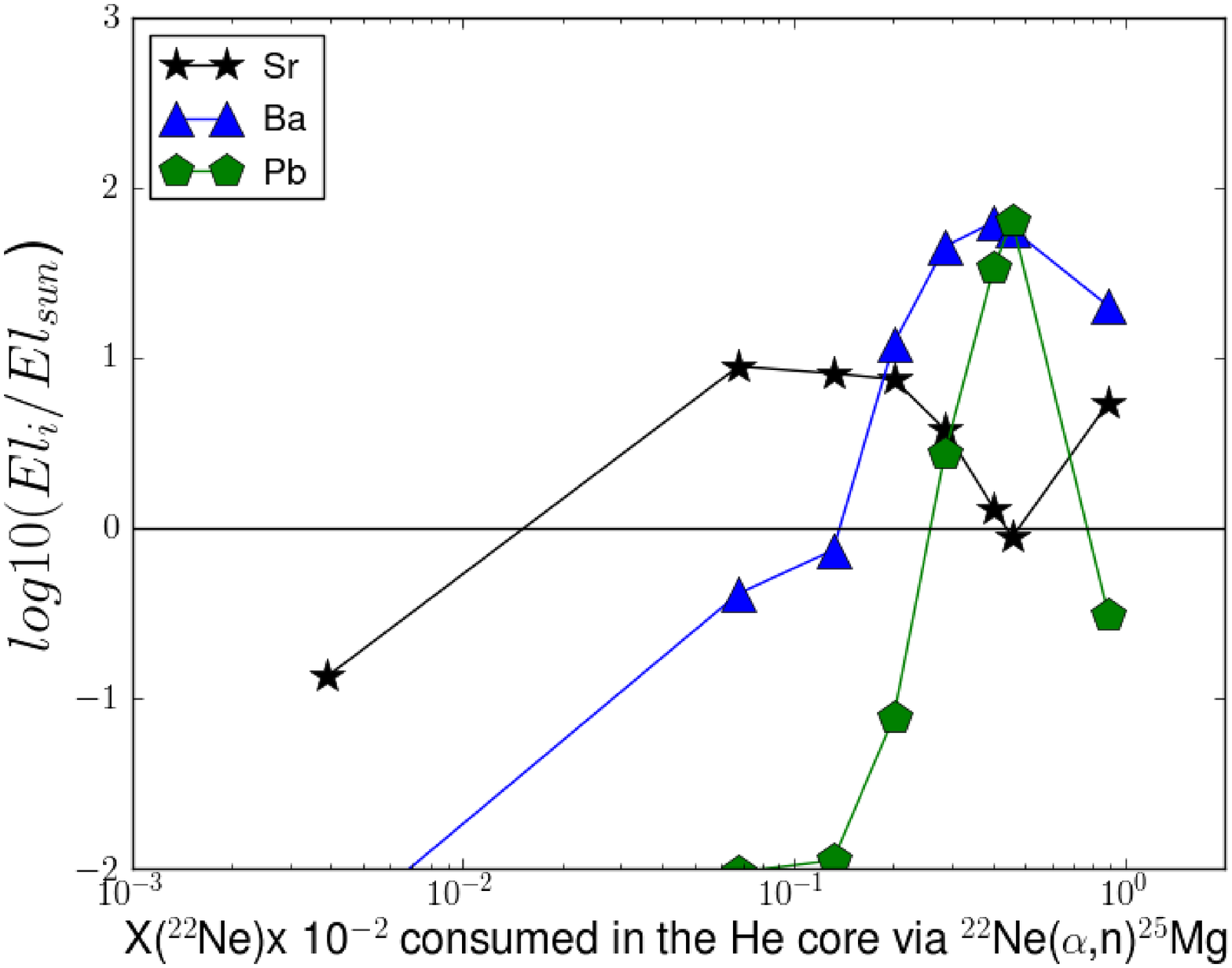}}}
\resizebox{8cm}{!}{\rotatebox{0}{\includegraphics{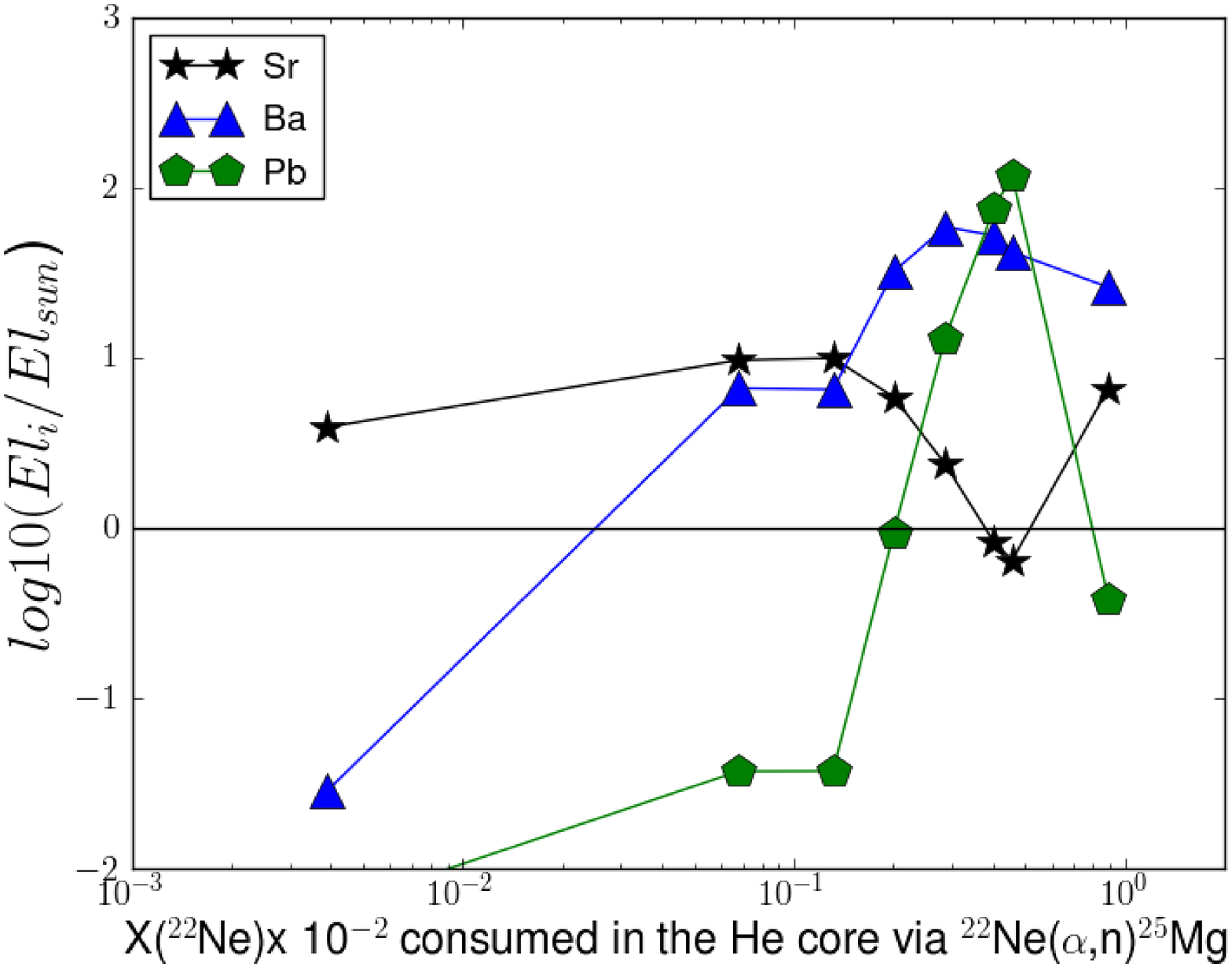}}}
  \caption{Left Panel $-$ The element abundances for Sr, Y, Zr, Ba and Pb are given at the end of C-burning
for different sets of initial compositions, taken from the He core ashes where different amounts of primary $^{22}$Ne is available
and consumed by the $^{22}$Ne($\alpha$,n)$^{25}$Mg \citep[25 M$_{\odot}$ star, {\rm [Fe/H]} = $-$3,][]{pignatari:08a}.
The $^{12}$C+$^{12}$C rate used is CF88,
temperature and density are $T$ = 1.0 GK and $\rho$ = 10$^5$ cm$^{-3}$ ($set1b$).
Right Panel $-$ As the Left Panel, but for $T$ = 0.65 GK and $\rho$ = 10$^4$ g cm$^{-3}$, and
using the CU $^{12}$C+$^{12}$C rate ($set2b$).
}
\label{fig:elements_low_metallicity}
\end{figure}

We have performed a set of calculations at [Fe/H] = $-$3.
Stars with this metallicity showing the LEPP component are observed in the
 Galactic halo \citep[e.g., HD 122563 at {\rm [Fe/H] = $-$2.7},][]{honda:06}.
A set of different initial abundances were taken from \cite{pignatari:08a},
at the end of convective core He$-$burning, considering a large spread of
primary $^{22}$Ne available in the He core due to rotational mixing.
Two different C$-$burning trajectories were used in the simulations:
a $set1$ trajectory (\S \ref{sec:c12c12a and c12c12p}, $T$ = 1.0 GK,
$\rho$ = 10$^5$ cm$^{−3}$, using the CF88 $^{12}$C+$^{12}$C rate,
hereafter $set1b$), which reproduced C shell conditions without the
cs-process component, and a $set2$ trajectory
(\S \ref{sec:c12c12a and c12c12p}, $T$ = 0.65 GK,
$\rho$ = 10$^4$ g cm$^{−3}$, using the CU $^{12}$C+$^{12}$C rate, hereinafter $set2b$), which did include the cs-process component. For complete fast rotating stellar model calculations at different metallicities, we refer to \cite{frischknecht:12}.
In Fig. \ref{fig:elements_low_metallicity} (left panel), the production of
elements located at the various neutron magic peaks
is shown for $set1b$ for different initial abundance distributions, defined by
the amount of $^{22}$Ne that is
consumed by $^{22}$Ne($\alpha$,n)$^{25}$Mg in the previous He core phase
($^{22}$Ne$_{\rm burnt}$).
The first case at $^{22}$Ne$_{\rm burnt}$ $\sim$ 4$\times$10$^{-5}$
uses an abundance distribution from a non-rotating star,
where the available $^{22}$Ne comes only from the initial CNO
abundances at [Fe/H] = $-3$, considering also
the $\alpha$-enhancement at this metallicity.
The other cases have initial abundances with an enhanced $s$-process
distribution, due to the primary $^{22}$Ne neutron source being activated in
fast rotating massive stars \citep[][]{pignatari:08a}.
In particular, the test at $^{22}$Ne$_{\rm burnt}$ $\sim$ 2$\times$10$^{-3}$
corresponds to the standard case considered at [Fe/H] = $-$3 in
\cite{pignatari:08a} (with the primary $^{22}$Ne around 1 \% by mass in the
He core).
We also used as initial distributions unpublished results obtained from the
same study, where different amounts of primary $^{22}$Ne available in the
He core.
The cases with $^{22}$Ne$_{\rm burnt}$ $\sim$ 4$\times$10$^{-3}$ correspond
to $\sim$ 2 \% of available primary $^{22}$Ne.
Finally, we also included an extreme case, where about 1 \% of $^{22}$Ne is
burnt by the
$^{22}$Ne($\alpha$,n)$^{25}$Mg in the He core.

As already mentioned by \cite{pignatari:08a}, in $set1b$ most of the
$s$-process yields are produced in the core He$-$burning phase,
and only a
partial modification occurs in the later shell C$-$burning phase.
Production of the Sr-Y-Zr peak is robust over a large range of primary
$^{22}$Ne$_{\rm burnt}$ ($\sim$ 7$\times$10$^{-4}$ $-$ 3$\times$10$^{-3}$).
Above 3$\times$10$^{-3}$, material at the Sr peak starts to be depleted,
feeding heavier elements along the $s$-process path.
For $^{22}$Ne$_{\rm burnt}$ $\sim$ 2$\times$10$^{-3}$ Ba starts rising, as well as Pb for $^{22}$Ne$_{\rm burnt}$ $\sim$ 3$\times$10$^{-3}$.
At $^{22}$Ne$_{\rm burnt}$ $\sim$ 5$\times$10$^{-3}$, Ba and Pb productions
reach their largest abundance.
Increasing the amount of $^{22}$Ne$_{\rm burnt}$ to
$\sim$ 10$^{-2}$ causes the Sr peak material to rise again, whereas
Ba and Pb decrease. To burn 1 \% of $^{22}$Ne via
$^{22}$Ne($\alpha$,n)$^{25}$Mg requires that the amount of primary $^{22}$Ne
is 5 \% or more.
Such a large abundance would make $^{22}$Ne acting mostly as a neutron
poison, consuming the neutrons produced.
Indeed, the production of Ba and Pb is limited not only by the amount of
iron seeds available, but also by the dual neutron source/neutron poison nature of $^{22}$Ne.
Notice that, however, such an amount of $^{22}$Ne is a
significant overestimate from present predictions of fast rotating massive
stars
\citep[][]{meynet:06,hirschi:08,frischknecht:12}.
Interestingly, this last case is different from Asymptotic Giant Branch stars
at low metallicity, where $^{22}$Ne becomes both a seed and a poison
for $s$-process nucleosynthesis up to lead. In the AGB case,
the main neutron source is the $^{13}$C($\alpha$,n)$^{16}$O reaction
and the neutron exposure is high enough to produce heavy elements starting
from light species \citep[][]{gallino:06}.

In Fig. \ref{fig:elements_low_metallicity} (right panel), we show the
same calculations for $set2b$, where also the cs-process
contribution has been considered, due to the activation of the
$^{13}$C($\alpha$,n)$^{16}$O.
Compared to the abundance at the He core exhaustion for the non-rotating case,
the Sr production factor compared to solar increases by about 40 times, up
to 3.9. Production factors of Ba and Pb increase by about 5 and 2 times,
 up to 0.028 and 0.006, respectively.
At both solar-like and [Fe/H] = $-$3 metallicities, the bulk of $s$-process
production is found at the Sr peak, with minor contributions to
heavier elements.
If we compare the Sr production factor obtained here and the one obtained with
the M25CU model and
half-metallicity (Table \ref{tab:sprocess_cshell_end_element_overabb_small}),
a secondary-like trend with the initial composition is observed, while light
neutron poisons are mostly primary.
Consequently the iron seeds force the nature of this $s$-process
component to be secondary, just like the $s$-process in fast rotators.

Concerning fast rotating massive stars, the production of Sr-Y-Zr is not
affected much by the cs-process component.
In fact the production of Sr has been almost saturated during the previous
core He$-$burning phase, since most of the Fe seeds
have already been consumed \citep[][Frischknecht et al. 2011]{pignatari:08a}.
As a result, the Sr-Y-Zr $s$-process yields from fast rotating massive stars
are robust over a large range of
$^{22}$Ne$_{\rm burnt}$ (i.e. the spread of initial stellar mass and
the efficiency of rotation-induced mixing),
within nuclear uncertainties of critical reaction rates like $^{22}$Ne+$\alpha$ \citep[][]{pignatari:08a},
$^{17}$O($\alpha$,$\gamma$)$^{21}$Ne
\citep[][]{hirschi:08} and $^{12}$C+$^{12}$C.
However, relevant uncertainties may be associated with other
nuclear reactions not considered here, and with the
treatment of rotation in stars.

The production of Ba and Pb is boosted by the cs-process, which eventually
feeds the Pb peak more efficiently for larger values of  $^{22}$Ne$_{\rm burnt}$.
 Indeed comparing results from $set1b$ and $set2b$,
Fig. \ref{fig:elements_low_metallicity}, shows that the maximum production
factors of Sr-Y-Zr and Ba are similar, whereas Pb is produced by a factor of 2
more in $set2b$. Additionally, in $set2b$ calculations $^{22}$Ne is only
partially destroyed, and also behaves like an efficient neutron poison.
Therefore the more primary $^{22}$Ne (and $^{25,26}$Mg, direct products via $\alpha$ capture on $^{22}$Ne)
there is available in the He core ashes, the less efficiently the cs-process
feeds the heavier elements.
In general, the maximum elemental ratio Pb/Ba that can be produced in the
present calculations for fast rotating massive stars is $\sim$ 3, including
the uncertainty related to the amount of primary $^{22}$Ne
available and/or consumed by the $^{22}$Ne($\alpha$,n)$^{25}$Mg, as well as the
uncertainty in the $^{12}$C+$^{12}$C rate.
Taking into account the present $^{22}$Ne($\alpha$,n)$^{25}$Mg uncertainty,
such conditions can be obtained when the primary $^{22}$Ne abundance in the
He core is about 2 \% for a 25 M$_{\odot}$
star \citep[about a factor of 2 more than predicted by][]{hirschi:08},
or for stars with larger initial mass, where $^{22}$Ne depletion via
 $\alpha$-capture is more efficient. According to stellar model
calculations by \cite{hirschi:08},
the production factors of Sr, Ba and Pb are 7.5, 12.0 and 0.08, respectively,
becoming 5.9, 31.9 and 0.9 when the contribution from the cs process is
included (Fig. \ref{fig:elements_low_metallicity}).
In this case, therefore, $s$-process elements are efficiently produced only up
to Ba, also taking into account the $^{12}$C+$^{12}$C uncertainty.

In \cite{pignatari:08a} it was shown that the $s$ process in fast rotating
massive stars due to primary $^{22}$Ne is a secondary process.  This was
confirmed by similar results obtained for a complete set of fast rotating
stellar models
\citep[][]{frischknecht:12}. As discussed before, the cs process is also
 secondary-like. In \S \ref{sec:c12c12sprocess} it was shown that
in principle the cs-component is a possible scenario for explaining the
LEPP in the solar system without failing
the weak $s$-process distribution. Assuming that the same constraints discussed
 in \S \ref{sec:c12c12sprocess}
are also valid for low metallicity, only a few massive stars would eject such
a component at [Fe/H] $\lesssim$ -3.

The LEPP signature could have been observed in stars such as HE 1327-2326, where
 [Fe/H]= $-$5.45 \citep[][]{frebel:05}, for which the efficiency of the
cs-process compared to other primary explosive processes in massive stars
proposed to feed the Sr-Y-Zr peak is supposed to be
lower \citep[][]{woosley:92,qian:07,farouqi:08,arcones:11}.
However, because at present only the upper limits of Ba and Eu are available
for this star, whether or not HE 1327-2326 is a LEPP star still requires confirmation.
A detailed comparison between stellar yields and spectroscopic observations
of metal-poor LEPP stars is beyond the scope of this paper, since complete massive
stellar calculations at low metallicity would be required.

In fast rotating massive stars, the occurrence of the cs-process would not
affect the conclusions of \cite{pignatari:08a} concerning nucleosynthesis at
the Sr-Y-Zr peak.
If such a component does exist however, a more efficient production
of Sr-Y-Zr is obtained also at lower rotation efficiencies, and for
fast rotators Ba peak elements, and possibly Pb, could be more abundant
in the final $s$-process yields.
The present calculations explored the possible impact of the
 $^{12}$C+$^{12}$C rate uncertainty on the $s$-process at low metallicity,
in non-rotating and fast rotating stars.
The analysis presented in this section is based on one-zone nucleosynthesis calculations and not on full stellar models.
However we did use realistic stellar conditions and explored the
impact of various uncertainties on theoretical abundance predictions,
which provides some guidance about $s$-process nucleosynthesis in
massive stars at low metallicity.
In the case that future nuclear physics experiments suggest a
higher $^{12}$C+$^{12}$C rate than the CF88 rate for central
C$-$burning conditions, a new set of complete stellar models at low
metallicity and different rotation efficiencies would be required in order
to revise the present estimates of $s$-process nucleosynthesis stellar yields in the early universe.

\section{Conclusions}
\label{sec:concl}

The aim of the present work was to explore the impact of the
$^{12}$C+$^{12}$C rate uncertainty in massive stars,
on pre-supernova nucleosynthesis and on the explosive $p$-process.
Our analysis was focused on one mass and
metallicity (25 M$_{\odot}$ and half solar metallicity)
and considered a large variation of the $^{12}$C+$^{12}$C rate: the lower
rate limit took into account possible hindrance effects in the carbon fusion
reaction, whereas the upper limit considered contributions from potential
subthreshold resonances at low temperature.

In all, we considered five different $^{12}$C+$^{12}$C rates:
the lower limit (CL), the standard rate of CF88, CF88 divided and
multiplied by a factor of ten (CF88d10 and CF88t10, respectively)
and the upper limit (CU). Other reactions critical for the energy generation during
stellar evolution were left unchanged. Up to central He exhaustion, the five
stellar models obtained with these rates
(M25CL, M25CF88d10, M25CF88, M25CF88t10, M25CU) share the same conditions.


With the lower $^{12}$C+$^{12}$C rates, CL and CF88d10, the $s$-process during
convective shell C-burning shows a slightly lower neutron exposure,
coupled with a higher neutron density peak, causing local effects at the branching
points along the $s$-process path (e.g., $^{79}$Se and $^{85}$Kr).
The production of long-lived isotopes $^{26}$Al and $^{60}$Fe during the
pre-explosive phase is also affected, as discussed by
\cite{gasques:07}. In particular, we confirm that $^{60}$Fe tends to be produced
 more with lower $^{12}$C+$^{12}$C rates.
However, results for $^{26}$Al are strongly affected by the stellar model used.
 In fact, if the convective C shell is switched off before the
core collapse starts, all of the $^{26}$Al has the time to decay. As a result
 the final stellar yields only contain explosive $^{26}$Al.
The model with the highest pre-explosive $^{26}$Al abundance in the C shell
material is M25CF88t10, since it is the only one to keep a convective C
shell until core collapse. Because the pre-explosive component of $^{26}$Al is
model-dependent, the final $^{26}$Al/$^{60}$Fe in SN ejecta
needs to be regarded cautiously.

For the higher $^{12}$C+$^{12}$C rates, CF88t10 and CU, the final $s$-process
yields tend to increase towards the Sr peak with respect to lighter
$s$-process elements. In the M25CF88t10 model, such an increase is within
20 \%, depending on the isotope considered, and is due to the overlap between
the first shell and the second convective carbon shell \citep[][]{bennett:10}.
In the M25CU model, the activation of the $^{13}$C($\alpha$,n)$^{16}$O in the C
core causes the formation of the $cs$-process component, which will also
dominate the final C-shell ejecta.
In agreement with paper I, the $s$-process yields at the Sr peak increase by
about two orders of magnitude.
It was shown that if only $\sim$ 2 \% of the $s$-process rich material
from massive stars carries the $cs$-process signature,
the weak $s$-process component from other stars would not be affected
and the $cs$-process could reproduce at least part of the LEPP
component in the Solar System.

We have provided a series of tests in which the strengths to the two dominant
nucleosynthesis channels of the carbon-fusion reactions
$^{12}$C($^{12}$C,$\alpha$)$^{20}$Ne and $^{12}$C($^{12}$C,p)$^{23}$Na have been
changed.
The impact of using a different $R_{\alpha}$/$R_p$ ratio changes according to
the total $^{12}$C+$^{12}$C rate.
For the calculations using CF88, the amount of neutrons available
(and the $s$-process efficiency) decreases with increasing proton channel
strength compared to the $\alpha$ channel. Indeed, $^{22}$Ne depletion in
C$-$burning conditions is dominated by
the ($\alpha$,n) and the (p,$\gamma$) channels which are respectively affected
by the amount of $\alpha$ and protons available.
The $cs$-process, on the other hand, drastically loses its efficiency with
increasing $\alpha$ channel strength, because less protons
are available to produce $^{13}$C via $^{12}$C(p,$\gamma$)$^{13}$N($\beta^+$)$^{13}$C.
For instance, using a ratio $R_{\alpha}$/$R_p$ = 0.95/0.05 (ten times higher
than the standard $R_{\alpha}$/$R_p$ = 0.65/0.35), the Sr abundance decreases by about a factor of 30. Small variations are obtained from
reducing the $\alpha$ channel strength.
Critical light species are also affected by $R_{\alpha}$/$R_p$ using the
 CU $^{12}$C+$^{12}$C rate:
$^{16}$O and $^{23}$Na decreases with a high $R_{\alpha}$/$R_p$ ratio,
while $^{20}$Ne increases. Therefore, uncertainty in the
$R_{\alpha}$/$R_p$ ratio may affect abundances of species that are the main
fuel for following evolutionary phases.

We also tested the impact of the uncertainty of the channel
$^{12}$C($^{12}$C,n)$^{23}$Mg on the $s$-process
calculations, assuming a range of probability between the standard rate (D77)
and D77 multiplied by 10 (D77t10).
First of all the impact of the neutron channel increases with increasing
 amounts of $^{12}$C available at the onset of shell C$-$burning. This in turn
 may depend on the $^{12}$C($\alpha$,$\gamma$)$^{16}$O rate during the previous
 core He$-$burning and convection prescription, as well as the
total $^{12}$C+$^{12}$C rate.
In the simulations reproducing the typical C shell conditions using the
CF88 $^{12}$C+$^{12}$C rate, it was seen that the neutron channel uncertainty
causes a propagation effect in the $s$-process distribution, peaked at the
$s$-only isotopes $^{80,82}$Kr, with an increase of more than a factor of two.
On the other hand in the last convective C shell of the M25CF88t10 model, 
uncertainty in the neutron channel only becomes important
in the final phase, associated with a temperature and density rise,
when the higher neutron density (being driven mostly by $^{12}$C($^{12}$C,n)$^{23}$Mg)
increases the abundance of the $r$-only isotopes
(e.g., $^{70}$Zn and $^{76}$Se) by 20-30 \%.

We have studied the impact of the $^{12}$C+$^{12}$C rate uncertainty on the
$p$-process yields, using standard $p$-process trajectories
from SN explosion \citep[][]{rapp:06}. In particular, with the exception of the
M25CU yields, the $p$-process calculations shows comparable results.
The $cs$-process component feeds the production of extremely abundant
$p$-process yields ($cp$-process), with average
$p$-only abundance 8.3 times higher than the standard $p$-process, and
reproducing also the Mo and Ru $p$-only abundances.
GCE calculations based on stellar models of different masses
and using updated SN explosions would be required to study the impact
of the $cp$-process on the chemical inventory of the $p$ nuclei.


Finally, we have explored the possible impact of the $^{12}$C+$^{12}$C rate for
 non-rotating and fast-rotating massive stars with
low metallicity ([Fe/H] = $-$3). It was found that at the considered metallicity
the $cs$-process is secondary, despite the primary neutron source
$^{13}$C($\alpha$,n)$^{16}$O and mainly primary neutron poisons.
The possible existence of the of the $cs$ component does not modify previous
conclusions regarding the $s$-process in fast
rotating massive stars, due to the primary $^{22}$Ne neutron source.
The main effects of the $cs$-process are a relevant abundance
production at the Sr peak, even without rotation, and the production of
elements heavier than Sr-Y-Zr, up to the Ba peak and
eventually up to Pb, in fast rotating massive stars.
However an efficient production of lead would occur for concentrations
of primary $^{22}$Ne in the He core which are
a factor of two higher than have previously been considered \citep[see for more details][]{pignatari:08a}.
Considering the $cs$-process contribution or, in other words,
the $^{12}$C+$^{12}$C rate uncertainty, a maximum ratio of Pb/Ba $\sim$ 3 was
 estimated for the $s$-process yields.
The production of more Pb compared to other lighter elements is limited by
the amount of primary $^{22}$Ne.  If there is too much of it available in the
He core, then its neutron poison efficiency reduces the $s$-process beyond
iron, or alternatively a $^{22}$Ne($\alpha$,n)$^{25}$Mg rate higher than the
present uncertainty would be required.


\acknowledgments NuGrid acknowledges significant support from NSF grants PHY 02-16783 and PHY 09-22648 (Joint Institute for Nuclear Astrophysics, JINA) and EU MIRG-CT-2006-046520. The continued work on codes and in disseminating data is made possible through funding from STFC (RH, UK), an NSERC Discovery grant (FH, Canada), and an Ambizione grant of the SNSF (MP, Switzerland). NuGrid computations were performed at the Arizona Advanced Computing Center (USA), the high-performance computer KHAOS at EPSAM Institute at Keele University (UK) as well as CFI (Canada) funded computing resources at the Department of Physics and Astronomy at the University of Victoria and through Computing Time Resource Allocation through the Compute Canada WestGrid consortium.
M.P. also thanks the support from Core project Eurogenesis.

\bibliography{all}

\clearpage


\begin{table}
\begin{center}
\tiny
\caption{The recommended, upper limit (CU) and lower limit rate (CL) is given for the $^{12}$C+$^{12}$C
fusion reaction.
}
\begin{tabular}{cccc}
\hline
T9 & \multicolumn{3}{c}{total rate}  \\
       &   recomm. & lower limit & upper limit  \\
\hline
1.00E-01  &  4.99E-52 &  2.25E-58 &  4.99E-52  \\
1.10E-01  &  1.30E-49 &  5.32E-55 &  1.30E-49  \\
1.21E-01  &  2.95E-47 &  1.21E-52 &  2.95E-47  \\
1.33E-01  &  5.66E-45 &  2.33E-50 &  5.66E-45  \\
1.46E-01  &  8.78E-43 &  1.97E-47 &  8.78E-43  \\
1.61E-01  &  1.21E-40 &  2.73E-45 &  1.21E-40  \\
1.77E-01  &  1.37E-38 &  1.19E-42 &  1.37E-38  \\
1.95E-01  &  1.39E-36 &  1.21E-40 &  1.41E-36  \\
2.14E-01  &  1.17E-34 &  3.07E-38 &  1.57E-34  \\
2.36E-01  &  8.90E-33 &  2.34E-36 &  6.46E-32  \\
2.59E-01  &  5.69E-31 &  3.74E-34 &  4.04E-29  \\
2.85E-01  &  3.49E-29 &  2.17E-32 &  1.53E-26  \\
3.14E-01  &  2.08E-27 &  2.32E-30 &  3.38E-24  \\
3.45E-01  &  1.34E-25 &  1.95E-28 &  4.55E-22  \\
3.80E-01  &  8.12E-24 &  1.35E-26 &  3.85E-20  \\
4.18E-01  &  3.98E-22 &  8.15E-25 &  2.16E-18  \\
4.60E-01  &  1.47E-20 &  3.63E-23 &  8.31E-17  \\
5.05E-01  &  3.98E-19 &  2.48E-21 &  2.25E-15  \\
5.56E-01  &  8.35E-18 &  1.48E-19 &  4.47E-14  \\
6.12E-01  &  1.42E-16 &  6.73E-18 &  6.68E-13  \\
6.73E-01  &  2.06E-15 &  2.33E-16 &  7.68E-12  \\
7.40E-01  &  2.86E-14 &  5.75E-15 &  7.01E-11  \\
8.14E-01  &  3.88E-13 &  1.08E-13 &  5.16E-10  \\
8.95E-01  &  5.04E-12 &  1.58E-12 &  3.13E-09  \\
9.85E-01  &  5.93E-11 &  2.00E-11 &  1.59E-08  \\
1.08E+00  &  6.72E-10 &  2.12E-10 &  6.87E-08  \\
1.19E+00  &  6.84E-09 &  2.24E-09 &  2.63E-07  \\
1.31E+00  &  6.53E-08 &  2.32E-08 &  9.02E-07  \\
1.44E+00  &  5.85E-07 &  2.35E-07 &  3.01E-06  \\
1.59E+00  &  4.93E-06 &  2.27E-06 &  1.12E-05  \\
1.75E+00  &  3.88E-05 &  2.05E-05 &  5.37E-05  \\
1.92E+00  &  2.85E-04 &  1.71E-04 &  3.17E-04  \\
2.11E+00  &  1.83E-03 &  1.30E-03 &  1.89E-03  \\
2.32E+00  &  1.16E-02 &  9.08E-03 &  1.18E-02  \\
2.56E+00  &  6.44E-02 &  5.69E-02 &  6.46E-02  \\
2.81E+00  &  3.29E-01 &  3.26E-01 &  3.30E-01  \\
3.09E+00  &  1.56E+00 &  1.71E+00 &  1.56E+00  \\
3.40E+00  &  6.25E+00 &  7.83E+00 &  6.25E+00  \\
3.74E+00  &  2.52E+01 &  3.50E+01 &  2.52E+01  \\
4.11E+00  &  8.48E+01 &  1.36E+02 &  8.48E+01  \\
4.53E+00  &  2.60E+02 &  4.90E+02 &  2.60E+02  \\
4.98E+00  &  7.32E+02 &  1.64E+03 &  7.32E+02  \\
5.48E+00  &  1.66E+03 &  4.73E+03 &  1.66E+03  \\
6.02E+00  &  3.45E+03 &  1.27E+04 &  3.45E+03  \\
6.63E+00  &  6.78E+03 &  3.18E+04 &  6.78E+03  \\
7.29E+00  &  1.12E+04 &  6.79E+04 &  1.12E+04  \\
8.02E+00  &  1.83E+04 &  1.36E+05 &  1.83E+04  \\
8.82E+00  &  3.06E+04 &  2.57E+05 &  3.06E+04  \\
9.70E+00  &  4.74E+04 &  4.13E+05 &  4.74E+04  \\
1.07E+01  &  7.73E+04 &  6.27E+05 &  7.73E+04  \\
\hline
\noalign{\smallskip}
\hline
\end{tabular}
\label{tab: c12c12rate_and_channels}
\end{center}
\end{table}


\begin{table}
\begin{center}
\caption{
For each of the stellar models presented in this work (25 M$_{\odot}$, $Z$ = 0.01, for different $^{12}$C+$^{12}$C rates, column 1), the following data are provided: helium core mass (M$^{75\%}_{\alpha}$, i.e., giving the mass coordinate above which the He abundance drops below 75 \%, column 2);
CO core mass (M$_{\rm CO}$, column 3); final total mass (M$_{\rm final}$, column 4); size convective carbon core (CC(M$_{\odot}$), column 5);
Ignition temperature and density of core C burning ($T_{\rm c}$ and
rho$_{\rm c}$, column 6 and 7), respectively;
number of convective C-burning shell episodes (column 8).}
\begin{tabular}{cccccccc}
\hline \hline
 Model & M$^{75\%}_{\alpha}$  &   M$_{\rm CO}$ &   M$_{\rm final}$  &   CC(M$_{\odot}$)  &  $T_{\rm c}$   &  $\rho_{\rm c}$ &   shells  \\
       & (M$_{\odot}$)        & (M$_{\odot}$) &  (M$_{\odot}$) &  (M$_{\odot}$)  &  (GK) & (g cm$^{-3}$) &  \\
\hline
M25CF88d10   &  9.354  & 6.205 &  14.370  &  radiative  &   0.894 &  2.129E+5 & 1 \\
M25CL        &  9.354  & 6.206 &  14.369  &  radiative  &   0.729 &  8.192E+4 & 1 \\
M25CF88      &  9.169  & 6.347 &  14.710  &  radiative  &   0.728 &  8.020E+4 & 1 \\
M25CF88t10   &  9.331  & 6.322 &  13.791  &  radiative  &   0.710 &  6.083E+4 & 2 \\
M25CU        &  9.275  & 6.550 &  13.780  &  4.12       &   0.559 &  1.544E+4 & 1 \\
\hline
\noalign{\smallskip}
\hline
\end{tabular}
\label{tab:modeldata_1}
\end{center}
\end{table}


\begin{table}
\begin{center}
\caption{For each of the stellar models presented in this work
(25 M$_{\odot}$, $Z$ = 0.01, different $^{12}$C+$^{12}$C rates,
see column 1), the lifetimes of all core burning stages and total
lifetime of the star
are given (in years): H-burning
($\tau_{\rm H}$, column 2); He-burning ($\tau_{\rm He}$, column 3); C-burning ($\tau_{\rm C}$, column 4); Ne-burning ($\tau_{\rm Ne}$, column 5); O-burning
($\tau_{\rm O}$, column 6); Si-burning ($\tau_{\rm Si}$, column 7) and total lifetime ($\tau_{\rm Total}$, column 8). Notice that all the models stop
before the end of central Si burning.}
\begin{tabular}{cccccccc}
\hline \hline
 Model & $\tau_{\rm H}$  &  $\tau_{\rm He}$ &   $\tau_{\rm C}$  &   $\tau_{\rm Ne}$  &  $\tau_{\rm O}$ &  $\tau_{\rm Si}$ & $\tau_{\rm Total}$  \\
\hline
M25CF88d10   &  6.631E+6  & 6.213E+5 & 2.248E+1 & 0.148 & 0.208 & 0.007 & 7.380E+6 \\
M25CL        &  6.631E+6  & 6.213E+5 & 1.726E+2 & 0.823 & 0.212 & 0.011 & 7.380E+6 \\
M25CF88      &  6.631E+6  & 6.213E+5 & 1.750E+2 & 0.498 & 0.301 & 0.007 & 7.380E+6 \\
M25CF88t10   &  6.631E+6  & 6.213E+5 & 4.256E+3 & 0.126 & 0.133 & 0.015 & 7.380E+6 \\
M25CU        &  6.631E+6  & 6.213E+5 & 2.152E+4 & 0.406 & 0.374 & 0.023 & 7.399E+6 \\
\hline
\noalign{\smallskip}
\hline
\end{tabular}
\label{tab:modeldata_2}
\end{center}
\end{table}


\begin{table}
\begin{center}
\caption{Initial abundances and abundances at He exhaustion for the present stellar model (25 M$_{\odot}$ star, and $Z$ = 0.01) are given for a sample of indicative species. For the initial abundance of $^{22}$Ne, the CNO species contribution to its abundance is taken into account. Finally, the neutron exposure $\tau_{n}$ is given at the bottom of the table.}
\begin{tabular}{ccc}
\hline \hline
$X_{\rm i}$       & initial  & Central He exhaustion       \\
\hline
   C  12 $\ldots$  &   1.748E-3 &  0.267      \\
   O  16 $\ldots$  &   4.379E-3 &  0.714      \\
   Ne 22 $\ldots$  &   1.001E-2 &  3.617E-3   \\
   Fe 54 $\ldots$  &   5.339E-5 &  5.463E-7    \\
   Fe 56 $\ldots$  &   8.692E-4 &  2.731E-4   \\
   Ge 70 $\ldots$  &   3.392E-8 &  1.373E-6    \\
   Se 76 $\ldots$  &   9.134E-9 &  1.896E-7    \\
   Kr 80 $\ldots$  &   2.132E-9 &  6.089E-8    \\
   Kr 82 $\ldots$  &   1.092E-8 &  1.509E-7    \\
   Sr 86 $\ldots$  &   4.215E-9 &  9.239E-8    \\
   Sr 87 $\ldots$  &   3.204E-9 &  5.231E-8    \\
   Sr 88 $\ldots$  &   3.613E-8 &  2.892E-7    \\
\hline
 & & $\tau_{n}$ (mbarn$^{-1}$)  \\
 & & 0.149 \\
\hline
\noalign{\smallskip}
\hline
\end{tabular}
\label{table:hecore}
\end{center}
\end{table}


\begin{table}
\begin{center}
\caption{Pre-explosive abundances in the convective C shell region of the models M25CF88d10, M25CL, M25CF88, M25CF88t10 and M25CU (column 2, 3, 4, 5 and 6 respectively) for a sample of selected species.}
\begin{tabular}{cccccc}
\hline \hline
 $X_{\rm i}$ & M25CF88d10 & M25CL  & M25CF88      & M25CF88t10        & M25CU        \\
\hline
 C  12  	& 4.546E-02  & 2.626E-02 & 2.589E-02 & 5.783E-03  & 1.651E-02  \\
 O  16  	& 6.011E-01  & 5.814E-01 & 5.998E-01 & 5.724E-01  & 6.168E-01  \\
 Ne 20  	& 2.861E-01  & 3.332E-01 & 3.234E-01 & 3.571E-01  & 2.951E-01  \\
 Ne 22  	& 9.342E-05  & 1.498E-04 & 1.132E-04 & 9.748E-05  & 2.022E-03  \\
 Na 23  	& 5.716E-03  & 8.421E-03 & 8.234E-03 & 7.744E-03  & 2.517E-02  \\
 Fe 54  	& 7.761E-08  & 8.764E-08 & 7.571E-08 & 8.567E-08  & 6.951E-08  \\
 Fe 56  	& 1.100E-04  & 1.175E-04 & 1.113E-04 & 1.140E-04  & 6.898E-05  \\
 Zn 70  	& 9.146E-08  & 1.693E-08 & 1.836E-08 & 4.729E-08  & 2.071E-08  \\
 Ge 70  	& 4.332E-06  & 3.881E-06 & 4.366E-06 & 4.062E-06  & 1.345E-05  \\
 Se 76  	& 7.816E-07  & 6.063E-07 & 6.862E-07 & 6.223E-07  & 6.505E-06  \\
 Kr 80  	& 5.154E-08  & 1.616E-07 & 1.605E-07 & 2.580E-08  & 3.372E-06  \\
 Kr 82  	& 4.377E-07  & 4.336E-07 & 4.767E-07 & 3.387E-07  & 1.266E-05  \\
 Sr 88  	& 4.196E-07  & 4.370E-07 & 4.660E-07 & 5.576E-07  & 7.740E-05  \\
 Y  89  	& 9.578E-08  & 8.965E-08 & 9.595E-08 & 1.046E-07  & 1.709E-05  \\
 Zr 96  	& 7.669E-09  & 2.214E-09 & 2.732E-09 & 2.178E-09  & 4.552E-08  \\
 Te124  	& 9.139E-10  & 1.005E-09 & 1.009E-09 & 7.768E-10  & 6.765E-08  \\
 Xe130  	& 1.087E-09  & 1.223E-09 & 1.211E-09 & 1.131E-09  & 4.824E-08  \\
 Xe134  	& 9.884E-10  & 1.893E-10 & 2.164E-10 & 5.124E-10  & 7.992E-10  \\
 Ba138  	& 3.196E-08  & 2.981E-08 & 3.026E-08 & 3.021E-08  & 1.322E-07  \\
\hline
 Al 26          & 1.690E-09  & 8.470E-09 & 6.728E-10 & 2.274E-06  & 1.367E-08  \\
 Fe 60          & 4.605E-06  & 1.132E-06 & 2.450E-06 & 3.326E-06  & 9.021E-08  \\
\hline
\noalign{\smallskip}
\hline
\end{tabular}
\label{tab:sprocess_cshell_end}
\end{center}
\end{table}


\begin{table}
\begin{center}
\caption{Pre-explosive element overabundances in the convective C shell region from C to Mo
for the models M25CF88d10, M25CL, M25CF88, M25CF88t10 and M25CU (column 2,3,4,5,6 respectively).
}
\begin{tabular}{cccccc}
\hline \hline
 $X_{\rm el}$/$X_{\odot}$ & M25CF88d10 & M25CL      & M25CF88    & M25CF88t10 & M25CU        \\
\hline
 C	 & 1.311E+01  & 7.575E+00  & 7.470E+00  & 1.668E+00  & 4.763E+00  \\
 N	 & 5.139E-04  & 8.662E-03  & 9.398E-03  & 5.553E-03  & 3.202E-02  \\
 O	 & 6.230E+01  & 6.025E+01  & 6.215E+01  & 5.932E+01  & 6.392E+01  \\
 F	 & 1.369E-02  & 2.631E-02  & 2.518E-02  & 3.063E-02  & 2.652E-01  \\
 Ne	 & 1.454E+02  & 1.694E+02  & 1.644E+02  & 1.816E+02  & 1.520E+02  \\
 Na	 & 1.429E+02  & 2.105E+02  & 2.059E+02  & 1.936E+02  & 6.293E+02  \\
 Mg	 & 7.433E+01  & 6.250E+01  & 5.587E+01  & 7.081E+01  & 5.192E+01  \\
 Al	 & 2.613E+01  & 2.974E+01  & 2.849E+01  & 2.782E+01  & 5.220E+01  \\
 Si	 & 6.938E+00  & 3.961E+00  & 3.366E+00  & 4.117E+00  & 1.979E+00  \\
 P	 & 1.146E+01  & 8.229E+00  & 8.065E+00  & 7.423E+00  & 1.240E+01  \\
 S	 & 3.506E-01  & 3.216E-01  & 3.161E-01  & 3.153E-01  & 2.560E-01  \\
 Cl	 & 2.498E+00  & 2.814E+00  & 2.818E+00  & 2.901E+00  & 3.259E+00  \\
 Ar	 & 3.586E-01  & 3.568E-01  & 3.575E-01  & 3.552E-01  & 3.280E-01  \\
 K	 & 2.368E+00  & 2.137E+00  & 2.172E+00  & 2.068E+00  & 2.111E+00  \\
 Ca	 & 2.273E-01  & 2.315E-01  & 2.282E-01  & 2.292E-01  & 1.836E-01  \\
 Sc	 & 1.154E+01  & 9.397E+00  & 9.780E+00  & 9.859E+00  & 1.180E+01  \\
 Ti	 & 9.323E-01  & 9.359E-01  & 9.576E-01  & 9.415E-01  & 1.798E+00  \\
 V	 & 5.069E-01  & 4.622E-01  & 4.823E-01  & 4.784E-01  & 1.069E+00  \\
 Cr	 & 3.439E-01  & 3.521E-01  & 3.474E-01  & 3.494E-01  & 2.950E-01  \\
 Mn	 & 1.032E-01  & 1.028E-01  & 1.013E-01  & 1.030E-01  & 7.756E-02  \\
 Fe	 & 2.260E-01  & 2.372E-01  & 2.292E-01  & 2.328E-01  & 1.335E-01  \\
 Co	 & 2.741E+01  & 2.589E+01  & 2.634E+01  & 2.521E+01  & 1.079E+01  \\
 Ni	 & 2.759E+00  & 2.740E+00  & 2.752E+00  & 2.715E+00  & 2.069E+00  \\
 Cu	 & 7.410E+01  & 7.078E+01  & 7.259E+01  & 6.934E+01  & 6.346E+01  \\
 Zn	 & 2.602E+01  & 2.623E+01  & 2.946E+01  & 3.074E+01  & 5.052E+01  \\
 Ga	 & 6.785E+01  & 6.604E+01  & 7.312E+01  & 6.709E+01  & 2.436E+02  \\
 Ge	 & 5.763E+01  & 4.726E+01  & 5.341E+01  & 5.323E+01  & 2.692E+02  \\
 As	 & 4.441E+01  & 3.353E+01  & 3.787E+01  & 3.913E+01  & 2.659E+02  \\
 Se	 & 2.852E+01  & 1.896E+01  & 2.179E+01  & 2.088E+01  & 2.248E+02  \\
 Br	 & 2.629E+01  & 1.554E+01  & 1.814E+01  & 1.876E+01  & 2.049E+02  \\
 Kr	 & 1.455E+01  & 1.308E+01  & 1.441E+01  & 1.344E+01  & 5.039E+02  \\
 Rb	 & 2.022E+01  & 1.065E+01  & 1.257E+01  & 1.875E+01  & 4.854E+02  \\
 Sr	 & 9.261E+00  & 1.182E+01  & 1.249E+01  & 1.365E+01  & 1.739E+03  \\
 Y	 & 7.791E+00  & 7.293E+00  & 7.805E+00  & 8.507E+00  & 1.390E+03  \\
 Zr	 & 3.394E+00  & 3.103E+00  & 3.327E+00  & 3.258E+00  & 6.575E+02  \\
 Nb	 & 5.012E+00  & 4.053E+00  & 4.407E+00  & 4.401E+00  & 8.198E+02  \\
 Mo	 & 1.224E+00  & 1.245E+00  & 1.316E+00  & 1.179E+00  & 2.866E+02  \\
\hline
\noalign{\smallskip}
\hline
\end{tabular}
\label{tab:sprocess_cshell_end_element_overabb_small}
\end{center}
\end{table}


\begin{table}
\begin{center}
\caption{Mass fraction abundances for a sample of selected species in the He-core ashes (column 2), when the final $^{12}$C is less than 2 \%
for $set1$ cases (columns 3 - where a05p95 means $R_{\alpha}$/$R_p$ = 0.05/0.95, columns 4 - where st means $R_{\alpha}$/$R_p$ = 0.65/0.35 and column 5 - where a95p05 means $R_{\alpha}$/$R_p$ = 0.95/0.05) and for $set2$ cases (column 6, 7 and 8, labels have the same meaning of $set1$ cases, and the symbol * is used to distinguish from $set1$). See the text for more details.}
\begin{tabular}{cccccccc}
\hline \hline
$X_{\rm i}$ & He ashes  & a05p95    & st	   & a95p05    & a05p95*   & st*       & a95p05*    \\
\hline
 C-12	& 2.673E-01 & 1.544E-02 & 1.544E-02 & 1.543E-02 & 1.999E-02 & 1.998E-02 & 2.000E-02  \\
 O-16	& 7.142E-01 & 5.791E-01 & 5.708E-01 & 5.658E-01 & 7.131E-01 & 6.478E-01 & 5.883E-01  \\
 Ne-20  & 5.026E-03 & 3.412E-01 & 3.639E-01 & 3.792E-01 & 1.241E-01 & 2.475E-01 & 3.419E-01  \\
 Ne-22  & 3.617E-03 & 1.166E-04 & 1.214E-04 & 1.123E-04 & 4.610E-03 & 3.887E-03 & 1.824E-03  \\
 Na-23  & 1.594E-04 & 1.970E-02 & 1.188E-02 & 4.171E-03 & 8.534E-02 & 3.898E-02 & 9.133E-03  \\
 Fe-54  & 5.463E-07 & 3.170E-07 & 2.430E-07 & 1.592E-07 & 1.079E-09 & 5.799E-10 & 3.359E-08  \\
 Fe-56  & 2.731E-04 & 2.117E-04 & 1.867E-04 & 1.526E-04 & 1.397E-05 & 1.060E-05 & 6.937E-05  \\
 Zn-70  & 1.040E-09 & 4.392E-09 & 7.682E-09 & 1.502E-08 & 2.886E-08 & 3.266E-08 & 8.577E-09  \\
 Ge-70  & 1.373E-06 & 2.206E-06 & 2.877E-06 & 4.258E-06 & 2.731E-05 & 2.989E-05 & 9.892E-06  \\
 Se-76  & 1.896E-07 & 2.822E-07 & 3.876E-07 & 6.253E-07 & 1.064E-05 & 1.256E-05 & 2.219E-06  \\
 Kr-80  & 6.089E-08 & 9.677E-08 & 1.341E-07 & 2.265E-07 & 5.510E-06 & 6.717E-06 & 8.797E-07  \\
 Kr-82  & 1.509E-07 & 1.475E-07 & 2.034E-07 & 3.903E-07 & 1.492E-05 & 1.860E-05 & 1.980E-06  \\
 Sr-88  & 2.892E-07 & 3.375E-07 & 3.610E-07 & 3.965E-07 & 2.490E-05 & 3.939E-05 & 1.287E-06  \\
 Y-89	& 5.994E-08 & 7.312E-08 & 8.011E-08 & 9.123E-08 & 3.781E-06 & 6.340E-06 & 2.000E-07  \\
 Zr-96  & 2.259E-10 & 1.835E-09 & 3.414E-09 & 6.825E-09 & 5.834E-11 & 5.055E-11 & 1.240E-10  \\
 Te-124 & 1.609E-09 & 1.413E-09 & 1.476E-09 & 1.574E-09 & 8.149E-09 & 1.192E-08 & 1.710E-09  \\
 Xe-130 & 2.056E-09 & 1.823E-09 & 1.724E-09 & 1.612E-09 & 7.133E-09 & 9.925E-09 & 1.762E-09  \\
 Ba-138 & 3.580E-08 & 3.812E-08 & 3.910E-08 & 4.033E-08 & 6.307E-08 & 7.268E-08 & 4.711E-08  \\
\hline
\noalign{\smallskip}
\hline
\end{tabular}
\label{tab:c12c12_ap_summary}
\end{center}
\end{table}


\begin{table}
\begin{center}
\caption{Mass fraction abundances for a sample of selected species in the
He-core ashes (column 2), and in the C shell when $^{12}$C is less than
2 \% for $set3$ cases: the standard case D77
(column 3 - where the $^{12}$C($^{12}$C,n)$^{23}$Mg rate by \cite{dayras:77}
is used), D77t5 (standard rate multiplied by 5, column 4) and D77t10
(standard rate multiplied by 10, column 5).
The average abundance variation of the $s$-only isotopes between iron and strontium
($^{70}$Ge, $^{76}$Se, $^{80,82}$Kr and $^{86,87}$Sr) is provided in the last line of the table for each case considered, normalized to the D77 case (labeled msv$_{\rm s}$ in the table).
}
\begin{tabular}{ccccc}
\hline \hline
 $X_{\rm i}$ & He ashes  & D77 & D77t5 & D77t10  \\
\hline
 C-12   & 2.673E-01 & 1.408E-02 &  1.408E-02   & 1.408E-02    \\
 Ne-20  & 5.026E-03 & 3.554E-01 &  3.532E-01   & 3.504E-01    \\
 Ne-22  & 3.617E-03 & 4.875E-05 &  6.408E-05   & 8.181E-05    \\
 Na-23  & 1.594E-04 & 9.919E-03 &  1.057E-02   & 1.139E-02    \\
 Fe-56  & 2.731E-04 & 1.786E-04 &  1.626E-04   & 1.451E-04    \\
 Zn-70  & 1.040E-09 & 9.198E-08 &  1.069E-07   & 1.267E-07    \\
 Ge-70  & 1.373E-06 & 3.361E-06 &  4.020E-06   & 4.827E-06    \\
 Se-76  & 1.896E-07 & 4.719E-07 &  5.946E-07   & 7.767E-07    \\
 Kr-80  & 6.089E-08 & 7.971E-08 &  1.174E-07   & 1.689E-07    \\
 Kr-82  & 1.509E-07 & 1.839E-07 &  2.583E-07   & 3.882E-07    \\
 Sr-88  & 2.892E-07 & 3.426E-07 &  3.487E-07   & 3.561E-07    \\
 Y-89   & 5.994E-08 & 8.013E-08 &  8.395E-08   & 8.826E-08    \\
 Zr-96  & 2.259E-10 & 7.393E-09 &  1.004E-08   & 1.387E-08    \\
 Te-124 & 1.609E-09 & 1.206E-09 &  1.377E-09   & 1.503E-09    \\
 Xe-130 & 2.056E-09 & 1.405E-09 &  1.390E-09   & 1.376E-09    \\
 Ba-138 & 3.580E-08 & 3.852E-08 &  3.872E-08   & 3.884E-08    \\
msv$_{\rm s}$ & $-$ & 1.000E+00 &  1.299E+00   & 1.772E+00   \\
\hline
\noalign{\smallskip}
\hline
\end{tabular}
\label{tab:c12c12_n_t9p1}
\end{center}
\end{table}


\begin{table}
\begin{center}
\caption{As in Tab. \ref{tab:c12c12_n_t9p1}, but for $set4$ cases, where calculations are based on a realistic C shell trajectory extracted from the M25CF88t10 stellar model. Notice that the case D77t10 is not considered, since the the simulations reach higher temperatures where the nuclear uncertainty of the
$^{12}$C($^{12}$C,n)$^{23}$Mg rate is lower. The symbol * in the labels is used to distinguish from the cases in Tab. \ref{tab:c12c12_n_t9p1}, based on different
conditions.
}
\begin{tabular}{cccc}
\hline \hline
 $X_{\rm i}$ & mixed He and C ashes  & D77*  & D77t5*  \\
\hline
 C-12   & 7.555E-02 & 5.792E-03 &  5.792E-03	\\
 Ne-20  & 2.791E-01 & 3.660E-01 &  3.653E-01	\\
 Ne-22  & 6.701E-04 & 8.997E-05 &  9.567E-05	\\
 Na-23  & 1.003E-02 & 9.882E-03 &  1.016E-02	\\
 Fe-56  & 1.223E-04 & 1.120E-04 &  1.099E-04	\\
 Zn-70  & 5.643E-09 & 8.376E-09 &  9.677E-09	\\
 Ge-70  & 3.650E-06 & 4.156E-06 &  4.266E-06	\\
 Se-76  & 5.652E-07 & 6.438E-07 &  6.648E-07	\\
 Kr-80  & 1.410E-07 & 6.868E-08 &  5.936E-08	\\
 Kr-82  & 5.154E-07 & 4.634E-07 &  4.445E-07	\\
 Sr-88  & 5.162E-07 & 5.724E-07 &  5.843E-07	\\
 Y-89   & 8.285E-08 & 8.156E-08 &  8.047E-08	\\
 Zr-96  & 3.718E-10 & 1.198E-09 &  1.455E-09	\\
 Te-124 & 9.948E-10 & 8.840E-10 &  8.667E-10	\\
 Xe-130 & 1.265E-09 & 1.199E-09 &  1.186E-09	\\
 Ba-138 & 2.925E-08 & 2.999E-08 &  3.015E-08	\\
msv$_{\rm s}$  & $-$ & 1.000E+00 &  9.695E-01  \\
\hline
\noalign{\smallskip}
\hline
\end{tabular}
\label{tab:c12c12_n_trajCF88p10}
\end{center}
\end{table}
\clearpage




\clearpage

%
%


%
%

\end{document}